\documentclass[12pt]{aastex}
\usepackage{psfig}

\begin{document}

%%%%%%%%%%%%%%
%Title Page %
%%%%%%%%%%%%%%

\lefthead{Chakrabarti et al.}
\righthead{Branch, Spur, and Feather Formation in Spiral Galaxies}

\title{Branch, Spur, and Feather Formation in Spiral Galaxies}

\author{
S.Chakrabarti\altaffilmark{1}, 
G. Laughlin\altaffilmark{2},
F.H. Shu\altaffilmark{3}}

\altaffiltext{1}{
Department of Astrophysics, University of California at Berkeley, Mail Code
3411, Berkeley, CA 94720 USA; sukanya@astro.berkeley.edu.}

\altaffiltext{2}{
UCO/Lick Observatory, University of California,  Santa Cruz, CA 95064;
laughlin@ucolick.org}

\altaffiltext{3}{
National Tsing Hua University, 101 Section 2 Kuang Fu Road,
Hsinchu, Taiwan 30013, ROC; shu@astron.berkeley.edu
}

\begin{abstract}

We use hydrodynamical simulations to investigate the response of
geometrically thin, self-gravitating, singular isothermal 
disks of gas to imposed rigidly rotating spiral potentials.  By minimizing reflection-induced
feedback from boundaries, and by restricting our attention to
models where the swing parameter $X \sim 10$, we minimize the swing amplification of
global normal modes even in models where Toomre's $Q_g \sim 1-2$ in the gas disk.
We perform two classes of simulations: short-term ones over a few
galactic revolutions where the background spiral forcing is large, and
long-term ones over many galactic revolutions where the spiral
forcing is considerably smaller.  In both classes of simulations, the 
initial response of the gas disk is smooth and mimics the driving
spiral field.  At late times, many of 
the models evince substructure akin to the so-called branches, spurs,
and feathers observed in real spiral galaxies.  We comment on the parts
played respectively by ultraharmonic resonances, reflection off internal features produced by
nonlinear dredging, and local, transient, gravitational instabilites within spiral arms
in the generation of such features.  Our simulations reinforce
the idea that spiral structure in the gaseous component becomes
increasingly flocculent and disordered with the passage of time, even when
the background population of old disk stars is a grand-design spiral.
We speculate that truly chaotic behavior arises when many overlapping
ultraharmonic resonances develop in reaction to an imposed spiral forcing that has
itself a nonlinear, yet smooth, wave profile.   
\end{abstract}


%%%%%%%
%Body%
%%%%%%%

\section{Introduction}

Nearly all spiral galaxies, even the grand designs in blue light, display
intricate features superimposed on the main spiral arms. Unruly branches,
spurs, and feathers invariably lend a ragged appearance to
the overall spiral structure.  Roberts (1969) showed that
two-armed galactic shocks can arise from the nonlinear response of the
gas to an ordered and steady spiral potential associated with a background of
disk stars.  Shu, Milione, \& Roberts (1973, hereafter SMR)
demonstrated that additional prominent, azimuthally nonsinusoidal, features
could appear as a consequence of ultraharmonic resonances with the two-armed driving potential,
and they suggested that self-gravity, not included in their formal calculations,
may enhance the intrinsically nonlinear response of the
gas.  Their suggestion is particularly attractive after the discovery that many spiral galaxies
possess grand-design spirals in infrared light despite looking quite flocculent
in their gas distributions and Population I stars (Block \& Wainscoat 1991;
Block et al. 1994; Block, Elmegreen, \& Wainscoat 1996).  Block \& coworkers,
however, give a quite different interpretation to their findings, and we shall return
in \S~7 to contrast their views with ours.

Woodward (1975) carried out one-dimensional
time-dependent numerical calculations which studied the azimuthal flow
of gas in both steady as well as time-varying spiral potentials.
He did not include the self-gravity the gas, and he found only the
strongest (4:1) ultraharmonic resonance discussed by SMR.  Moreover, he found that
numerical viscosity inhibited the development of secondary shocks.
Woodward concluded that unless self-gravity of the gas is significant, spiral
substructure could not be explained in terms of ultraharmonic
resonances.  Fortunately, gravitation is universal, and the self-gravity
of the cold component of the interstellar gas cannot be ignored in real disk galaxies
on the scale of the spiral arms. 

Given the dissipative nature of interstellar gas, it is natural to associate
some of the fragmented features
in the overall spiral structure with corotating but sheared disturbances that arise from
local gravitational instability (Goldreich \& Lynden-Bell
1965) or as a steady response to an imposed point mass (Julian \& Toomre 1966).  
Piddington (1973) noted that feathers and
spurs are often seen in combination, as if they are manifestations
of a single phenomenon.
By examining the pitch angles and widths of
spurs and feathers in conjunction with kinematic arguments, Elmegreen (1980) argued 
that they can be identified as density waves.  The detailed
analysis of the morphology of NGC 1566 by Elmegreen \& Elmegreen (1990)
led to the association of several optical features with resonances, including
two ultraharmonic resonances (see also
Visser's 1980 use of SMR's code to identify the long spiral branches seen
between the main spiral arms of M81 with the 4:1 ultraharmonic resonance).
Thus, the substructure that we observe in real galaxies
may be a hybrid phenomenon, i.e., a mixture of the transient
swing-amplification of ``shearing bits and pieces'' (Toomre 1981, 1990, 1991),
especially near the corotation circle,
and nonlinear structures produced by the processes discussed in this paper.   

Balbus \& Cowie (1985) studied the transient gravitational instability
of local quasi-axisymmetric disturbances in the linear regime, but against a background of
nonlinear flow represented by the Robert's (1969) shock solutions
for the steady (non-self-gravitating) response of gas to the forcing of a background spiral potential.
(The linear studies of Goldreich \& Lynden-Bell 1965 and Julian \& Toomre 1966 have
base states that are time-independent and axisymmetric.)
Balbus \& Cowie considered the expanding shear flow as the gas leaves an arm region well-displaced
from the corotation circle (another difference with the earlier studies),
and they proposed a simple modification of Toomre's (1964) $Q_g \rightarrow Q_{\rm sp}$
criterion as applied to gas disks to characterize the onset of local Jeans instability
behind spiral arms.  We return to this modified value of $Q_{\rm sp}$ in our discussion of feathers
in \S~7.

Balbus (1988) extended the study by investigating all wavenumber
directions in the disk plane, and discovered {\it two}
preferred directions for initial wavenumbers for the temporary
growth of gravitational instabilities: roughly parallel and perpendicular to the 
main spiral arm.  He suggested that periodic, closely spaced, spurs may develop in the former,
while the latter favors nonperiodic spur formation.  We shall henceforth
use the name ``feathers'' for both types of features produced by local
gravitational instability of a transient kind.  We reserve
the name ``spurs'' for distinctly leading spiral features, which do {\it not} form
by this mechanism according to our interpretation.  These
spurs protrude from 
the main arms and have shorter azimuthal extent than the ``branches'',
which in our interpretation are due to the ultraharmonic resonances.

Kim \& Ostriker (2002, hereafter KO)
carried out local magnetohydrodynamic (MHD) simulations and found that magnetic fields are
crucial for the formation of feathers (called by them as ``spurs'') when
the Toomre $Q_g$ parameter for the gas
is not small.  The magneto-Jeans instability arises more easily than the purely hydrodynamic
calculations (for which the $Q_{\rm sp}$ criterion works reasonably well)
because magnetic fields destroy the stabilizing effects
of potential-vorticity conservation behind galactic shocks (see also Lynden-Bell 1966
and Elmegreen 1993). The net result is to form a wing of feathers that jut out individually,
more-or-less perpendicularly, on the downstream side of the shock front at the main spiral arms.

In this paper, we carry out two-dimensional, global, hydrodynamic simulations that
incorporate the self-gravity of the gas.  We work within the context
of the singular isothermal disk (SID) to understand the
dynamical formation of spiral substructure in self-gravitating, purely
hydrodynamical systems.  However, in our physical interpretations,
we shall identify the formation of feathers in low Toomre-$Q_g$ systems as the
equivalent (inside spiral arms) of an MHD system with high effective $Q_g$ values.  This is important because we shall find that low $Q_g$ gas-disks with
relatively high forcing cannot be evolved for
many galactic rotations in hydrodynamic simulations
without developing catastrophic increases in the surface
density in many regions.  In high $Q_g$ disks,
we can carry the hydrodynamic simulations forward for the much longer times that are necessary
to accumulate resonant influences and to see the side-effects of nonlinear dredging.
The result is a much richer variety of nonlinear substructure, including spurs and hints of
chaotic behavior akin to flocculence -- but no feathers.  Thus, we speculate that
more complete, whole-disk, MHD simulations of moderately high effective $Q_g$ gas disks
will find branches, spurs, and feathers as possible substructures within a single
self-consistent simulation, with flocculence a possible end-product if the system
becomes chaotic through the effects of overlapping ultraharmonic resonances.

The dynamics of SIDs, with their self-similar surface density
($\Sigma \propto 1/\varpi$) and flat rotation curves ($v_{\rm rot}={\rm constant}$)
have been intensively studied in the forty years following
Mestel's (1963) groundbreaking article. They are reasonable analogs for disk galaxies,
while their self-similarity and overall simplicity lend them to detailed 
analytic and semi-analytic investigations. The properties of SIDs in the context 
of spiral galaxies have been reviewed and explored by Shu et al (2000, hereafter S00),  
who studied their linear stability properties, both in the context of their 
susceptibility to nonaxisymmetric bifurcations, as well as their ability 
to promote swing over-reflection of wave trains impinging on the corotation
circle. In the current work, we extend the S00 analysis to investigate the 
nonlinear response of gaseous SID structures to spiral gravitational 
perturbations arising from an imposed rigidly rotating spiral pattern 
present in the stellar component of a disk galaxy. 

\section{Partial SIDs}

A massive spiral disturbance sliding through unperturbed gas is
bound to cause complicated unrest.  Numerical simulations are thus an 
ideal tool for investigating
the nonlinear response of a self-gravitating,
differentially rotating, razor-thin, gas disk to a steady spiral
forcing potential. The simulations presented here adopt
a so-called partial SID as the equilibrium
reference state. We define the partial fraction of the disk, $F$, as the
amount of mass in the gas disk, with the remainder, $1-F$, approximated as a rigid
stellar component.  Our imposed two-armed trailing-spiral potential, rotating uniformly
at a pattern angular speed $\Omega_p$,
can be thought of as arising from this stellar component. Therefore, in
our gas-dynamical models, the stellar disk manifests itself both through
a time-independent, axisymmetric, gravitational field, and through the non-axisymmetric
perturbing potential. One is equally free to imagine that the
radial gravitational field arises from the more realistic combination of
a true stellar disk and a nonreactive, spherically symmetric, dark-matter halo, 
whereas the spiral forcing arises from a normal-mode structure
endemic to the true stellar disk.  In the latter, more realistic, scenario,
the spiral forcing may have relatively small amplitude relative to
the total axisymmetric, radial gravitational field, and yet represent
a nonlinear perturbation in the infrared surface brightness of the true
stellar disk.  Thus, we are motivated to consider both linear and nonlinear
wave profiles for the planform of the spiral force field.

\subsection{Basic Equations}

In polar coordinates $(\varpi, \phi)$ and time $t$, we denote the mass
per unit area of the gas and stellar components of a completely flattened distribution of matter
as $\Sigma_{g}(\varpi,\phi,t)$ and $\Sigma_{*}(\varpi, \phi, t)$ respectively. 
We model both as isothermal fluids, with the constant stellar dispersive speed being
$c_*$ and the gaseous isothermal sound speed being $c_g$. We denote $u$ and $j$ with the appropriate
subscripts as the $\varpi$ component of the fluid velocity and the $z$ component
of the specific angular momentum.   With these definitions, the equations of
continuity and momentum for the gas are
written
\begin{equation}
\partial\Sigma_{g}/\partial t+\frac{1}{\varpi}\frac{\partial(\varpi\Sigma_{g}{u_g})}
{\partial \varpi}+\frac{1}{\varpi^{2}}\frac{\partial(\Sigma_{g} j_g)}{\partial\phi} =0 \, ,
\end{equation}
\begin{equation}
\partial u_g/\partial t + u_g\frac{\partial u_g}{\partial\varpi}
+\frac{j_g}{\varpi^{2}}\frac{\partial u_g}{\partial\phi} 
-\frac{j_g^{2}}{\varpi^{3}}=
-\frac{c_g^2}{\Sigma_{g}}\frac{\partial\Sigma_{g}}{\partial\varpi} 
-\frac{\partial {\cal V}}{\partial\varpi} \, ,
\end{equation}
\begin{equation}
\partial j_g/\partial t + u_g\frac{\partial j_g}{\partial\varpi} 
+\frac{j_g}{\varpi^{2}}\frac{\partial j_g}{\partial\phi}=
-\frac{c_g^2}{\Sigma_{g}}\frac{\partial\Sigma_{g}}{\partial\phi}
-\frac{\partial {\cal V}}{\partial\phi} \, .
\end{equation}
A similar set holds for the disk of stars if we were to allow them
actively to respond to the collective gravitational field.  For the purposes of the
present paper, however, we consider both the axisymmetric and spiral
distributions of the stellar component to be rigidly given by
external considerations. 

In equation (4), ${\cal V}$ is the combined gravitational potential
of the gas and the stellar component.  It is given by Poisson's integral:
\begin{equation}
{\cal V}(\varpi,\phi,t)= -G \oint d\psi \int_{0}^{\infty} \frac{\Sigma(r,\psi,t)\, rdr}
{[r^{2}+\varpi^{2}-2r\varpi\cos(\psi-\phi)]^{1/2}} \, ,
\end{equation}
where $\Sigma(r,\psi,t)$ is the sum of the gaseous 
and stellar surface densities at the source point in the disk.

According to S00, the surface density, rotation angular velocity, and epicyclic frequency,
of a full, single component, axisymmetric SID
have the following properties:
\begin{equation}
\Sigma =\frac{c^{2}}{2\pi G\varpi}(1+D^{2}) \, ,
\end{equation}
\begin{equation}
\Omega=\frac{cD}{\varpi} \, ,
\end{equation}
\begin{equation}
\kappa=\sqrt{2}\frac{cD}{\varpi} \, ,
\end{equation}
where $c$ is a dispersive velocity or isothermal sound speed, and $D$
is a dimensionless rotation parameter.  S00's formalism
is easily extended to two partial disks in which
we attribute the fraction $(1-F)$ of the full gravity in the axisymmetric state to a
rigid stellar component and another fraction $F$ to an active gaseous disk: 
\begin{equation}
\Sigma_{*}=\frac{(1-F)c^2(1+D^2)}{2\pi G\varpi} , \qquad
\Sigma_{g}=\frac{Fc^{2}(1+D^{2})}{2\pi G\varpi} .
\end{equation}
The coefficients in the above relations
are chosen so that the surface density (and gravitational field)
remains the same as for the full disk in equation (5).

We denote the angular velocity of the stars and gas, respectively, by
$\Omega_* = cD_*/\varpi$ and $\Omega_g = cD_g/\varpi$, with associated
epicyclic frequencies that are $\surd 2$ larger.  Then
radial force balance for the stellar and gas disks in their equilibrium states
can be expressed by
\begin{equation}
c_*^2+c^2D_*^2 = c_g^2+c^2D_g^2 = c^2(1+D^2).
\end{equation}
Since the full disk has no independent meaning, we are free to
choose $c \equiv c_*$ and $D \equiv D_*$.  Expressing $c_g$ as a fraction
$f$ of $c_*$ now, we get from the above, with $c_g = fc_*$, that
\begin{equation}
D_g^2 = 1+D_*^2-f^2.
\end{equation}

Since the rotation and epicyclic frequencies of the gas disk are
$\Omega_{g}=c_{*}D_{g}/{\varpi}$ and $\kappa_{g}=\sqrt{2}\Omega_{g}$,
Toomre's (1964) axisymmetric-stability parameter, $Q_{g}$ for the gas becomes
\begin{equation}
Q_{g} \equiv \frac{\kappa_{g}c_{g}}{\pi G\Sigma_{g}} 
=Q_{*}(D_{g}/D_{*})\frac{(1-F)f}{F} \, ,
\end{equation}
whereas the corresponding value for the dynamically inactive $(1-F)$ fraction of the
disk reads,
\begin{equation}
Q_{*} \equiv \frac{\kappa_{*}c_{*}}{\pi G\Sigma_{*}}
=2\sqrt{2}\frac{D_{*}}{(1-F)(1+D_{*}^{2})} \, .
\end{equation}

For comparison, the full disk has an associated $Q$:
\begin{equation}
Q=2\sqrt{2}\frac{D_*}{1+D_*^{2}}.
\end{equation}
Thus, for a full SID, $Q$ reaches a maximum of 
$\sqrt{2}$ at a model $D_*=1$ and decreases on either side to unity
for models where $D_*=\sqrt{2}\pm 1$.  These values of $D_*$
do not corrrespond to very rapidly rotating fiducial systems,
so one might think Toomre's asymptotic stability analysis that depends
on $D_* \gg 1$ might be called into question.  The situation can be partially saved by invoking
a rigid massive dark-matter halo to provide some of the radial gravity
attributed above to a flat disk of stars.  But it also turns out
that an exact stability analysis, even for a full disk, yields remarkable agreement
with the Toomre criterion, insofar as its predictions
concerning fragmentation into rings go (see S00).  On the other hand,
overall radial collapse occurs in a fundamental way because a disk rotates too
slowly (small $D_*$), so it cannot be well described by looking only
at the $Q$ parameter.  In \S~2.2 therefore,
we re-examine the question of the radial collapse of a partial gas disk
from the formal perspective of an exact stability analysis (but keeping
the ``stellar'' disk fixed).

In any case, $Q_g$ yields an accurate measure of the response of the
active gas to less-than-galactic scale perturbations, whether axisymmetric or
nonaxisymmetric.  Since the partial gas disk derives 
much of its support from the rigid stellar component (which can include
here a massive dark-matter halo), its 
$Q_g$ surpasses that of the full disk unless $f$ is too small or
$F$ is too  close to unity.  Indeed, it is
well known that the addition of a such a rigid component or
halo will serve to stabilize the response of the active component of a disk
(e.g. Ostriker \& Peebles 1973). Increasing the fraction of
support from the rigid halo also increases the ratio 
\begin{equation}
X={2\over{m}}{D_{g}^{2}\over{F(1+D_{*}^{2})}} \, ,
\end{equation}
of the azimuthal wavelength to the critical wavelength at which ring
fragmentation first occurs. For $X>3$, the swing amplification
mechanism (Toomre 1981) loses its effectiveness.  In the partial-disk
evolutionary simulations presented in this paper, we adopt $F=0.1$, leading to values
of the swing parameter 
$X\sim 10$.  The swing amplification of corotating disturbances
is therefore effectively minimized in our simulation models.

\subsection{Stability of Partial Disk}

The Toomre-$Q$ values for a full (as given by equation 13) and a partial SID (as given by equation 11) as
functions of $D_{\star}$ are depicted in Figure 1.
The full and partial SIDs shown in this figure correspond, respectively, to
$F=f=1$ and $F=f=0.1$.  Note that since the Q-values for SIDs 
(both full and partial) are constrained by the rotation parameter $D_{\star}$,
it is not possible to get arbitrarily large values for $Q$.

As we have already remarked,
Toomre's stability analysis gives only asymptotically accurate results for a single active component.
For completeness, we repeat the exact
analysis of S00 for a partial disk when it is the only active component
in the system.  The more complex, linear stability analysis needed when both the
gas and star disks are active has been treated by Lou \& Shen (2003).  We refer the
interested reader to their paper for details.

When an active gas disk is perturbed by a small disturbance, we can linearize
the fluid equations (1) - (3) around the basic underlying state.  We
can then look for perturbation solutions that are periodic in $t$ and 
$\phi$, e.g.
\begin{equation}
\Sigma_1 = S(\varpi) e^{im(\omega t-m\varphi)} \, ,
\end{equation}
and similarly for $u_{1}$, $j_{1}$, and ${\cal V}_{1}$. 
The pattern speed of the disturbance is given
by $\Omega_{p} = \omega/m$.
Upon substituting these
trial solutions into the linearized fluid equations, and eliminating $u_{1}$
and $j_{1}$, one arrives at a second-order integro-differential equation
which describes a global, self-consistent, self-gravitating, spiral density 
perturbation of the disk
(Lin \& Lau 1979):
$$
(\omega-m\Omega) S + {1\over \varpi}{d\over d\varpi}\left\{
{\varpi \Sigma_g \over (\omega-m\Omega)^2-\kappa^2}\left[
-2\Omega {m\over \varpi}+(\omega-m\Omega){d\over d\varpi}\right]\Phi\right\}
$$
\begin{equation}
-{m\Sigma_0\over \varpi [(\omega-m\Omega)^2-\kappa^2]}\left[
(\omega-m\Omega){m\over \varpi}-{\kappa^2\over 2\Omega}{d\over d\varpi}
\right] \Phi = 0.
\end{equation}

If we set $\omega=0$ in order to find the marginal conditions for the swing over-reflection
of spiral density waves when $m\neq 0$, or for axisymmetric collapse and fragmentation into rings
when $m=0$ (see S00),
substitution of  the reference state values for the partial SID for the surface density, angular
and epicyclic frequencies into the above equation leads to:
\begin{equation}
\left[ -S + {F \over{ D_{g}^{2} (m^2-2) }} \left({ m^{2} \over{\varpi}
} -
2{d\over{d \varpi}} - \varpi{d^{2} \over{d \varpi^{2}}} \right)
\left( {f^{2} \varpi \over{F}}S + {(D_{g}^{2}+f^{2}) \over {2 \pi G}} V
\right) \right]=0 \, 
\end{equation}
where $V$ and $S$ are related through the linearized 
version of Poisson's equation
\begin{equation}
V (\varpi) = -G\oint d\chi \int_0^\infty
{S(r) \cos(m\chi)\,rdr \over [r^2+\varpi^2-2r\varpi\cos\chi]^{1/2}}.
\end{equation}
The above two equations admit scale-free solutions of the form
(Syer \& Tremaine 1996, Lynden-Bell \& Lemos 1993)
\begin{equation}
S(\varpi) = s \varpi^{-3/2} e^{i\alpha \ln \varpi} \, 
\end{equation}
\begin{equation}
V(\varpi) = v \varpi^{-1/2} e^{i\alpha \ln \varpi} \, 
\end{equation}
with $s$, $v$, and $\alpha$ equal to constants. The pitch angle $i$ of such logarithmic spirals
is a constant and given by the formula $\tan i = m/\alpha$. 
While the perturbation solutions are scale free, they have
radial amplitudes which decline more rapidly with $\varpi$ than
the underlying equilibrium. 
Hence, all but infinitesimally
small disturbances of the form given by equations (19) and (20) will
achieve arbitrary amplitudes sufficiently
near the origin, leading to a formal breakdown of the perturbative
analysis. This breakdown is generally averted by cutting out a hole
in the center of the disk to represent an inactive
central bulge (Zang 1976, Evans \& Read 1998).

Substituting the scale-free solutions into equations (17) and (18), we obtain the 
conditions of marginal stability from the solutions of
\begin{equation}
-1+ {1 \over { D_{g}^{2} ( m^{2} -2 )}} \left( m^{2} + \alpha^{2} + {1
\over{4}}
\right) \left( f^{2} - F (f^{2}+D_{g}^{2}) {\cal N}_{m}(\alpha)
\right)=0\ \, .
\end{equation}
Rearranging, we find
\begin{equation}
D_{g}^{2}= { (m^{2} + \alpha^{2} + {1\over{4}})(F f^{2} {\cal
N}_{m}(\alpha) - 1)
\over{2 - m^{2} - (m^{2} + \alpha^{2} + {1\over{4}})F {\cal N}_{m}(\alpha)
}} \, ,
\end{equation}
where ${\cal N}_{m}(\alpha)$ is the Kalnajs (1971)
proportionality
relation for the logarithmic spiral potential-density pairs:
\begin{equation}
v=-2\pi G {\cal N}_{m}(\alpha)s \, ,
\end{equation}
with
\begin{equation}
{\cal N}_{m}= {1\over{2}} { \Gamma [(m + 1/2 + i\alpha)/2] \Gamma [(
m + 1/2 - i\alpha)/2] \over {
\Gamma [(m + 3/2 + i\alpha)/2] \Gamma [(m + 3/2 - i\alpha)/2] } }  \, .
\end{equation}

The curves of marginal axisymmetric
stability are then calculated by setting $m=0$, 
and choosing values for $f$ and $F$. Figure 2a shows cases where
$F=f$, in increments of 0.1 from $F=f=0.1$ to $F=f=0.9$.
For reference, the full disk
branches with $F=f=1.0$ are also shown. As the stellar and dark-matter components
of the potential 
become increasingly dominant within the sequence of partial SIDs, the 
axisymmetric collapse branch disappears rapidly.  Thus, we only have
to worry about the fragmentation branch for all practical purposes,
and the condition of marginal stability for that branch is quite accurately
approximated by Toomre's $Q_g$ criterion because
increasingly wavy planforms are required
to trigger such instabilities in low $F$ systems.  We also show the self-consistency curves in Figure 2b for $m=2$ disturbances in increments of 0.1 from 
$F=f=0.1$ to $F=f=1.0$.  Our high $Q_{g}$ 
disks (with associated rotation parameters cited in \S ~5) will not support zero frequency
spiral waves of any $\alpha$.  However, our low $Q_{g}$ disks will
support $m=2$ disturbances in the range of $\alpha=0-2$.  As described
in \S ~3, we consider two types of spiral planforms in our
simulations, namely, the linear logarithmic spiral planform,
and the nonlinear SYL planform.  The logarithmic
spiral planform has associated $\alpha=6.15$, which lies
outside the range that would be swing amplified even if it were
to propagate into the central regions undamped with similar
wavenumber.  However, since the SYL
planform does have contributions from wavenumbers in the range of 
$\alpha=0-2$, these wavenumbers could have been amplified had
they penetrated and reflected from the central regions.  We effectively
reduce feedback from the potential amplification of disturbances with these wavenumbers by our
implementation of sponge boundary conditions as described in \S ~5.   Thus, we have minimized SWING amplification not only as described by
Toomre's {\it local} $X$ criterion, but also {\it global} amplification which
would have been possible had we allowed disturbances with suspect
wavenumbers (those that are prone to
amplification) to travel unimpeded into the central regions.

\section{Model Parameters}

We have performed a number of numerical simulations to investigate
the disk gas response to an imposed spiral field.  In these simulations,
we denote the strength of the spiral forcing as $\mathcal{F}$.  
This is the maximum amplitude of the perturbing spiral 
field as a fraction of the equilibrium
centrifugal acceleration.  The dimensionless model parameters
for the short-term and long term simulations are summarized in Table 1
and 2 respectively.  
Column 1 labels each run, and Column 2 gives the
$Q_{g}$ for the unperturbed gas disk.
Column 3 lists the partial disk fraction $F$. 
We adopt $F=f$ in all of the runs. Column 4 lists the values of the
spiral forcing, $\mathcal{F}$, as a percentage of $v_{\rm rot}^{2}/\varpi$. 
The quantity $c_{g}/v_{\rm rot}$ is expressed as a percentage in column 5.
We fix the geometric planform of the imposed spiral forcing by one
of the two fixed methods discussed below.
For each planform, then, only the four parameters listed in columns
2-5 are varied for our models over the restricted range listed in Table 1.
We note, however, that the quantity $Q_g \equiv \kappa_gc_g/\pi G\Sigma_g$ is uniform only at the
initial instant of time.  Later, it varies spatially because of the induced spiral
structure, and  temporally because of the nonlinear 
dredging of $\Sigma_g$ produced by systematic radial drifts in the presence
of galactic shocks.  Henceforth, when we cite values for $Q_g$,
we are referring to the value for the unperturbed gas disk.

Models H1-Hn are steadily forced by a two-armed logarithmic spiral, with
pitch angle $i=18^\circ$.  Model HSYL is steadily
forced by a two-armed spiral planform adopted from 
Shu, Yuan, \& Lissauer (1985; hereafter SYL). These authors
reported the development of an inviscid theory for nonlinear, self-gravitating, ``long,'' spiral
density waves that utilizes the asymptotic assumption of tightly wound
disturbances, but which allows for large
amplitude density waves in the true stellar disk with $\Sigma_{*1}/\Sigma_{*0}\sim1$. 
After employing this ordering, they derived a nonlinear integral equation
which describes self-consistent large-amplitude spiral disturbances
in the underlying equilibrium disk of ``stars.'' In the linear regime, their
integral equation can be rigorously reduced to an ordinary
differential equation for the planform of the disturbance.

SYL derived a nonlinear dispersion relation and an angular momentum conservation
relation in the far wave zone where the WKBJ approximation is valid. Since
a direct attack on the full integral equation did not yield a solution,
SYL invented a heuristic ordinary differential equation (ODE) that
reduces to the linear limit when forcing amplitudes are small, and
which produces the proper nonlinear dispersion and amplitude relations
in the far wave zones. The ODE pragmatically captures
the essential properties of the intractable integral equation.
SYL further demonstrated that when numerical solutions of their ODE are
back-substituted into the full integral equation, the equation is satisfied
fairly accurately, even in the region of resonant coupling near the inner
Lindblad resonance. 
We obtain our so-called ``SYL profile'' for the
rigidly rotating disturbance in the ``stellar'' component of
our partial SIDs by numerically integrating SYL85's heuristic ODE (SYL85.82)
Gray-scale images of the resulting spiral planform are shown in Figure 3,
with the x and y axes in terms of the radial coordinate
$\xi$.

The SYL profile is a more open spiral than the logarithmic spiral because
it is formally derived for ``long'' waves (relevant for the Saturn's
rings application of interest to SYL), whereas the pitch angle of
$i$ = 18$^\circ$ has been chosen arbitrarily to correspond more to the ``short''
waves that are believed to dominate the structure of most normal-spiral galaxies.
But this is not the most important difference between the two planforms in the present context.
The wavy logarithmic radial profile is combined with an angular dependence
(see eqs. [15] and [19]) that makes it a pure sinusoid in the azimuthal direction.
The same is not true of the SYL planform; by construction,
it contains angular harmonics above the fundamental (SYL).  Thus, our
use of both types of planforms allows us to contrast nonlinear response
due to linear forcing (as in the case of the logarithmic planform),
and nonlinear response due to nonlinear forcing (as in the case of the
SYL planform).  We will discuss the relevance of the nonlinear SYL planform
to ultraharmonic resonance phenomena in Section 4.

The self-similarity of SIDs implies that physical scales are fixed only after
we specify dimensional values for the rotation speed $v_{\rm rot}$ of the gas
and the pattern speed $\Omega_p$ for the imposed spiral forcing.
For definiteness, we choose $v_{\rm rot}=$246 km s$^{-1}$, and $\Omega_{p}$ equal to 21.5 and 
11.5 km $\mbox{s}^{-1}$ $\mbox{kpc}^{-1}$, respectively,
for the $Q_{g}$=1.3 and 2.48 disks. (Note: 11.5 km $\mbox{s}^{-1}$ $\mbox{kpc}^{-1}$ is the pattern speed
recommended by Lin, Yuan, \& Shu 1969 for the spiral structure of our own Galaxy.) 
For $F=f=0.1$, we then get
$\Sigma_{g}$ = 22 $M_{\odot}$ pc$^{-2}$ (10 kpc/$\varpi$).
To be representative of published observational data
(e.g., Wong \& Blitz 2002,
and Helfer et al 2003), a coefficient for $\Sigma_g$ half as large might have been better.  
This suggests that we should have chosen $F=f=0.05$.
The $F=0.1$ ratio applies as a rough summary of observed gas to observed stars in many
galaxies.  But we are really comparing gas to stars plus dark-matter halo.
The latter, projected into a disk, typically equals the contribution of the observed stars
in supporting the rotation curve within the optical spiral structure.
Thus, we might have more realistically set $F=f=0.05$.
With the latter choice, we get $c_g=$ 5.6 km s$^{-1}$ and 11 km s$^{-1}$, repectively,
for the $Q_g$ = 1.3 and 2.48 disks of Table 1.
These values are roughly compatible with the 1D velocity dispersion $\sim 7$ km s$^{-1}$
of H I clouds (Heiles 2001), while the random motions of molecular clouds are somewhat smaller.
However, we used the higher values of gas surface density and sound speeds,
with the same mean stability measure $Q_g$,
to hasten the development of self-gravitating substructure in the simulations.
Alternatively, we may imagine that these faster time scales refer to
an earlier epoch in the universe when galaxies were more gas rich than at the present epoch.
For future detailed comparisons of time scales and length scales with observed features
in nearby galaxies, we would recommend pairs more like $F = f = 0.05$ rather than the adopted $F=f=0.1$
of this paper.
 
With the above understanding, the short-term (H-series) simulations were
run for a total of 2.5 revolutions at 10 kpc (637 Myr), and the long-term
(L-series) simulations were run for a total of 15 revolutions (3.8 billion years).
L1, L2, and L3 are forced by the $i=18^{\rm o}$ logarithmic spiral, whereas
the long-term simulations LSYL1, LSYL2, and LSYL3 are forced by the
spiral planform adopted from SYL. 

\section{Ultraharmonic Resonances:  Resonance Conditions}

We follow the slightly nonlinear part of the analysis of SMR to
locate the positions of resonances.  In such an analysis, SMR
found the $n$-th harmonic response to an $m$-armed logarithmic spiral
to formally diverge when
\begin{equation}
{{\Omega_p-\Omega}\over \kappa} = \pm {1\over m}\sqrt{n^{-2}+ x},
\end{equation}
where
\begin{equation}
x\equiv m^2c_{g}^{2}/\varpi^2\kappa^{2}\sin^2 i.
\end{equation}

For the zero pressure case ($x=0$), these resonances
occur where the gas meets the $m$-armed stellar wave at $1/n$ times the epicyclic 
frequency.  A nonzero sound speed for the gas shifts the
radial position of formal resonance.  The case $n$ = 1
corresponds to the inner and outer Lindblad resonances (for $\Omega$ greater than
and less than $\Omega_p$, respectively).  The case $n$ = 2 is the
first ultrahamonic, etc.  The first ultraharmonic lies closest
in radial distance to the Lindblad resonance; as $n$ is increased beyond 2.
higher ultraharmonic resonances approach closer to the corotation radius
where $\Omega = \Omega_p$.

The prevailing nomenclature in the literature
(not one that we particularly like) of the ``4:1
resonance'' refers to case $m=2$ and $n=2$.  
The 4 refers to $mn$ (this combination appears togther in eq.~[25] if $x=0$)
and the 1 refers to the forcing by
a pure sinusoid, the only case considered by SMR.
Whenever the forcing waveform itself contains a $j$-th harmonic above the fundamental,
however, resonances are possible where $mj$ replaces
$m$ in equation (25), with $m$, $n$, and $j$ all positive integers.
The SYL planform allows for such additional resonances,
since by construction it contains a full Fourier decomposition
of forcing harmonics above the fundamental (SYL).  Thus, for a given
forcing amplitude relative to the axisymmetric gravitational field,
the SYL planform is nonlinear whereas the
logarithmic spiral planform is linear.  Our use of the SYL planform
allows us examine the nonlinear response of the gas due to
a nonlinear forcing.  

At small amplitudes, resonances are sharply located at unique radii.
However, at finite forcing amplitudes, resonances acquire finite widths
(Artymowicz \& Lubow 1992), and resonance overlap can take place (see Figure 9 of SMR).
In particular, SMR found that (their version of) the Robert's (1969) equations failed
to yield steady solutions at forcing amplitudes slightly larger than when
resonance overlap occurs, but they failed to realize that this might signal the
onset of chaos (because chaos theory was not well developed then).
We note that the nonlinear nature of the SYL forcing planform
allows for potentially more opportunities of resonance overlap,
and therefore a greater tendency to chaotic behavior.  We speculate that
such a condition could lead to greater flocculence in the resulting gaseous
spiral structure.

At finite forcing amplitudes, SMR found a relatively extended
radial region of secondary compression corresponding
to the $n = 2$ ultraharmonic resonance.  This bifurcates the main ($m=2$)
spiral arms, which makes the 4:1 resonance
especially suitable for producing branches.  The more restricted spatial
response of the higher ultraharmonic resonances suggested
that they would generate only relatively
short spurs or feathers.  SMR suggested that the inclusion
of the self-gravity of the gas may enhance the formation of
spiral substructure via these resonances.  One of the primary
objectives of this paper is to test this suggestion by SMR.
 
We obtain the resonance radii (for linear forcing) for our models from equation (27) and present 
them in Tables 3 and 4.  In particular, these resonance
radii are the same for models H1-H3, L1, and LSYL1 
i.e., the low-$Q$ models.  They are also
the same for models H4-H6, and L2-LSYL3, i.e., the high-Q models.  

\section{Numerical Procedure}

Our numerical simulations use a two-dimensional hydrodynamics code
based on the second-order van Leer advection scheme 
described by Stone \& Norman (1992). We use a grid of 256 evenly
spaced azimuthal zones and 256 logarithmically spaced radial
zones which run from an inner radius $\xi_{\rm in}=0.2$ to an 
outer radius $\xi_{\rm out}=5.0$.  The calculations are performed in the non-rotating frame.
We have re-run several of our
simulations with 512x512 zones and find no significant changes in
the resulting dynamics unless highly chaotic nonlinear phases are reached.  The simulations are reported in terms of a radial coordinate $\xi$, in which $\xi$=1
is understood to correspond to 10 Kpc.

The gravitational potential ${\cal V}_g$ arising from the gas within
the computational domain is computed using the cylindrical grid
FFT algorithm described in Binney \& Tremaine (1987). We do
not apply an artificial viscosity. We find that when shocks develop
in the simulations, the intrinsic viscosity of the numerical method
smooths out post-shock oscillations.

In our numerical computations, we work in a system of units
with $G=1$.  In these units, all of our equilibrium models have constant $v_{\rm rot}=0.16$,
and $\Sigma_{g}=0.0004 / \xi$. The isothermal gas sound speed, $c_{g}$
is taken to be either $c_{g}=0.046 \, v_{\rm rot}$, which yields $Q_{g}=1.3$,
or, alternately, $c_{g}=0.088 \, v_{\rm rot}$, which yields $Q_{g}=2.48$
(see Tables 1 and 2).  The stellar and gas rotational parameters, $D_{\star}$ and $D_{g}$ for the $Q_{g}=2.48$ disk are 0.53 and 1.13 respectively.  For the $Q_{g}=1.3$ disk, $D_{\star}$ and $D_{g}$ are 1.89 and 2.13 respectively.    To convert to physical coordinates, we take 10 kpc to correspond to $\xi=1$ and one revolution at that distance to 
correspond to 250 Myr.  

The initial equilibrium is established by 
balancing the analytically prescribed centrifugal force $v_{\rm rot}^{2}/\xi$, 
the pressure gradient $-dP/d\xi$, and the FFT-estimated gravitational force
$-dV_{\rm gas disk}/d\xi$, with an additional radial force, $F_{\rm ext}$,
\begin{equation}
F_{\rm ext}(\xi)=\frac{v_{\rm rot}^{2}}{\xi}+\frac{dP}{d\xi}+\frac{dV_{\rm gas disk}}{d\xi} \, ,
\end{equation}
that is subsequently maintained at a constant value for the entire simulation.
$F_{\rm ext}(\xi)$ is understood to arise from (1) the potential gradient due to
the underlying axisymmetric stellar disk component, and (2) additional radial 
force to account for the gravitational attraction of gas interior and exterior
to the computational grid, and for the systematic softening introduced
by the FFT gravity solver.

As shown by Zang (1976, and summarized by Toomre 1977), the imposition
of a sharp cut-out at the inner disk edge causes reflection of incoming trailing
wave trains into leading wave trains and the possible excitation of unstable global modes.
We thus strive to minimize radial reflection by implementing sponge boundary
conditions at the radial edges of the computational domain. In the
${n_{d}}_{i}=18$ radial zones interior to $\xi=0.25$, and in the 
${n_{d}}_{o}=4$ radial zones exterior to $\xi=4.7$, we smoothly impose
an admixture of the equilibrium solution $x_{\rm equil}(\xi)$ (derived from
equations 5-7) into the hydrodynamically computed variables $x_{\rm hydro}(\xi)$.
That is, at the inner edge we have,
\begin{equation}
x(\xi_{j})=({j-1 \over{ {n_{d}}_{i}-1 }}) x_{\rm hydro}(\xi_{j}) +
({ {{n_{d}}_{i}-j} \over{ {n_{d}}_{i}-1 }}) x_{\rm equil}(\xi_{j}) \, .
\end{equation}
This damping is applied at a cadence ${t_{d}}_{i}=\Delta\xi({n_{d}}_{i})/c_{g}$
at the inner edge, and ${t_{d}}_{o}=\Delta\xi({n_{d}}_{o})/c_{g}$ at the outer
edge, where $\Delta\xi(j)$ is the radial width of zone $j$. This cadencing
was not used in S00, where wave packet simulations were run for much shorter
times to study over-reflection at the co-rotation radius.

The effectiveness of the radial boundary conditions and 
the minimization of SWING amplification is shown in Figure 4a and Figure 4b
which chart the progress of a simulation in which the $Q_{g}=1.3$
and $Q_{g}=2.48$
equilibrium disks are perturbed with a transient leading spiral potential
(which is shut off after time $t=t_{0}$).

\begin{equation}
{\cal V}_{1_{\rm ext}}(\xi, \varphi,t)=v_{o} \, {\rm ln}({\xi \over{10}})
e^{-(\xi-\xi_{0})^2/h^2}\sin{( \pi t / t_{0} )}
e^{i(a {\rm ln}\xi + \omega t -m \varphi)} \, ,
\end{equation}
with $v_{0}=7\times10^{-5}$, $\xi_{0}=2.5$, $h=0.5$, $t_{0}=78.4.$, and $a=6.0$.
The five panels on the left show the real and imaginary components of the
$m=2$ response to the the transient perturbation (scaled by $\xi^{1/2}$).
Time increases from bottom to top. The entire sequence covers $n \approx 11$
full rotations at $\xi=1$.  The bottom two plots show the
disturbance in the disk while the perturbation is still active.  In the
$Q_{g}=1.3$ disk, the leading spiral propagates radially in both
directions, but
is only minimally amplified relative to the initial disturbance, and is largely absorbed,
as desired, at the grid edges (see S00's study of overreflection of leading
disturbances in $Q_{g} \sim 1.3$ disks, which do show significant amplification, as a comparative example).  In the $Q_{g}=2.48$ disk, the initial
disturbance undergoes even less amplification than in the $Q_{g}=1.3$ disk,
and is largely absorbed at the edges.  However, we begin to see a minimal amount of
numerical noise at the inner boundary after about 9 revolutions.  Note that 9 revolutions at $\xi=1.0$ corresponds to
$n \approx 45$ revolutions at the inner boundary.  Thus, signals
at the inner boundary have been transmitted more frequently than
at $\xi=1.0$.  Since the numerical noise is more evident
at the inner boundary than at the outer boundary, and more apparent
in the $Q_{g}=2.48$ disk relative to the $Q_{g}=1.3$ disk, we 
hypothesize that this numerical artifact scales with the
frequency of signal transmissions, which is higher
both at the inner boundary and in the $Q_{g}=2.48$ disk (due to its faster
signal speed).  This numerical noise 
saturates at the levels seen in Figure 4a and 4b for the $Q_{g}=1.3$ disk
and $Q_{g}=2.48$ disks, respectively.  Simulations run out to 
$n \approx 15$ revolutions at $\xi=1.0$ show no increase in the level
of noise at the inner boundary, but a continued presence of 
this numerical artifact, which is slightly more heightened in
the $Q_{g}=2.48$ disk.

\section{Results}

Our results divide naturally into two categories: the short-term
evolution (over several hundred Myr) of the models in Table 1,
and the long-term evolution (over a few Gyr) of the models in Table 2.  
We will use the terms branches, spurs,
and feathers to describe the substructure that forms
during the simulations.  Our use of these terms is primarily visually
motivated, but knowing the conditions in the numerical simulations that
give rise to these features allows us to give the physical interpretations
for their origin mentioned in \S~1. 
By branch formation visually, we refer to the emergence of arm-like
structure that spans an azimuthal range comparable to the spiral
arm itself.  Branches wind in the same sense as the main arms, and
appear as a bifurcation of the main spiral arms.    
By spur formation, we refer to the appearance of structures
protruding from the arms that wind in the sense opposite to that of the
main arms, i.e., spurs are leading structures.  These spurs
are often short and stubby in appearance, as in the short-term
simulations, and generally shorter in azimuthal extent
than the branches.  Feathers look like branches, but
are shorter in length and have density 
contrasts only of order unity.  Figures 5-16 display the surface density,
along with azimuthal cuts, of the logarithm of the density response.  The images shown are in 
terms of the radial coordinate $\xi$, as defined in the numerical procedure section.
For both the short and long-term
simulations, we applied the spiral potential steadily throughout
the duration of the simulation.  The spiral potential is turned on
to full strength over a third of a revolution and
held steady after reaching maximum strength.  For the low $Q_g$ disk, corotation (CR) is at $\xi$=1.14,
and the inner Lindblad resonance (ILR) and outer Lindblad resonance (OLR) are
at $\xi$=0.33 and $\xi$=1.93 respectively; for the high-$Q_g$ disk,
CR is at $\xi$=2.13, and the ILR and OLR are at
$\xi$=0.623 and $\xi$=3.63 respectively.  Both the logarithmic
and SYL spiral planforms are applied between the ILR and OLR, while smoothed with a gaussian
of the form $e^{{-(\xi-2})^2/0.5}$.     
 
\subsection{Short-Term Evolution: Formation of Substructure}

The top panels in Figures 5-10  depict the density response at  
successive times in the x-y plane, where $x=\xi \cos(\phi)$ and $y=\xi \sin(\phi)$.
The second panel shows 
corresponding azimuthal cuts of the fractional density
response, i.e, $\Sigma_{\rm frac}=\Sigma(\xi,\phi,t)/\Sigma_{o}(\xi)$.
Azimuthal cuts are shown at the 4:1 ultraharmonic resonance radius,
along with cuts close to the resonance condition.

The third time snapshot of Model H1 shows clear
branch formation. The branch appears as a bifurcation of the main
spiral arms and has a smaller pitch angle than the main arms.
The tighter winding of this emergent branch
is in accord with the predicted effect of the second ultraharmonic as described in the
Artymowicz \& Lubow's (1992) semi-analytical study, and with results from SPH simulations carried
out by Patsis, Grosb{\ae}l, \& Hiotelis (1997). The second time snapshot shows that when the
secondary compression emerges, it is
strongest relative to the main spiral arms at the $n=2$ resonance radius.
Figure 1e shows that the amplitude of the secondary compression 
is nearly equal to the main spiral peak at $\xi=1.56$. 
The off-resonance $\xi=1.7$ cut indicates 
the presence of a secondary disturbance, but it is not as pronounced 
relative to the main spiral peaks at that radius. 
The final time snapshot shows that the bifurcation has become most
pronounced at $\xi=1.7$, although a secondary compression continues to be seen at the
4:1 resonance radius.  Possible reasons for the amplitude evolution
of the secondary peaks at the resonant radius will be explored in the discussion section.

The second time snapshot of model H2 shows a secondary
compression which is slightly more pronounced at the 4:1 ultraharmonic radius
than at $\xi=1.7$.  In the final time snapshot, H2 is deluged 
by the growth of substructure.  Along with the pair of branches
that is associated with the 4:1 resonance, we see a second bifurcation corresponding to the 6:1 
ultraharmonic ($n=3$).  The azimuthal cuts also hint at a pair of smaller compressions.  
H3, the non-self-gravitating analogue of model H2,
develops a pair of clear branches at late times, but the density contrasts are
smaller than H2's by an order of magnitude. As before, when the branch forms,
the azimuthal cut at the 4:1 resonance radius
has a stronger secondary peak than the
cut at $\xi=1.7$.  Model HSYL, is stable
to spur and branch formation and reaches arm-interarm contrasts of $\sim 4$
late in the simulation. After 2 pattern rotations, 
model H5 develops two, clear bisymmetric spurs
with amplitudes $10\%$ those of the main spiral
arms.  At the last time snapshot, the spurs are most pronounced at 
$\xi=1.3$, which is slightly displaced from 4:1
resonant radius, $\xi=1.17$ (note that the high-$Q_{g}$ 
models have different resonant radii).  H6, which is the non-self-gravitating
analogue of H5, does not develop substructure.

In summary, the non-self-gravitating and self-gravitating models with low $Q_g$,
namely models H1-H3, showed the development of branch-like secondary compressions.
In contrast, model H5, with self-gravity and high $Q_g$, formed spur-like
structures. The high-$Q_g$ model (H6) with no self-gravity showed no secondary structure.
The high-$Q_g$ SYL model (HSYL) with self-gravity did
not develop substructure during the extent of the short-term simulation.
However, as the next section on long-term evolution shows,
high-$Q_g$ models with self-gravity will not remain viable over
many revolutions if the spiral forcing $\cal F$ is higher than 5\%.  
While a secondary compression did result in the low $Q_g$, non-self-
gravitating model, only the self-gravitating models displayed
more than one secondary compression.  

Some of the models in Table 1 reached
extremely high density contrasts near the end of the 2.5 revolution simulations.
For example, H2 and H5 attained $\Sigma_{1}/\Sigma_{0}>100.$, which is remarkable,
given the decidedly small amplitude of the background forcing. In order for a disk
to remain viable over the long term with an isothermal equation of state,
lower forcing amplitudes are required.
We note that our short and long term simulations can alternately
be described as the strong and weak forcing simulations, respectively.
The short term simulations have higher forcing amplitudes
and evince the growth of substructure on shorter timescales, whereas the weak
forcing simulations require several revolutions to display
the growth of substructure.

\subsection{Long-Term Evolution}

Figures 11-16 show the time evolution of simulations run for 15 revolutions.
The main arms of L1 can be seen to bifurcate at the second time snapshot; 
this arm splitting is more pronounced at $\xi=1.7$ than at the 4:1 ultraharmonic radius.
At the last time snapshot, a secondary compression continues to be 
seen at the 4:1 resonant radius, and although the final structure is disorganized,
L1 remained stable, with density contrasts $\Sigma_{1}/\Sigma_{0}\sim 1$.
Model LSYL1 has developed 
a pair of branches in the second time snapshot, and has a flocculent appearance at
the end of the simulation.
LSYL1 maintained $\Sigma_{1}/\Sigma_{0} \sim 1 $.
Model L2 did not show branch formation and remained stable throughout.  Model LSYL2,
which differs from L2 only in the shape of the forcing spiral also remained stable.
An incipient branch is seen in the last time snapshot. The growth
of leading structures protruding from the main arms can be seen in the last timesnap for Model L3.  These
structures are more pronounced at radii slightly displaced from the
4:1 resonant radius. 
L3 remained stable for nearly the entire simulation, reaching $\Sigma_{1}/\Sigma_{0}=100$
after 14 revolutions.  Model LSYL3 developed strong spurs.
They are pronounced at $\xi=1.3$ in the second time snapshot. 
The last snapshot shows LSYL3 to have been inundated by the growth of substructure. By the time $\Sigma_{1}/\Sigma_{0}$
reached 100, nonlinear dredging had removed much of the gas from the resonant region.

In summary,
spiral forcing amplitudes of 1.3-1.5$\%$ maintain a steady-state response in the $Q_{g}$=1.3 disks. 
The resulting spiral structure shows less organization than when the forcing
amplitudes are high. The branches are strongest at radii slightly displaced
from the 4:1 ultraharmonic radius.
For the $Q_{g}$=2.48 disk, a 5$\%$ forcing admitted stability when the logarithmic
spiral planform was used, whereas the gas became unstable when forced with the
SYL spiral. As in the case of the short term simulations, the high-$Q_g$ disks display the
development of leading structures, i.e., spurs, in contrast to the low-$Q_g$ disks which manifest
the growth of branches.  A 3.5$\%$ forcing for the $Q_g=2.48$
disk results in a smooth, steady-state response for both spiral planforms.
It is apparent that the SYL spiral, in contrast to the logarithmic spiral, 
produces somewhat higher density contrasts and causes the growth of more 
substructure.

\section{Discussion of Results}

Subsequent to the identification of the ultraharmonic resonances by 
SMR, Artimowicz \& Lubow 1992 (AL), studied the nature of these resonances in finer detail.  AL's 
insightful semi-analytical treatment of the dynamics of ultraharmonic resonances
in the non-self-gravitating case showed that ultraharmonic waves of order
$n$, which have $n$ times the wavenumber as the main arms, arise due to 
quadratic velocity stress terms and produce an angular momentum flux that is fourth-order
in the perturbing potential, in contrast to the second-order flux that
arises for the Lindblad resonances.  These results demonstrate that the first ($n=2$)
ultraharmonic resonance produces a bifurcation of the main spiral arms
with a smaller pitch angle than the main arms.  AL also gave a clear
description of the higher order resonances.  They found that the 
characteristic wavelength of the driving decreases as $\propto 1/n$ for 
the higher order resonances, and that resonances of high $n$ are weakened
due to a combination of several effects, namely, the dependence of the response
on a higher power of the driving force (assumed small), and the increasing 
importance of torque-cutoff effects.  Finally, AL's analysis revealed
that the ultraharmonic wave acquires amplitude in a region that is centered
about the location of the resonant radius, and not precisely at the
resonant radius.

The azimuthal cuts presented in Figures 5-16 were made at the 4:1 ($n=2$)
ultraharmonic radius along with a cut close to the resonant radius.
By inspection, we find
that the growth of branches and spurs, while
initiated at the 4:1 resonant radius, acquired pronounced amplitude at
radii slightly displaced from the resonance. This displacement agrees with
AL's analysis.  AL's semi-analytical analysis cannot however 
show the time development of the resonant response. This task
is best accomplished by simulations. Toomre (1969) showed 
that linear spiral disturbances propagate radially at the usual group velocity of the
wave.  It is likely that the nonlinear analog of this radial propagation is
responsible for the time variation we observe in the amplitude of the
secondary compressions near the resonant radii.
The radial spreading of the secondary compressions in model H1, for instance, between
the second and last time snapshot, occurs approximately over the sound crossing time
between the 4:1 ultraharmonic radius and $\xi=1.7$.  Our results also differ from AL's semi-analytic
analysis by explicitly including global self-gravity. AL argued,
on the basis of the conservation of angular momentum flux (e.g. Goldreich \& Tremaine 1979),
that ultraharmonic waves in a gas disk with only pressure will carry the same flux
as in a self-gravitating disk.  Our self-gravitating models, in contrast to the
non-self-gravitating models, display a much greater depletion of gas in the resonant regions.
Thus, we find that the angular momentum flux carried
by the ultraharmonic waves is highly sensitive to the self-gravity, probably because
it changes the effective amplitude of the resultant waves.

In contrast to the branches that formed for low $Q_g$ models, many high $Q_g$ models with
high forcing displayed leading structures, or spurs, after several revolutions.
The low $Q_g$ models will not remain viable for several revolutions
if the forcing is comparable to that used to drive the high $Q_g$ disks.  It is
important to note that only those high $Q_g$ disks that were run for several 
dynamical times displayed spurs.  For instance, HSYL
is stable to spur formation at about 500 Myr.  However, the long-term analogs of HSYL,
like LSYL3, begin to show leading structures after about 1 Gyr.  Moreover,
the appearance of these structures is also highly sensitive to the level
of the forcing.  L2, a high $Q_g$ disk with a forcing of 3.5\%, remains
stable to spur formation.  L3, also a high $Q_{g}$ disk, but with a forcing of
5\%, shows clear spur formation after about 3 Gyr.  In addition,
the appearance of these features is also observed to depend
on the type of spiral planfrom used to force the disk.  LSYL2, which is analogous to L2 except that it 
is forced by the nonlinear SYL planform (at the same forcing amplitude
as L2)
displayed incipient spurs at the last timesnap while L2 remained
stable througout.  These observations, namely the
sensitive dependence on strength of forcing, duration of simulation,
and nonlinearity of planform,
suggest that the traditional linear interpretation of leading
features as partially reflected trailing waves off $Q$-barriers (which exist
only near the corotation circle of our simulations) or by propagation through the galactic
center (which is prevented by sponge boundary conditions in our simulations),
may be more properly understood in some circumstances
as arising from nonlinear effects.  

What are some of these effects?
When models are run at relatively large forcing
amplitudes for many dynamical times, nonlinear dredging occurs because associated galactic
shocks interior to CR produce a gradual inspiraling of the gas
(Roberts \& Shu 1972).  The gas tends to pile up against the
nearest resonance strong enough to produce such shockwaves.  Such
a pile-up in the local surface density then yields a barrier
against the inward group motion of trailing gaseous spiral waves that are
continuously being generated by the background forcing.  The trailing spiral
waves partially reflect from the barrier as leading spiral waves
and partially transmit across it as trailing spiral waves of reduced amplitude.
Gas dredging is particularly apparent in the grey scale images of H5, L3, and LSYL3.  
Reflection of trailing spiral waves off the sharp features just beyond
the 4:1 ultraharmonic radius can
then produce leading features, or spurs, while the transmitted trailing spiral
waves do indeed have reduced amplitude in comparion with the incident trailing
waves farther out in radius.

KO also carried out an extensive study, by performing MHD simulations, on the formation of 
substructure in spiral galaxies.  Their 
analysis focused on the formation of feathers
via the magneto-Jeans instability.  In contrast with our
simulations of low $Q_g$ disks, their published
purely hydrodynamical models were stable to feather formation (although they
clearly did simulations of unstable systems).
There is no contradiction between their results and ours.  KO heeded
Balbus \& Cowie (1985) suggestion that $Q_{\rm sp}=Q_g(\Sigma_0/\Sigma_{\rm max})^{1/2}$
is the appropriate parameter
that characterizes the gaseous response to an applied perturbation.
The square root appears rather than the more naive linear factor for the
surface density compression behind a galactic shock because of the
increase in the local postshock shear from the unperturbed state.  KO
first carried out 1D asymptotic calculations to construct steady spiral
shock configurations that were quasi-axisymmetrically stable. 
In terms of the local $Q_{\rm sp}$ parameter, their 2D simulations were
informed by these 1D quasi-axisymmetrically stable model to satisfy $Q_{\rm sp}>0.9$.
They then found that their unmagnetized hydrodynamical models were
stable to feather formation.  Our 2D models are not restricted to the
range $Q_{\rm sp}>0.9$, and thus could develop transient local gravitational instabilities
that gave rise to feathers.  KO's magnetized models did 
form prominent feathers for cases where simple $Q_{\rm sp}$ considerations 
would have predicted stability. Evidently, magnetization destabilizes high $Q_{\rm sp}$ disks by
the mechanism of shear reduction in the magneto-Jeans instability, as
discussed in the linear regime by Lynden-Bell (1966), Elmegreen (1993), and Kim \& Ostriker (2001).
However, the MHD stability criterion is not easy to state analytically,
so we will be satisfid with the loose notion that instability arises when
an effective $Q_g$ crosses some threshold value.  In any case, because the wing of
feathers behind the primary shock front were all of roughly equal strength
in KO's local simulations, such feathers could not be born from ultraharmonic
resonances of different forcing capability.

In a sense, therefore, our low $Q_g$ hydrodynamic simulations are mimicking the
feathering behavior of relatively high effective $Q_g$ magnetohydodynamic simulations.  It is important to note that the usual $Q_g$ does not characterize the signal speed of compressive wave
propagation in the MHD context.  Thus, the effective $Q_{g}$ of magnetized
disks, due to the MHD modifications of the sound speed, should be
considered when evaluating the response of the disk.
Only high effective $Q_g$ disks can avoid runaway catastrophes that prevent the simulations
from being carried out long enough to accumulate the ultraharmonic resonant encounters
that explain branching and to develop the secular drifts that explain
spur formation.  Thus, we anticipate that moderately high $Q_{g}$ simulations,with the inclusion of frozen-in magnetic fields, can produce models that simultaneously
exhibit branches, spurs, and feathers. 

Another question that motivated this study is whether a steady
ordered driving field can produce, via overlapping nonlinear effects,
a disordered response in the gas.  Repeated passages of gas through the spiral arms
at ultraharmonic periodicities will cause
secondary compressions in the form of branches and spurs that become especially pronounced 
for our self-gravitating models.  Model H2 and LSYL3, in particular, illustrate a divergent
growth of substructure that ultimately causes the systems to destabilize.  We cannot, however,
attribute this apparently chaotic effect to the resonance-overlap mechanism as we cannot 
rule out the growth of substructure via purely numerical effects in H2 and LSYL3.  We have
performed a high-resolution run (512x512) for H2 and find that even though
it is initially convergent, the agreement with the lower-resolution run breaks down towards
the end of the simulation.   However, the high-resolution run that we performed for
L1 maintains agreement with the lower-resolution run throughout.  Thus,
we can say with greater confidence that the growth of substructure in L1
is not due to numerical artifacts, whereas we cannot say this definitively
for models like H2 where surface density contrasts reached $\sim 100$.  The unequivocal demonstration of this phenomenon in a simulation, i.e.,
flocculence arising purely from overlapping nonlinear
effects, will necessary
require ruling out the possibility that it arose from numerical artifacts.    

We have noted previously that for the same forcing amplitude, the SYL
spiral planform induces the growth of more substructure than the
logarithmic spiral planform.  This happens despite the fact that the more
open spiral winding of the SYL planform couples less naturally to the
spiral response of the gas, which does not easily sustain disturbances
with large radial wavelength.   For instance, we note that LSYL2 shows incipient
spur formation at the last timesnap, while L2 is stable to
the formation of substructure.  We also noted in the section on ultraharmonic
resonances that the nonlinear nature of the SYL planform allows for potentially
more occurrences of resonance overlap than the linear logarithmic-spiral
planform.  The positive correlation of
a more chaotic or flocculent response for the SYL planform thus
does lend credence, but not yet demonstrated assurance,
to the idea that a steady, smooth driving force field can induce,
over the long run, a disordered and even flocculent response in the interstellar
gas clouds (and the population I stars born from them)
via the mechanism of overlapping ultraharmonic resonances.

\subsection{Caveats and Future Work}

It is important to emphasize that since we performed whole-disk simulations
with an isothermal equation of state, we can study the formation
of secondary spiral structures that grow via the ultraharmonic
resonances, but we cannot study the fragmentation of such
structures to become self-gravitating bodies (as KO did).  The reason for this
is two-fold.  Our global simulations
do not have the dynamic range necessary to follow the development
of strong feathering that lead to local fragmentation.  Differential dredging could lead
to similar problems locally.  Moreover, global
simulations admit multiple reflections off radial irregularities in the system;
these can lead to accidental resonant cavities that create growing
quasi-modes.  In this fashion, some self-gravitating models with large spiral
forcing may have developed high
density contrasts in the main spiral arms that tend to destabilize the
system as a whole.  Model H2, for instance, illustrates the result
of higher order resonances, but also reaches excessively high
density contrasts. 

Our simulations (and KO's) are also limited by the isothermal equation
of state. No damping mechanism other than radial drift prevents the gas from
shocking repeatedly.  Self-regulation from star formation
and supernovae explosions may play an integral role in the realistic scenario
in holding off local runaway collapse by increasing local $Q_{g}$ values.  Another
line of improvement in the context of purely hydrodynamical
models, as suggested by KO, is a more realistic treatment of the
true interstellar medium (ISM), with
particular attention paid to its microscopic
properties.  Shu et al. (1972) made an early attempt in this direction by treating
the ISM as two thermally stable phases (Field, Goldsmith, \& Habing 1969)
forced by an external spiral field.  

The present understanding of the ISM as a highly turbulent and multiple phased medium
compels a greater scrutiny and a finer treatment of the interplay
among many physical effects (McKee \& Ostriker 1977).
In particular, the ubiquitous 3D turbulence observed in the ISM may 
serve to stabilize the disk even when the gas is driven by a large
spiral field.  Local 3D simulations developed to study magnetic and 
hydrodynamic self-gravitating flows indicate that some systems remain stable 
for conditions where a 2D treatment would have indicated
instability (Kim, Ostriker, \& Stone 2002)  

Finally, we note that the intent of the section on long-term simulations
was to consider models repesentative of long-lived spirals that have existed in relative
isolation for many revolutions. We found that the repeated passage of
gas through regions of ultraharmonic resonance induce in these models
secondary structures akin to the branches, spurs, and feathers observed
in the population I stars and gas of many external disk galaxies.
The short-term simulations of high $Q_{g}$ disks at higher relative
amplitudes of background spiral forcing yielded much cleaner grand-design
structures in the interstellar gas.  On the basis
of numerical simulations by other workers starting with Toomre \& Toomre
(1972), a similar ``cleaning up'' of the otherwise messy appearance (in blue light) of
normal galaxies may be accomplished by the violence of a close tidal encounter
with another galaxy.

On the other hand, Block et al. (1994) point out the existence
of many disk galaxies that have orderly grand-design spiral patterns when observed in
the infrared, yet appear quite messy, and even flocculent, when observed
in optical or blue light.  They suggest that the existence of so many
infrared grand-design spirals cannot be the result of occasional tidal encounters, but
must almost certainly represent the quasi-stationary spiral normal-mode
long postulated by C. C. Lin and collaborators to be present in the background of 
old disk stars (Lin \& Shu 1964; Lin et al. 1969;
Lin \& Lau 1979; Bertin \& Lin 1966).  However, Block \& coworkers
(Block \& Wainscoat 1991, Block et al 1994, 1996) also suggested
that the contrasting lack of order in the population I component, and by
inference, in the interstellar gas, must mean that the dynamics of
the gas and old disk stars are decoupled.  Our long-term simulations demonstrate
that this conclusion need not follow.  The gas of a
disk galaxy can be driven gravitationally
by an orderly spiral pattern existing in the infrared background
of disk stars, and yet the nonlinear response can be disorderly and even chaotic,
especially if the ordered driving lasts long enough to establish the nonlinear superposition
of several overlapping ultraharmonic resonances.  But better calculations are needed
before we can confidently establish that truly flocculent galaxies can be constructed by
the mechanisms of nonlinear dynamics alone. 

\section{Conclusion}

Impressive displays of spiral substructure in the form of inter-arm branches, 
protruding spurs, and feathering
between the main arms imparts a scabrous and disordered
appearance to the large-scale structure of many spiral galaxies.  We have performed
a number of hydrodynamical simulations to study the growth of these structures
in the presence of various ultraharmonic 
resonances.  We find that self-gravity is a primary catalyst in heightening the 
strongest ultraharmonic, i.e., the 4:1 resonance, which produces a bifurcation of the main
arms.  Moreover, we see that self-gravity is crucial for the growth of substructure
via the higher-order resonances.  In our simulations, we strove to isolate the effects of the resonance 
mechanism by  purposely minimizing the roles of the swing amplifier and reflection-induced
feedback from inner and outer boundaries (but not ones produced by
internal dredging).  Our short-term simulations with large forcing amplitude generate
a vigorous response in the gas, eliciting the growth of branches and feathers in the low $Q_g$ cases.  
The long-term simulations, which had
lower forcing amplitudes and were run over a timescale of several Gyr, illustrated 
initially a smooth response in the gas that resembled the driving spiral field, but which over time,
accumulated the effects of dredging and ultraharmonic resonances.  Prominent leading spurs as offshoots
from the secondary commpressions and dredging associated with ultraharmonic resonances
are the main surprise of such long-term simulations.  When combined with the speculation that
high $Q_g$ hydromagnetic simulations mimic the feathering capability of low $Q_g$ hydrodynamic
simulations, these results reinforce the idea that the gaseous response in disk galaxies
to an ordered background spiral gravitational field becomes increasingly disordered with time,
with the appearance of numerous branches, spurs, and feathers atop the main grand-design spiral arms.
Background forcing planforms with a sufficiently nonlinear azimuthal dependence may
even generate so many overlapping subharmonic resonances that
the entire gas structure becomes chaotic and flocculent.

In the cleaner spiral galaxies, as noted in Elmegreen \& Elmegreen (1990),
if a number of the ultraharmonic resonances can be matched 
with observed features, the determination of the pattern speed of the stellar
background would be on a firmer footing.  However, any design study
of a particular galaxy that attempts to match observed features with a given resonance must
also endeavor to disentangle the effects of the two other primary mechanisms of structure
formation in self-gravitating systems, namely, SWING amplification and reflection-induced
feedback.  As we have noted, 3D simulations that model the ISM more
realistically are needed to further understand the details of the growth of substructure
via resonance phenomena.  Finally, the addition of a responsive stellar component would serve
to demonstrate the thesis (Roberts \& Shu 1972; see also AL for a related point concerning
the damping associated with ultraharmonic resonances)  that galactic shocks would extract energy
and angular momentum from the stellar wave of a sign as perhaps to saturate
its intrinsic tendency to grow as an unstable normal mode of the system.
Such a program would fulfill the hope and vision embodied 
by the original hypothesis of quasi-stationary spiral structure proposed by Lin \& Shu (1964). 

We are grateful to the referee for helpful criticisms that improved the presentation of this
paper.  S.C. would also like to thank L. Blitz, J. Graham, W.T. Kim, M. Krumholz, C. McKee,
J. Sellwood, A. Toomre, and especially C. Yuan for useful discussions and insightful criticisms.
We also thank S. Dawson, C. Heiles, and E. Rosolowsky for suggestions on improving the
visual presentation of the data.  This research was funded in the United States
by the National Science Foundation and NASA, and in
Taiwan, by the National Science Council and Academia Sinica.

\begin{figure}
\centerline{\psfig{file=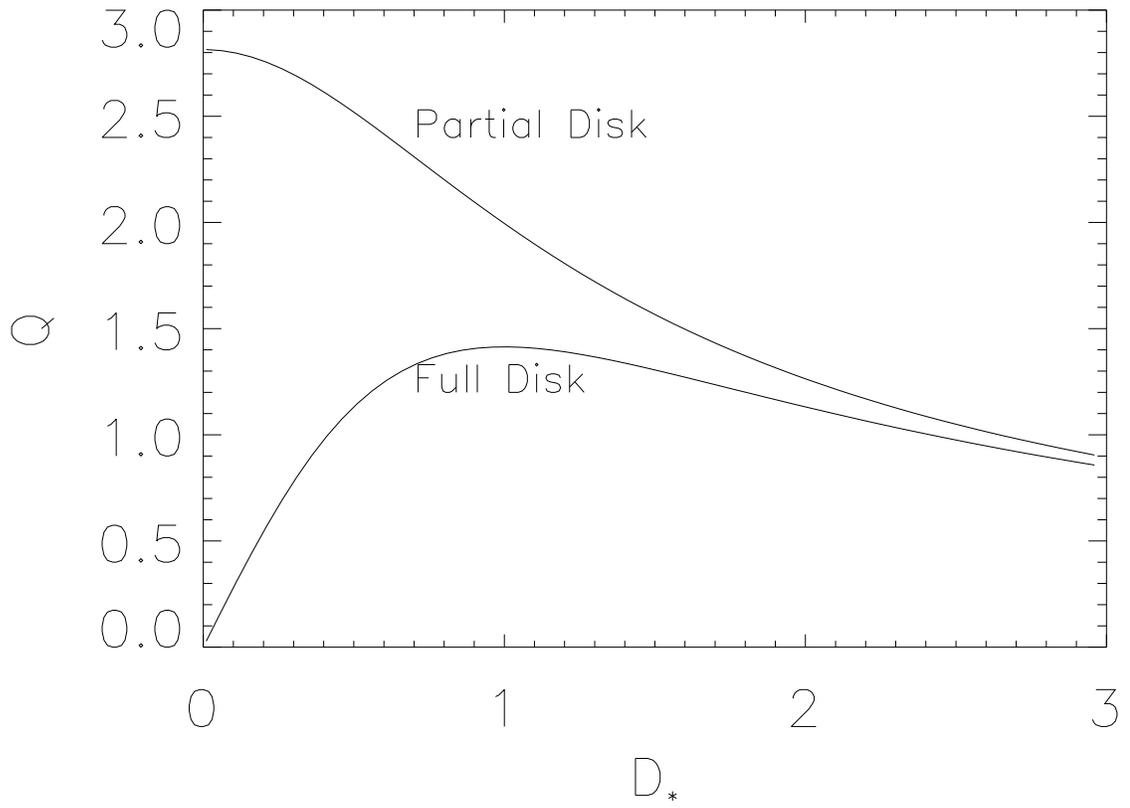}}
\caption{$Q$ vs $D_{\star}$ for full disk overplotted with $Q_{g}$ vs $D_{\star}$ for partial disk}
\end{figure}

\begin{figure}[h] \begin{center}
%\centerline{\psfig{file=f2a.eps,height=3.in,width=3.in}
%\psfig{file=f2b.eps,height=3.in,width=3.in}}
\end{center}
\caption{(a)Curves of marginal stability for axisymmetric (ringlike) disturbances
in partial SIDs. {\it Heavy solid lines} delineate the collapse and
ring fragmentation regions for a full ($F=f=1.0$) SID. Also
shown are computed curves of marginal stability for $F=f=0.9$ through
$F=f=0.1$ in decrements of $F=f=0.1$. (Note that the curves showing
the collapse branches for $F=f<0.7$ are outside the limits of the
plot.)  Our models with $Q_{g}$=2.48 and $Q_{g}$=1.3 fall in the range
of oscillating disturbances. (b)Self-consistency curves for nonrotating m=2 spiral disturbances.
The bottom curve is for F=f=1.0 as in S00.  The upper nine
curves run through F=f=0.9 to F=f=0.1}
\end{figure}

\begin{figure}[h] \begin{center}
%\centerline{\psfig{file=f3a.eps,height=3.in,width=3.in}
%\psfig{file=f3b.eps,height=3.in,width=3.in}}
\end{center}
\caption{(a) The logarithmic spiral profile, (b) The SYL spiral profile}
\end{figure}	

\begin{figure}[h]\begin{center}
\centerline{\psfig{file=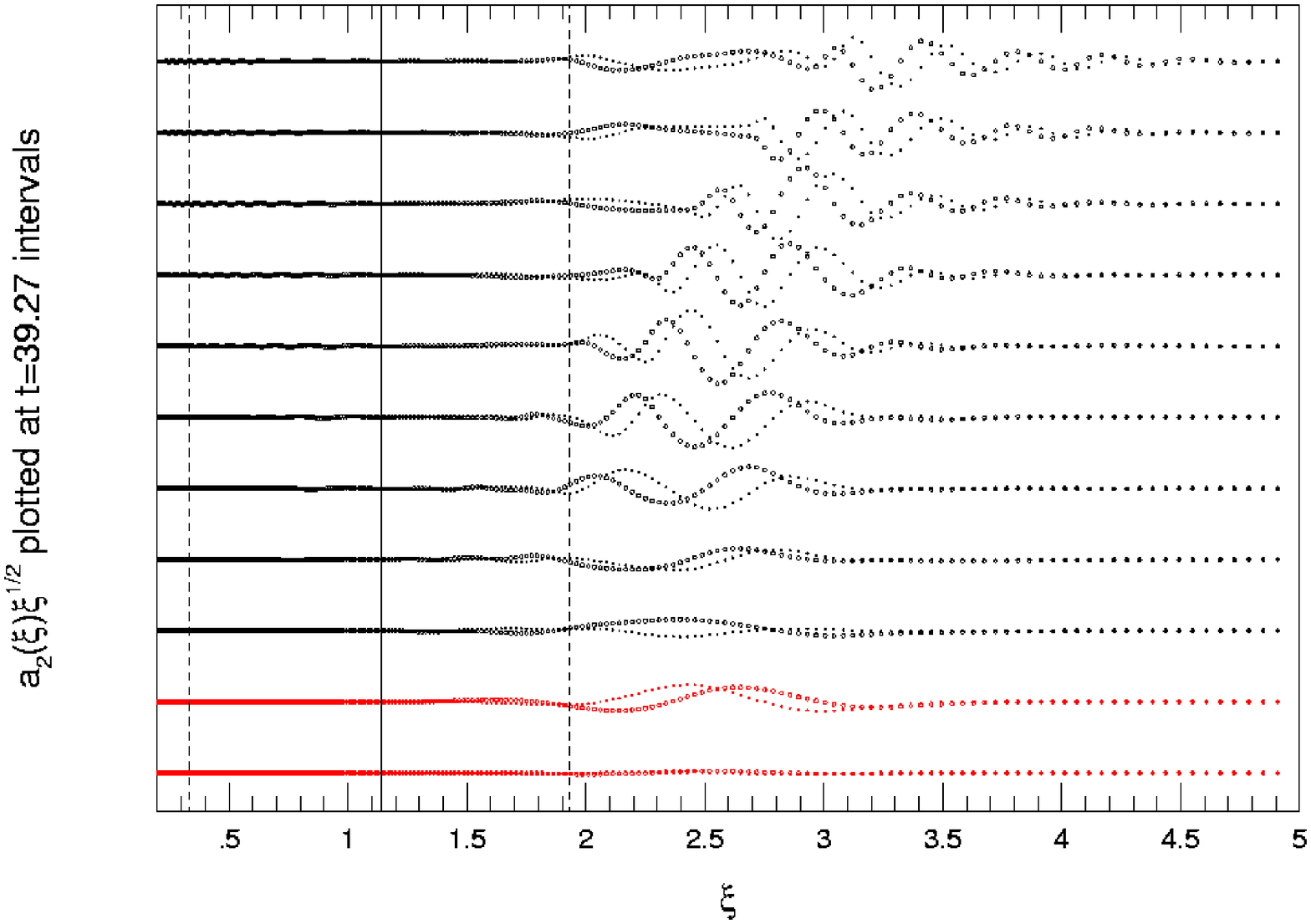,height=2.5in,width=2.5in}
\psfig{file=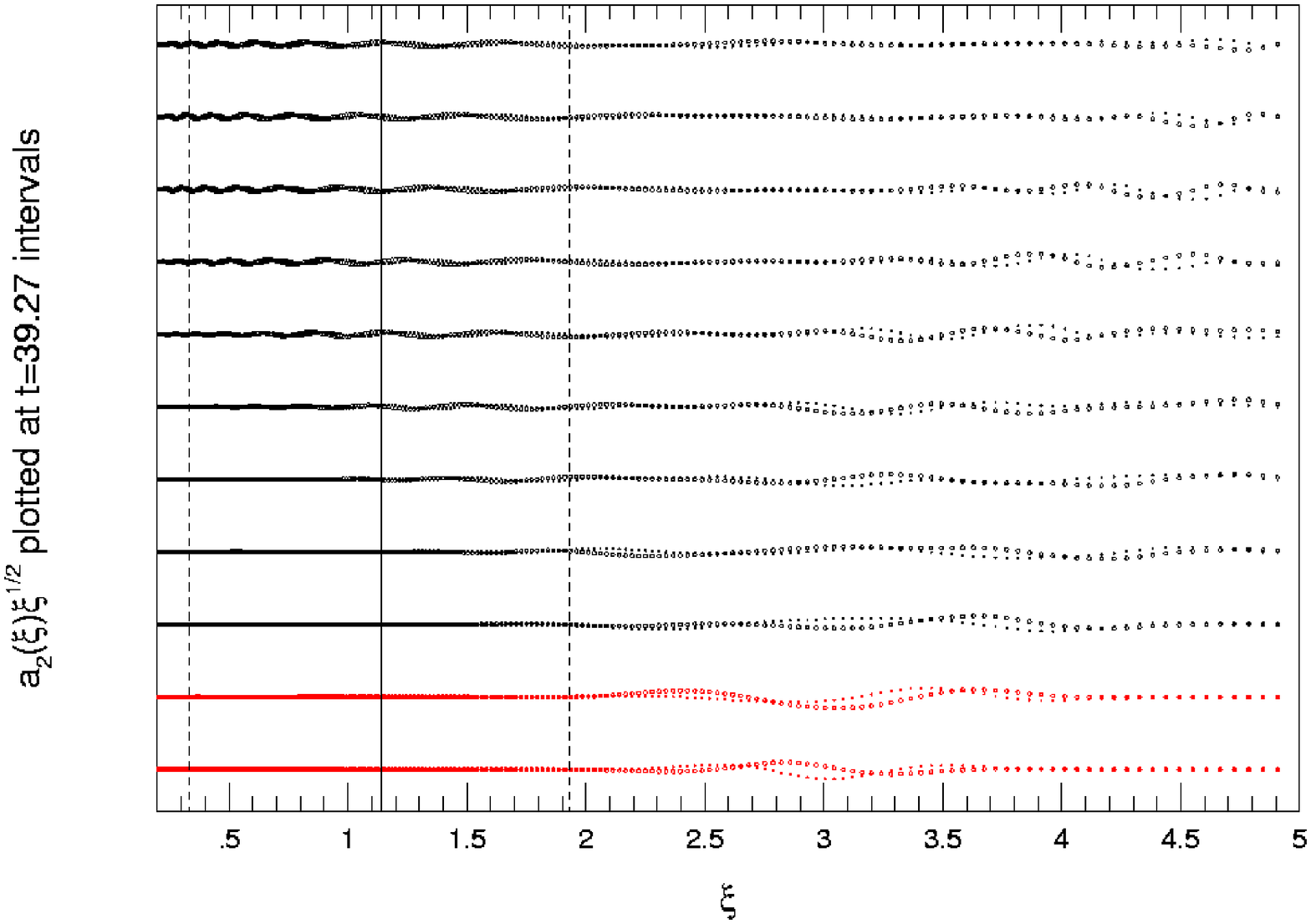,height=2.5in,width=2.5in}}
\end{center}
\caption{(a) Response of the gas disk upon applying a transient leading spiral
perturbation in the $Q_{g}=1.3$ disk.
Times are given in machine units (at $\xi=1$, one full revolution
requires $t=39.2$) (b) Response of the gas disk upon applying a trasient
leading spiral perturbation in the $Q_{g}=2.48$ disk}
\end{figure}		

\begin{figure}[h] \begin{center}
%\centerline{\psfig{file=f5a.eps,height=2.5in,width=2.5in}
%\psfig{file=f5b.eps,height=2.5in,width=2.5in}
%\psfig{file=f5c.eps,height=2.5in,width=2.5in}}

\centerline{\psfig{file=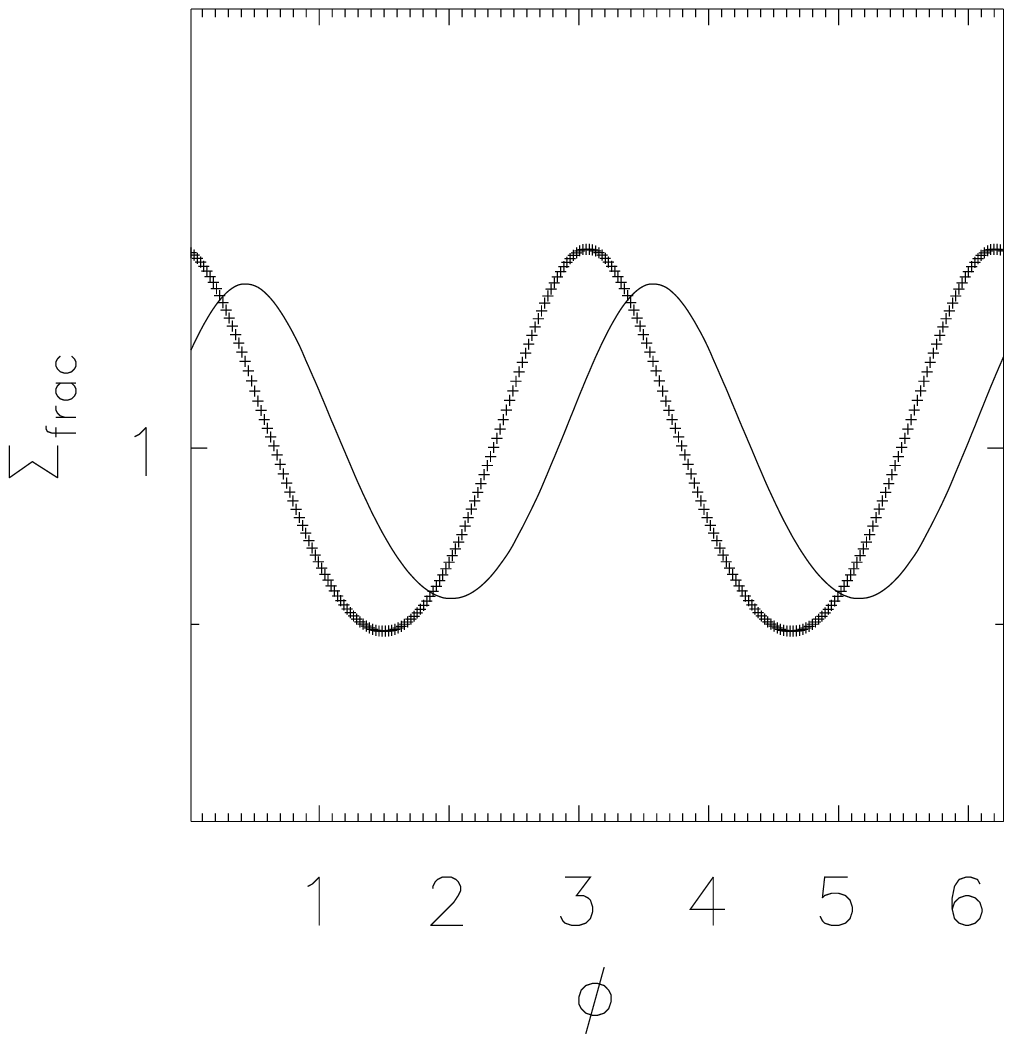,height=2.5in,width=2.5in}
\psfig{file=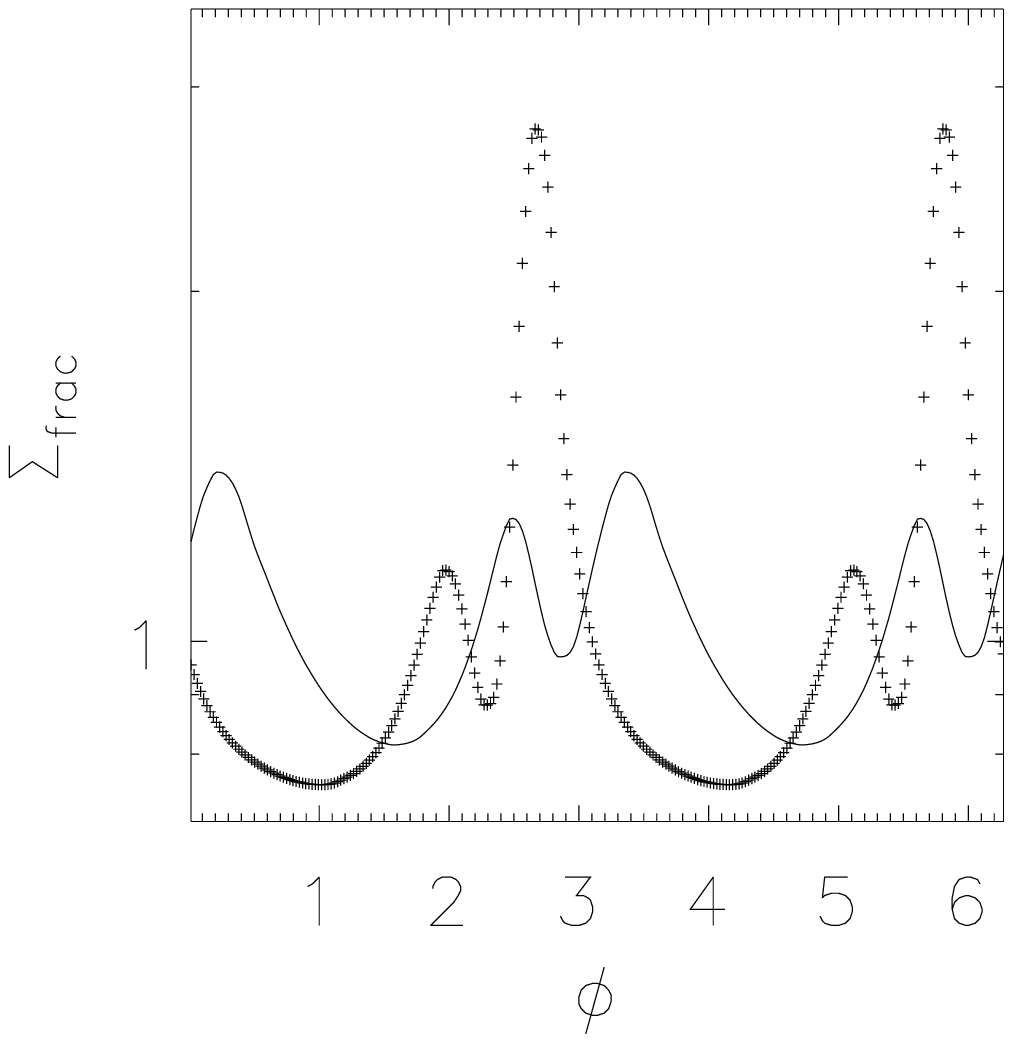,height=2.5in,width=2.5in}
\psfig{file=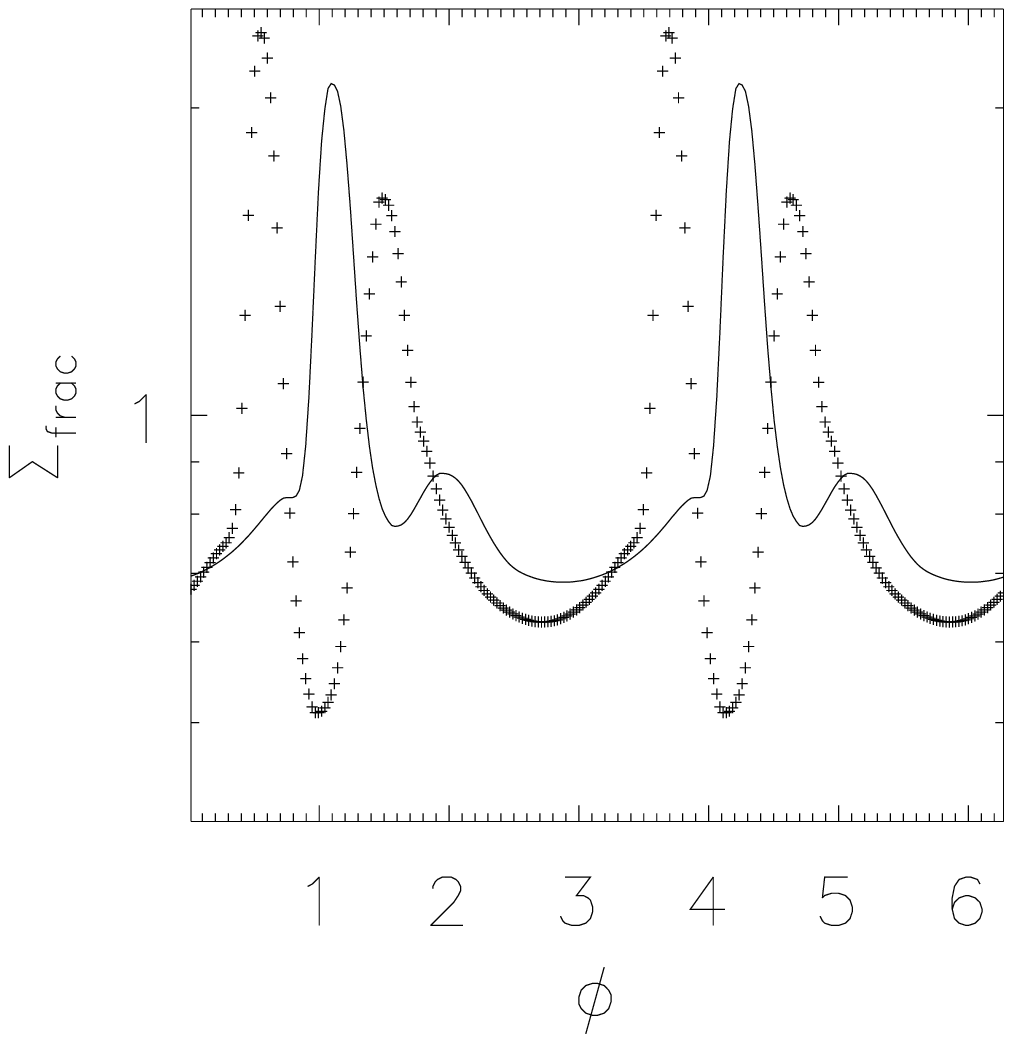,height=2.5in,width=2.5in}}

\end{center} 
\caption{(a)-(c)Snapshot of model H1 at 95 Myr, 318 Myr, and 477 Myr,
(d)-(f)Azimuthal cuts at $\xi$=1.7 and $\xi$=1.56 at times 95 Myr, 318 Myr, and 477 Myr.  Solid lines denote the $\xi$=1.56 cut and
the dotted lines are the $\xi$=1.7 cut.  Emergent branches are marked by arrows.  (a)-(c) included as jpg files.  See listed website
for higher resolution embedded ps figures} 
\end{figure}
 
\begin{figure}[h] \begin{center}
%\centerline{\psfig{file=f6a.eps,height=2.5in,width=2.5in}
%\psfig{file=f6b.eps,height=2.5in,width=2.5in}
%\psfig{file=f6c.eps,height=2.5in,width=2.5in}}

\centerline{\psfig{file=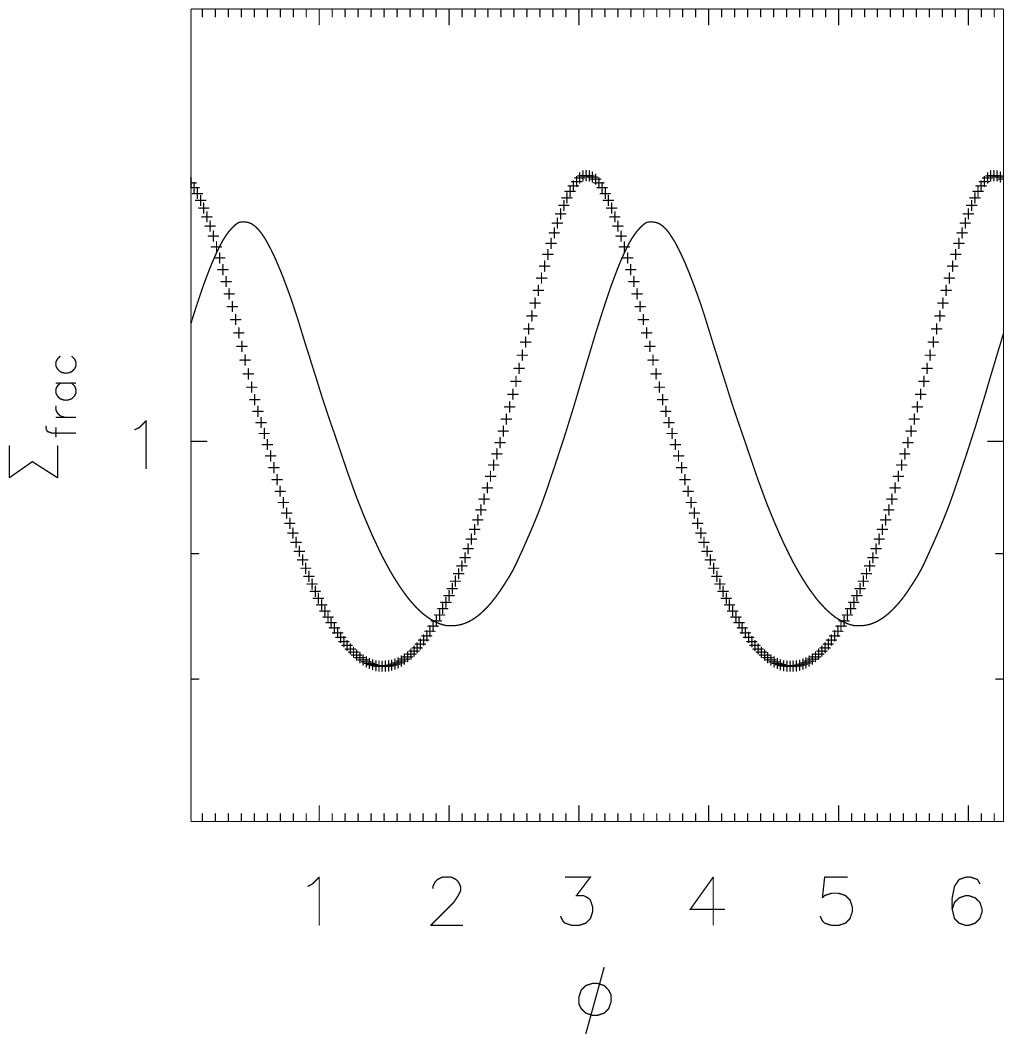,height=2.5in,width=2.5in}
\psfig{file=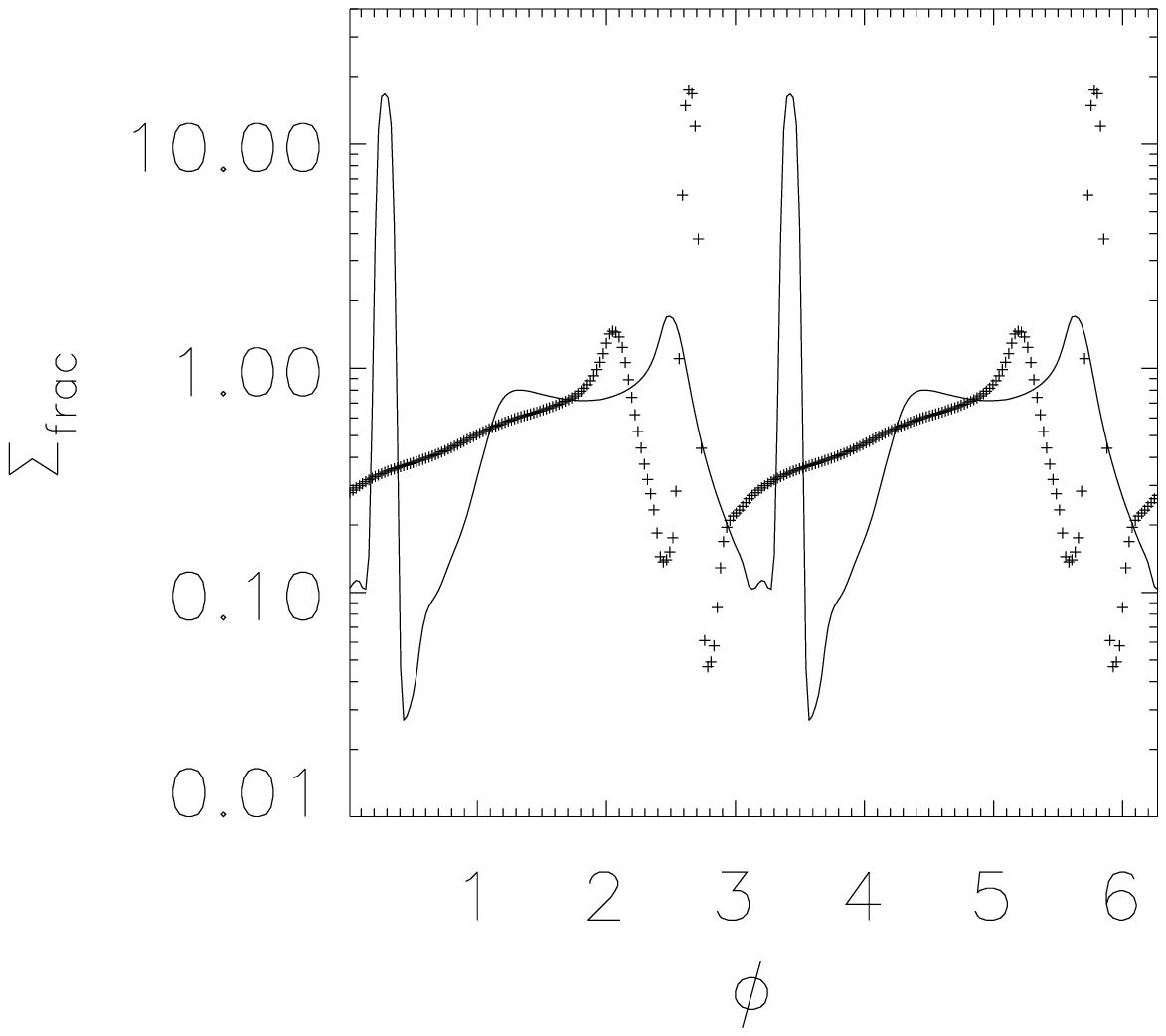,height=2.5in,width=2.5in}
\psfig{file=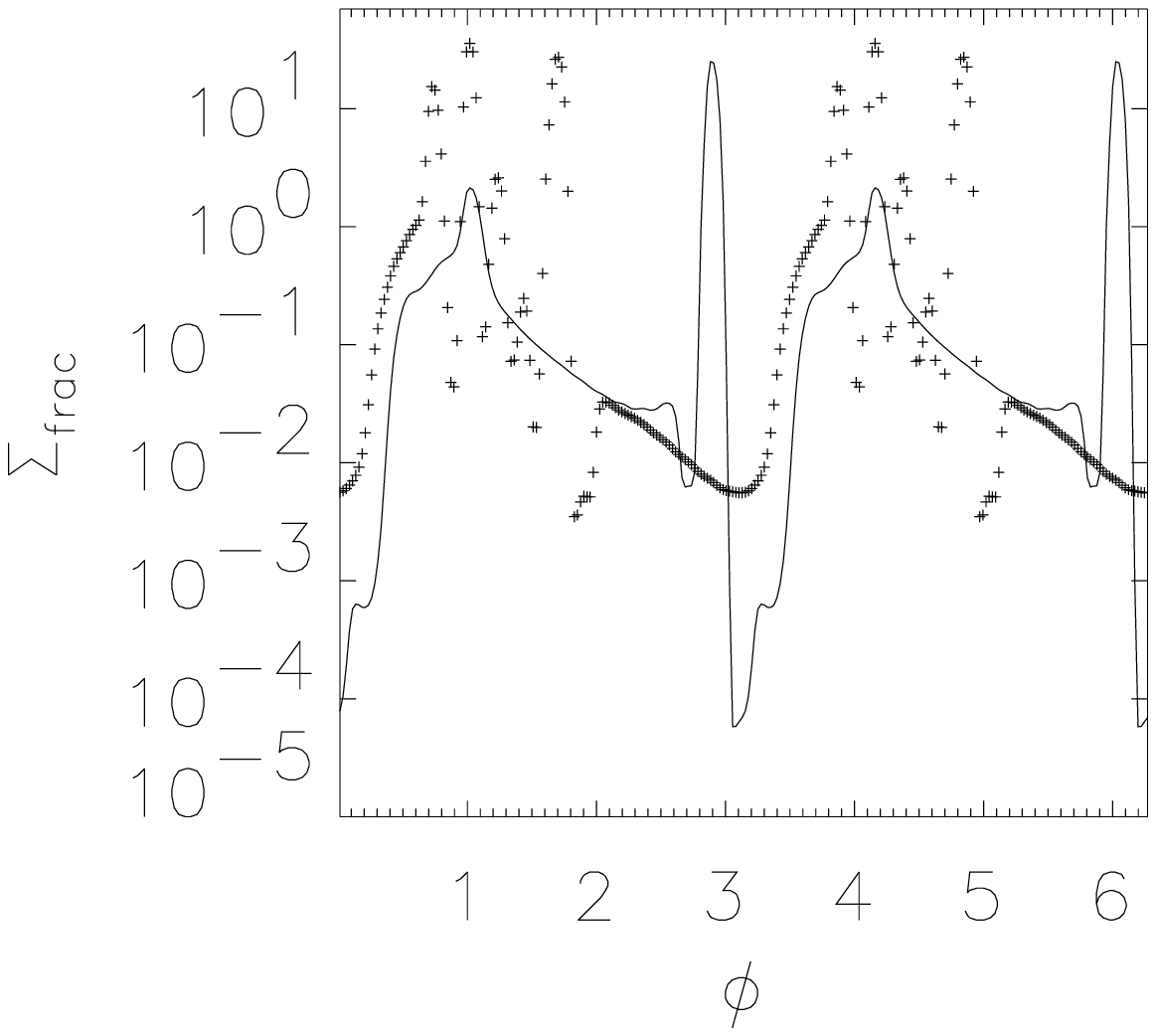,height=2.5in,width=2.5in}}

\end{center} 
\caption{(a)-(c)Snapshot of model H2 at 95 Myr, 318 Myr, and 477 Myr,
(d)-(f) Azimuthal cuts at $\xi$=1.7 and $\xi$=1.56 at 95 Myr, 318 Myr, and 477 Myr}
\end{figure}

\begin{figure}[h] \begin{center}
%\centerline{\psfig{file=f7a.eps,height=2.5in,width=2.5in}
%\psfig{file=f7b.eps,height=2.5in,width=2.5in}
%\psfig{file=f7c.eps,height=2.5in,width=2.5in}}

\centerline{\psfig{file=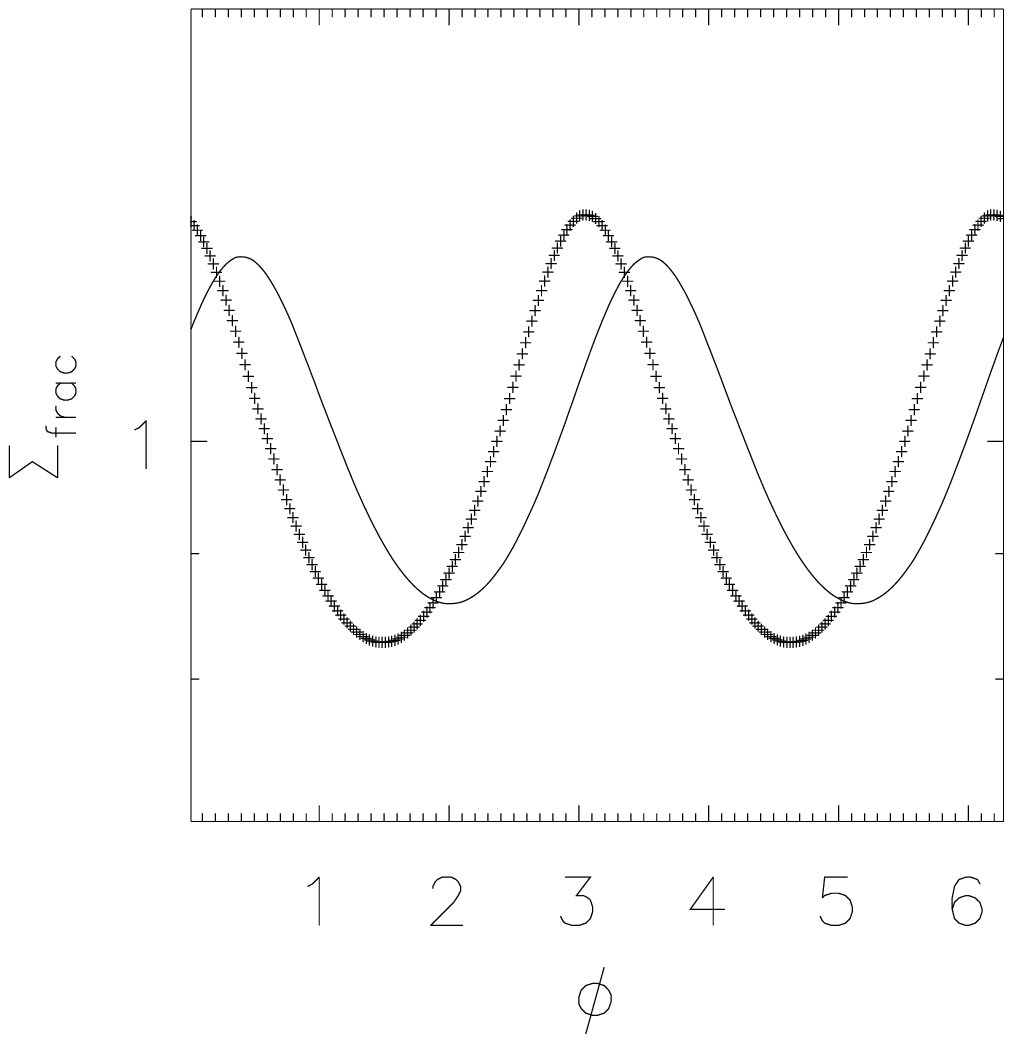,height=2.5in,width=2.5in}
\psfig{file=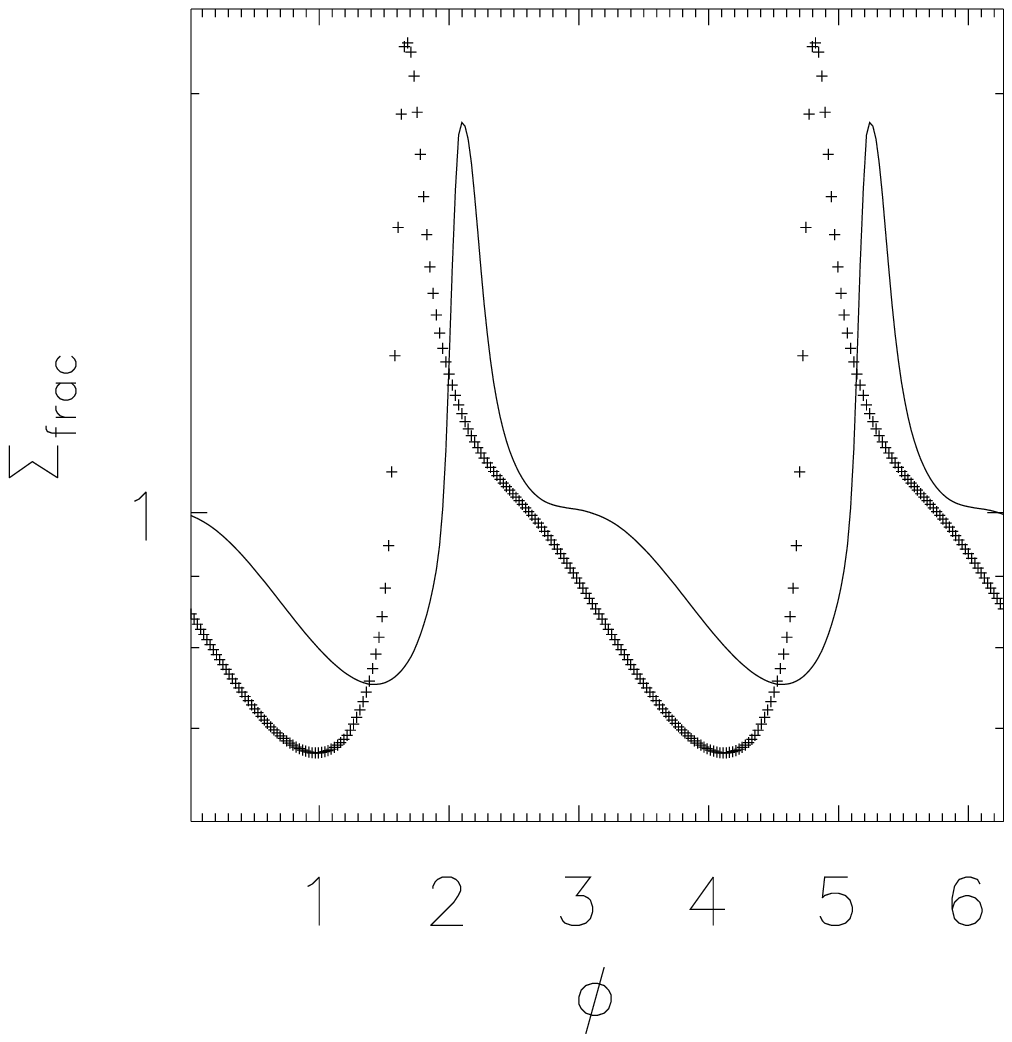,height=2.5in,width=2.5in}
\psfig{file=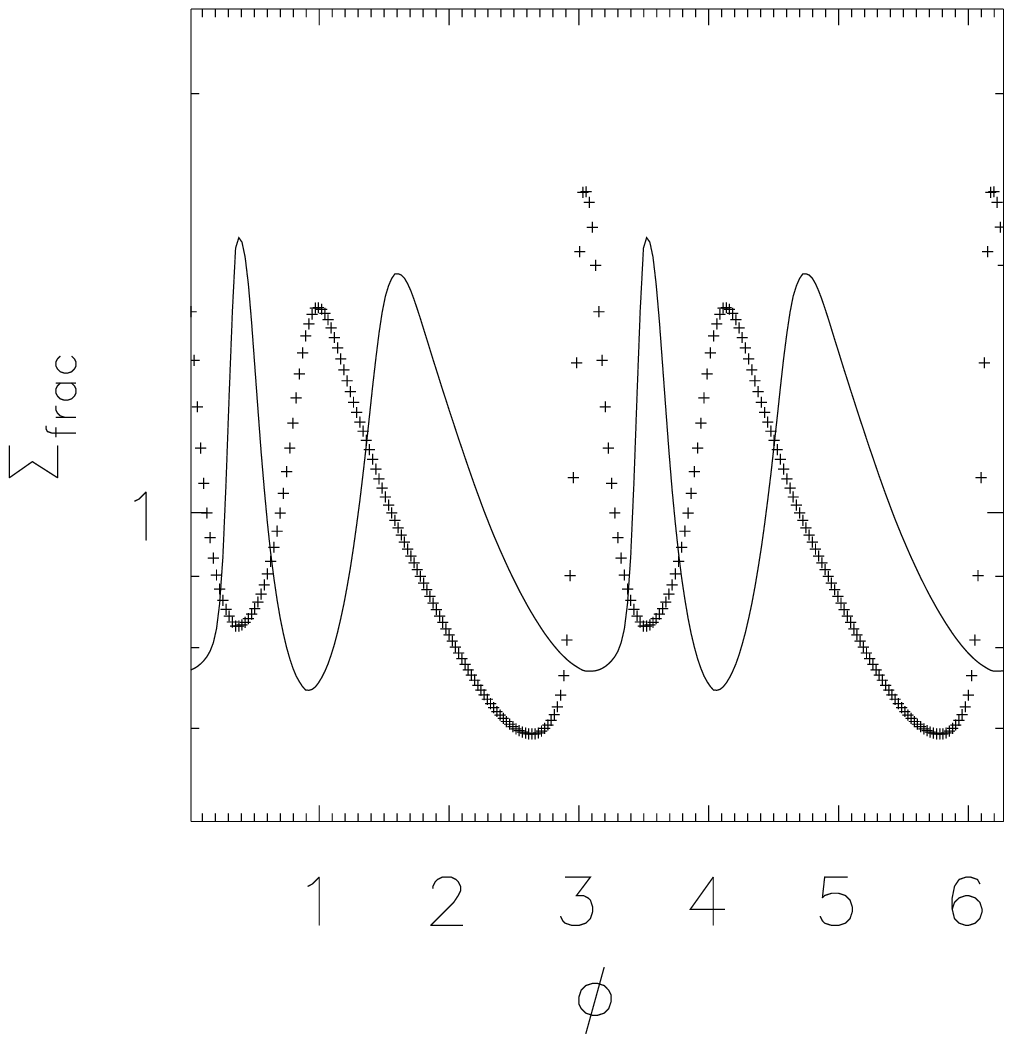,height=2.5in,width=2.5in}}

\end{center} 
\caption{(a)-(c)Snapshot of model H3 at 95 Myr, 318 Myr, and 477 Myr,
(d)-(f)Azimuthal cuts at $\xi$=1.7 and $\xi$=1.56 at 95 Myr, 318 Myr, and 477 Myr}
\end{figure}

\begin{figure}[h] \begin{center}
%\centerline{\psfig{file=f8a.eps,height=2.5in,width=2.5in}
%\psfig{file=f8b.eps,height=2.5in,width=2.5in}
%\psfig{file=f8c.eps,height=2.5in,width=2.5in}}

\centerline{\psfig{file=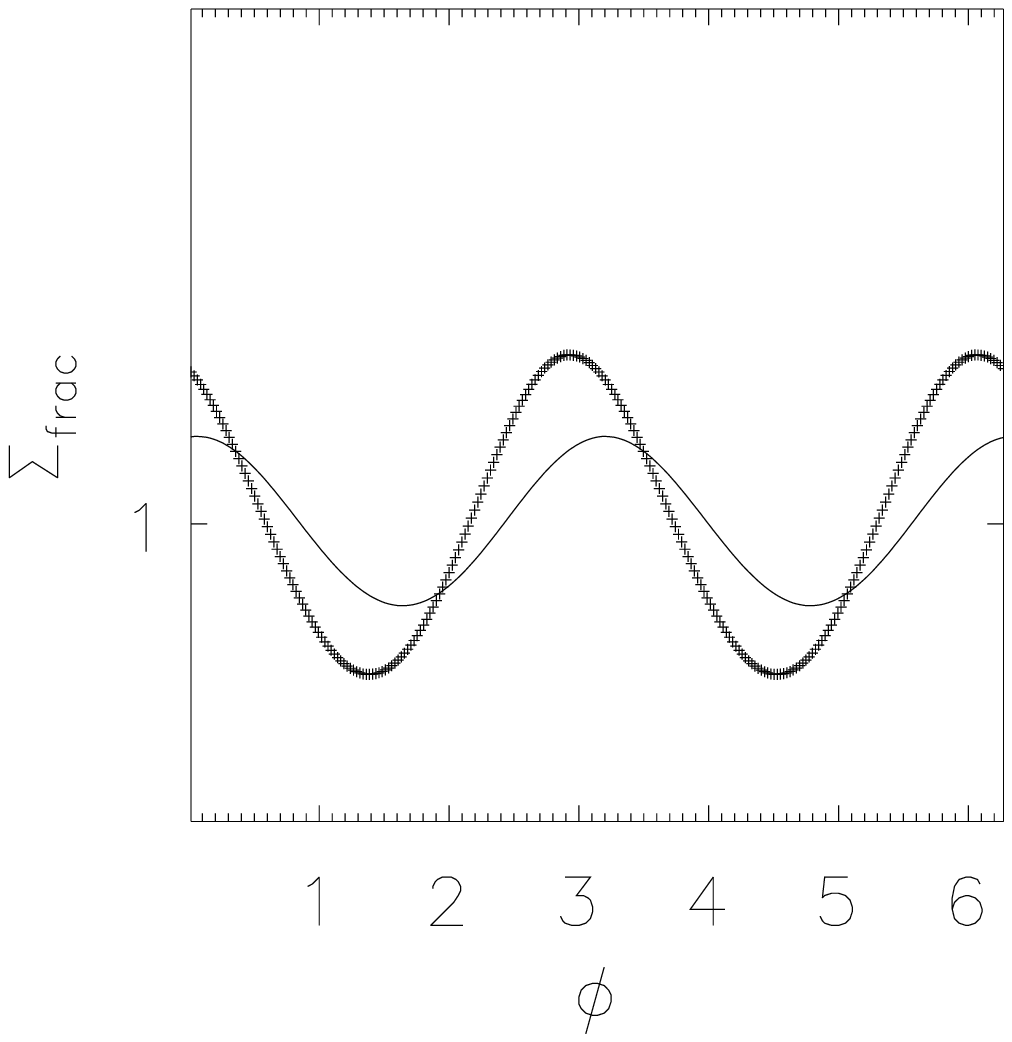,height=2.5in,width=2.5in}
\psfig{file=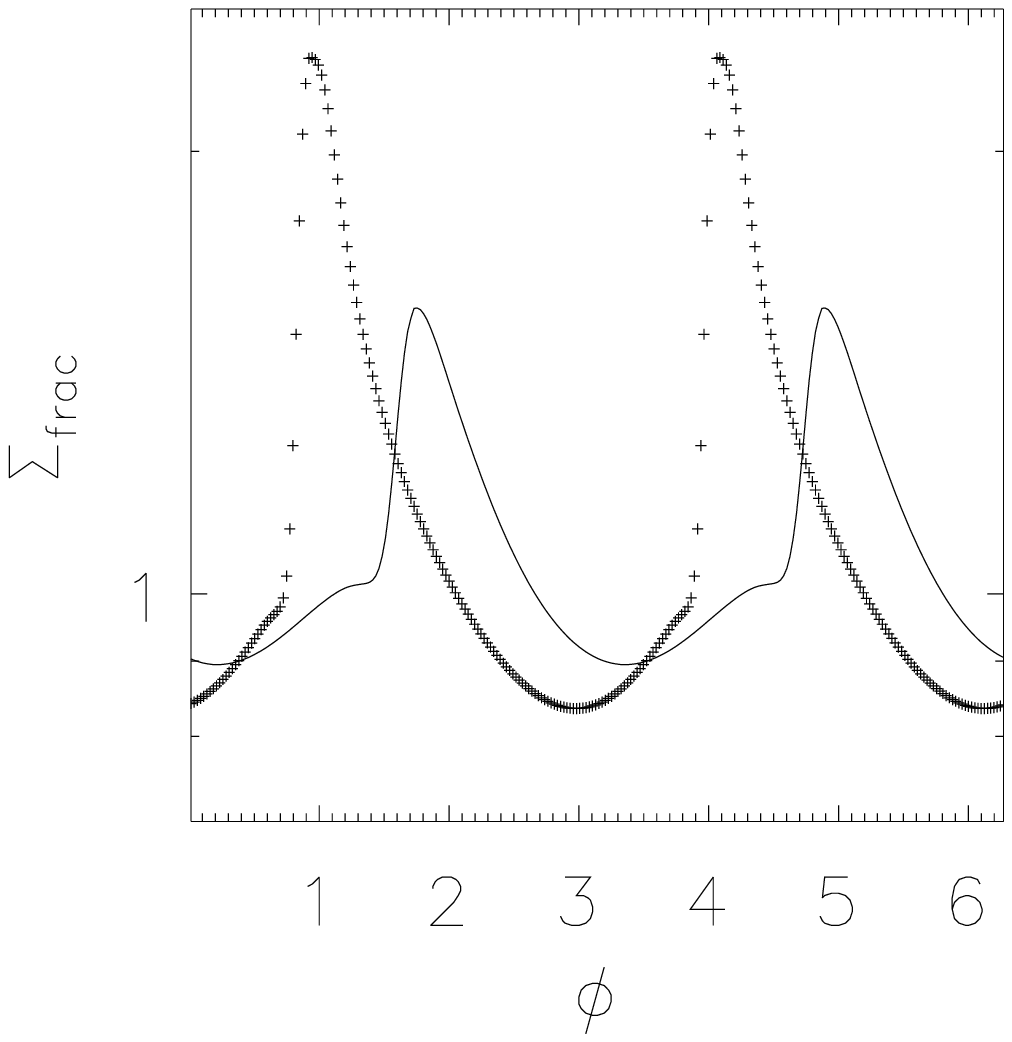,height=2.5in,width=2.5in}
\psfig{file=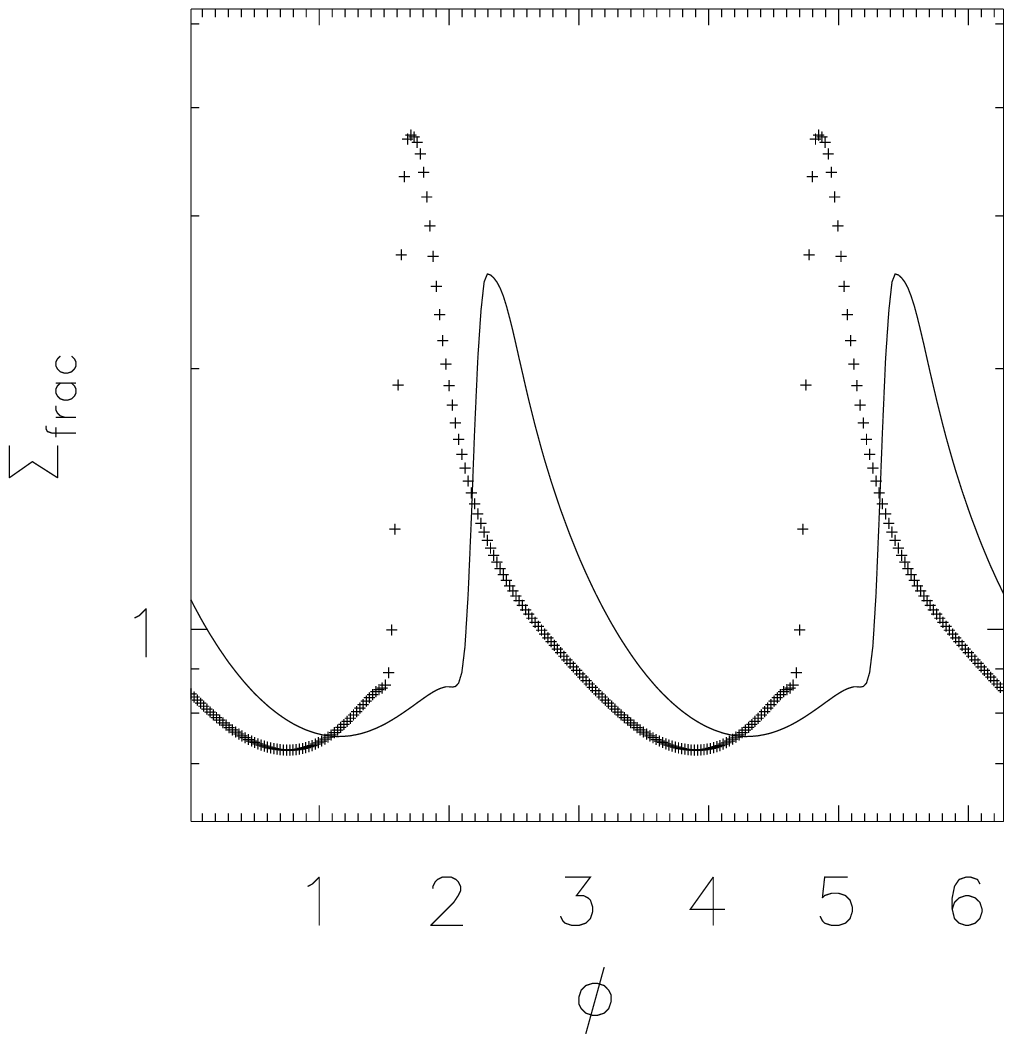,height=2.5in,width=2.5in}}

\end{center} 
\caption{(a)-(c) Snapshots of model HSYL at 95 Myr, 318 Myr, and 477 Myr,
(d)-(f) Azimuthal cuts at $\xi$=1.3 and $\xi$=1.17 at 95 Myr, 318 Myr, and 477 Myr}
\end{figure}

\begin{figure}[h] \begin{center}
%\centerline{\psfig{file=f9a.eps,height=2.5in,width=2.5in}
%\psfig{file=f9b.eps,height=2.5in,width=2.5in}
%\psfig{file=f9c.eps,height=2.5in,width=2.5in}}

\centerline{\psfig{file=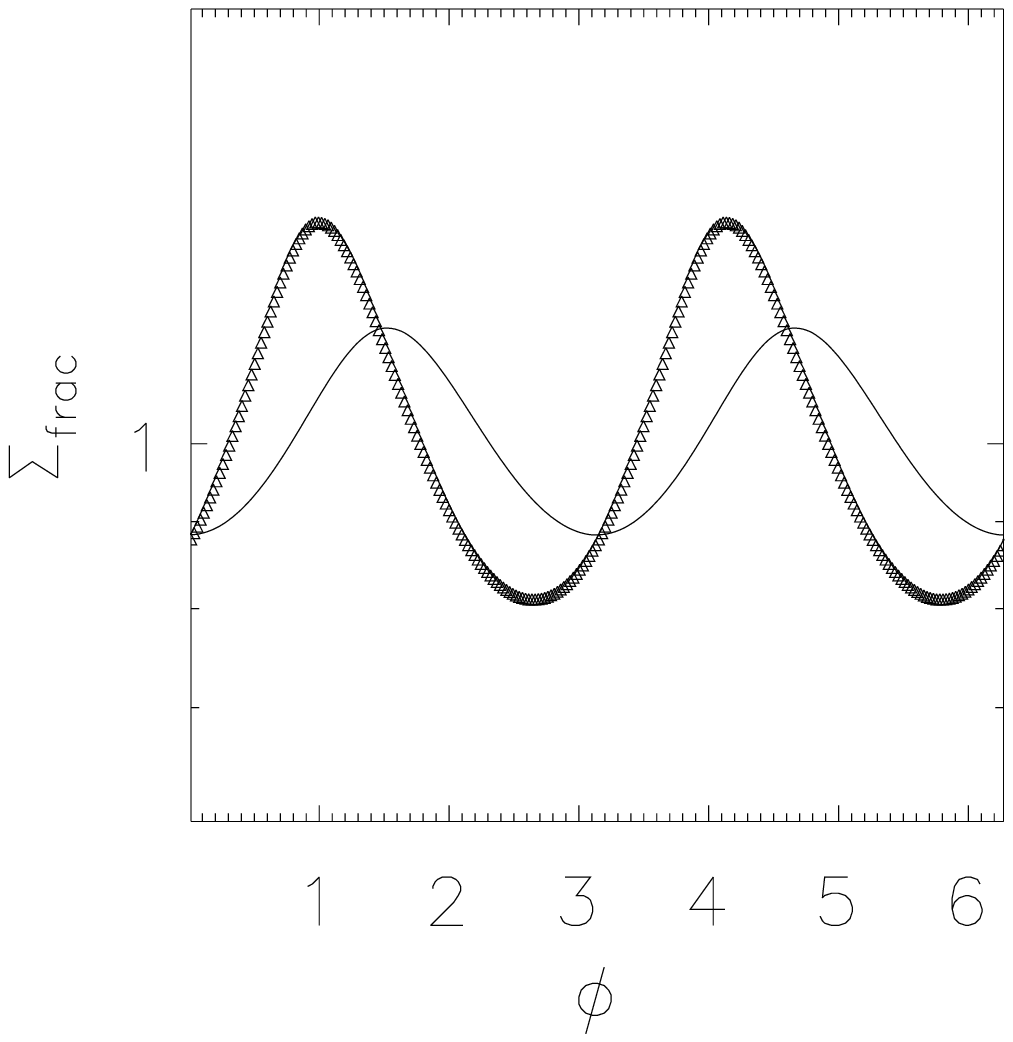,height=2.5in,width=2.5in}
\psfig{file=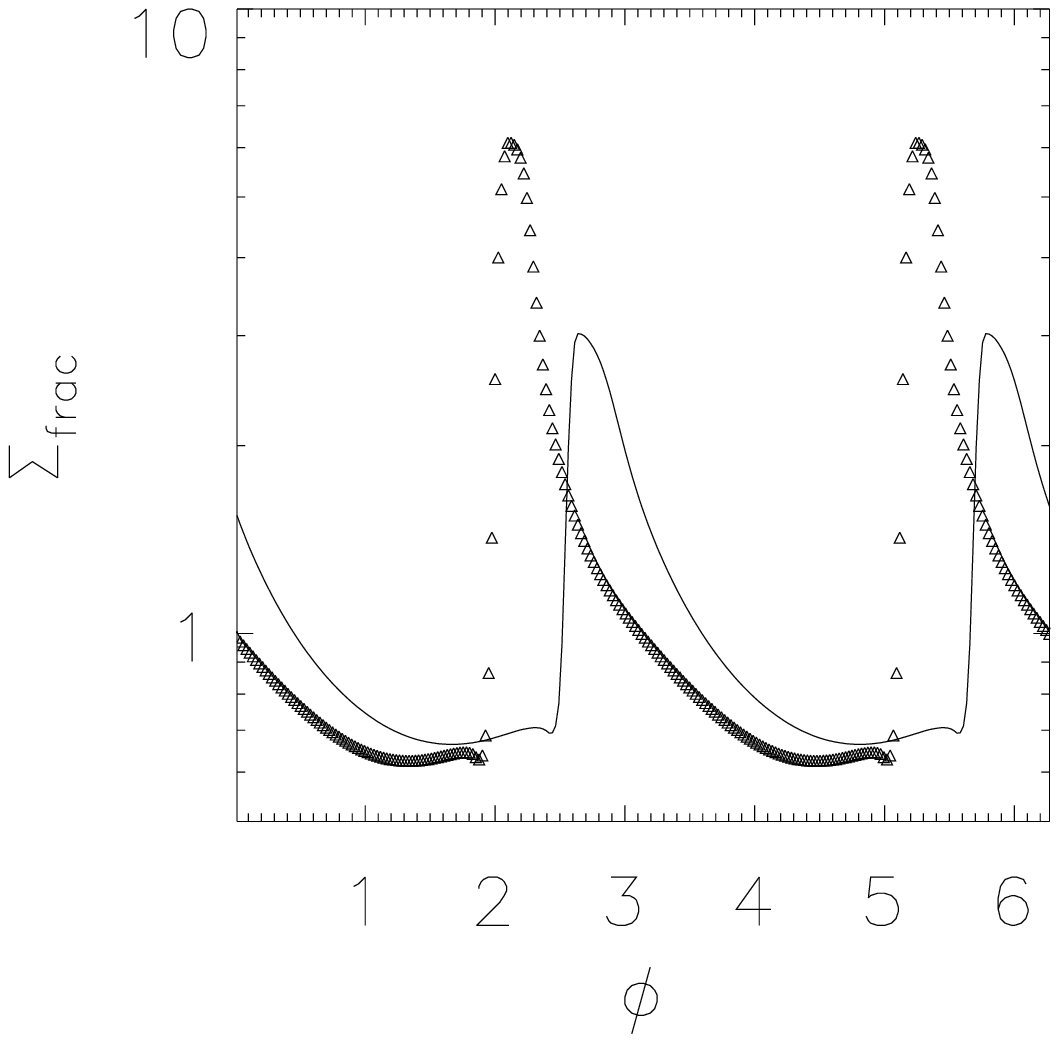,height=2.5in,width=2.5in}
\psfig{file=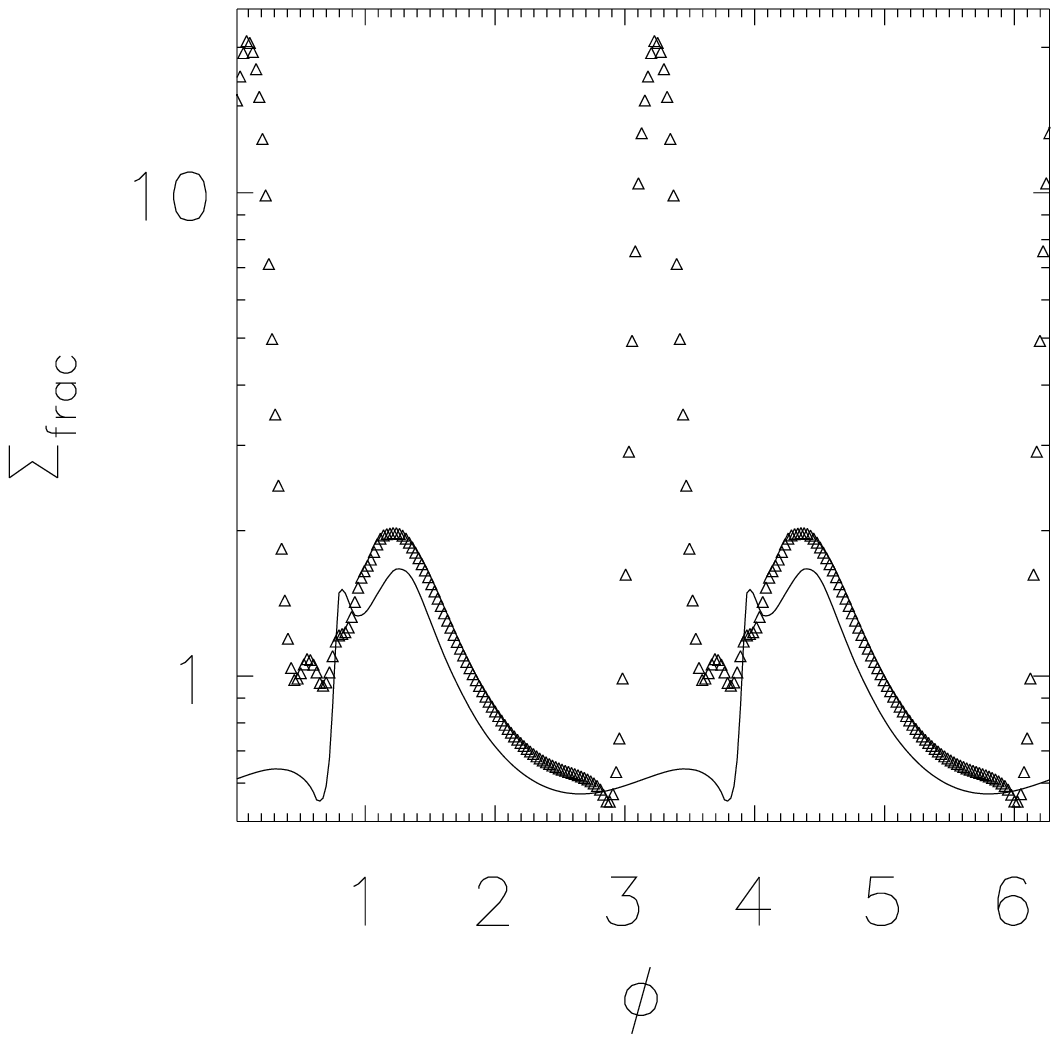,height=2.5in,width=2.5in}}

\end{center} 
\caption{(a)-(c) Snapshot of model H5 at 95 Myr, 318 Myr, and 477 Myr,
(d)-(f) Aziumthal cuts at $\xi$=1.3 and $\xi$=1.17 at 95 Myr, 318 Myr, and 477 Myr.  Emergent spurs are marked by arrows}
\end{figure}

\begin{figure}[h] \begin{center}
%\centerline{\psfig{file=f10a.eps,height=2.5in,width=2.5in}
%\psfig{file=f10b.eps,height=2.5in,width=2.5in}
%\psfig{file=f10c.eps,height=2.5in,width=2.5in}}

\centerline{\psfig{file=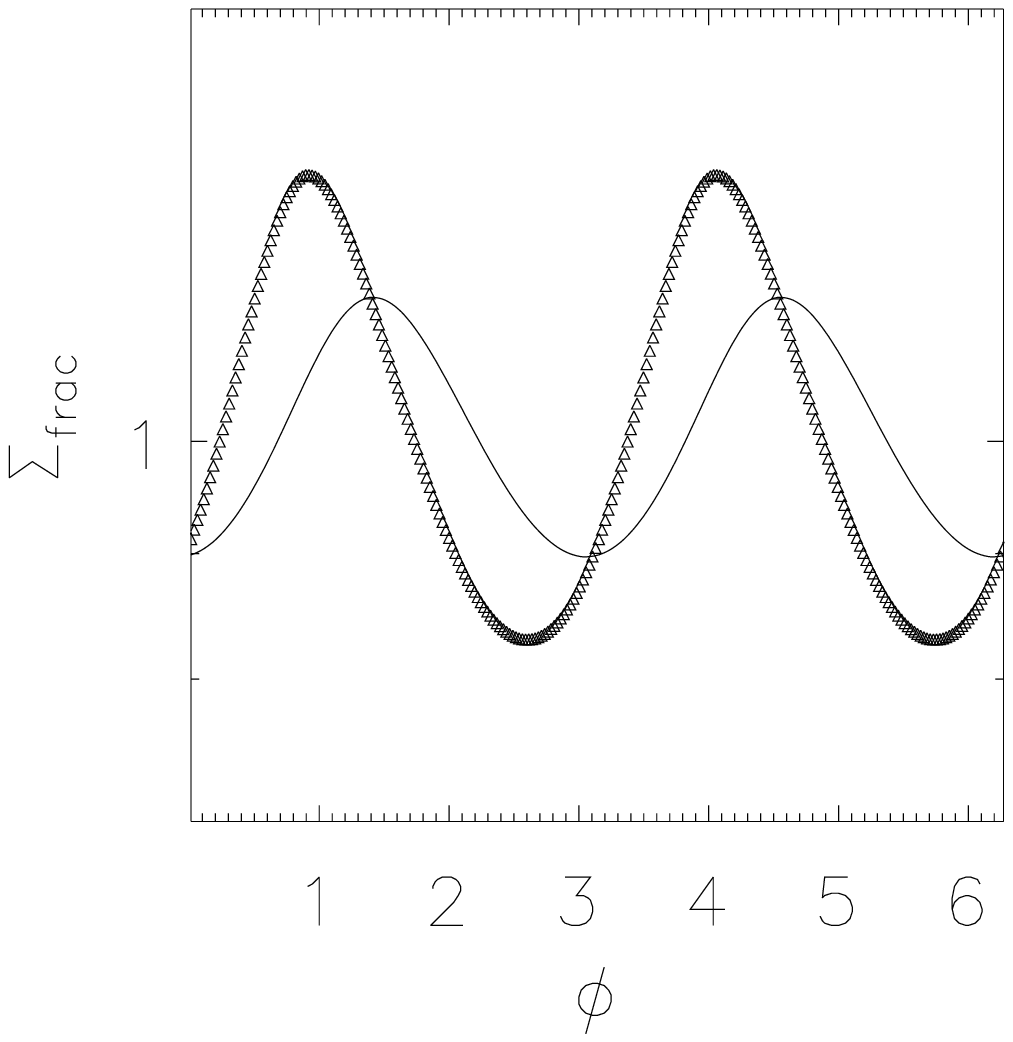,height=2.5in,width=2.5in}
\psfig{file=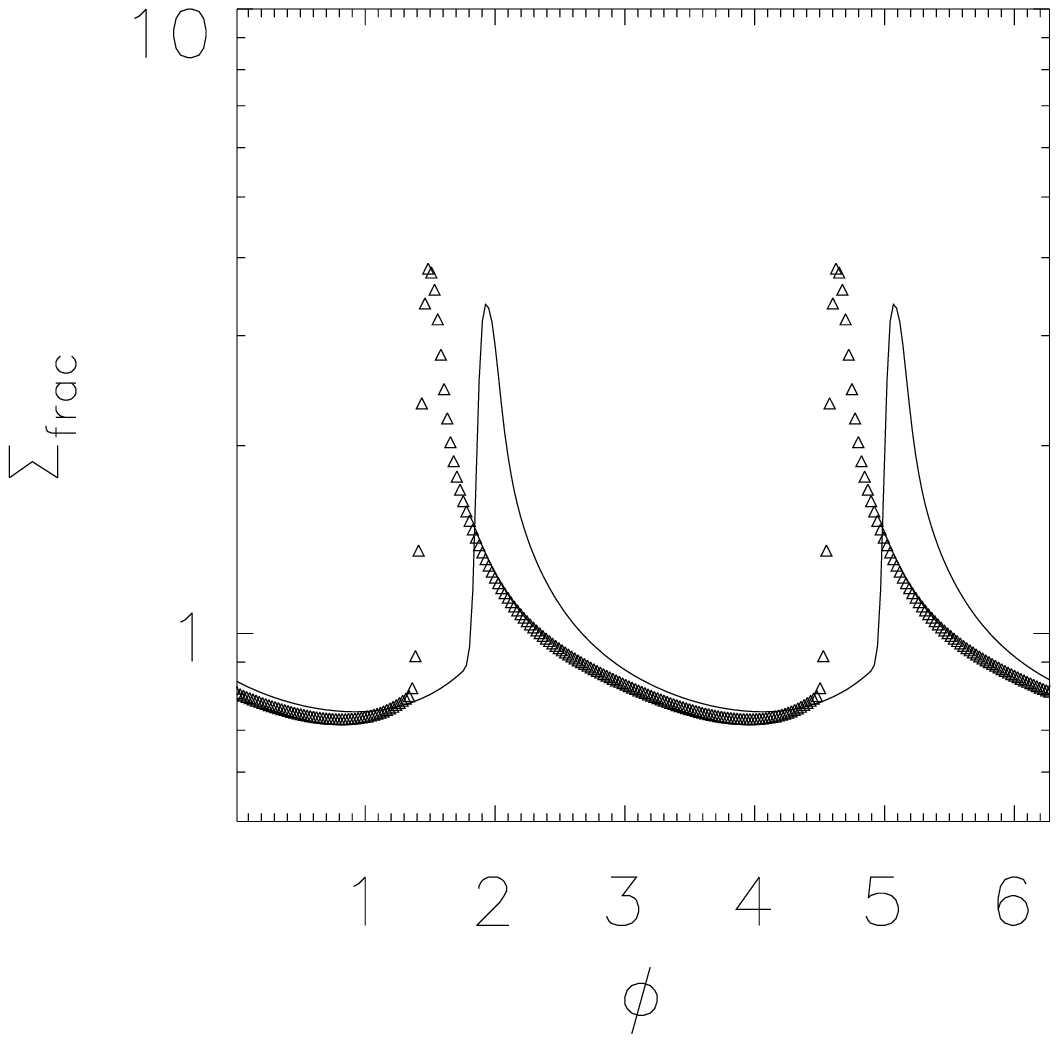,height=2.5in,width=2.5in}
\psfig{file=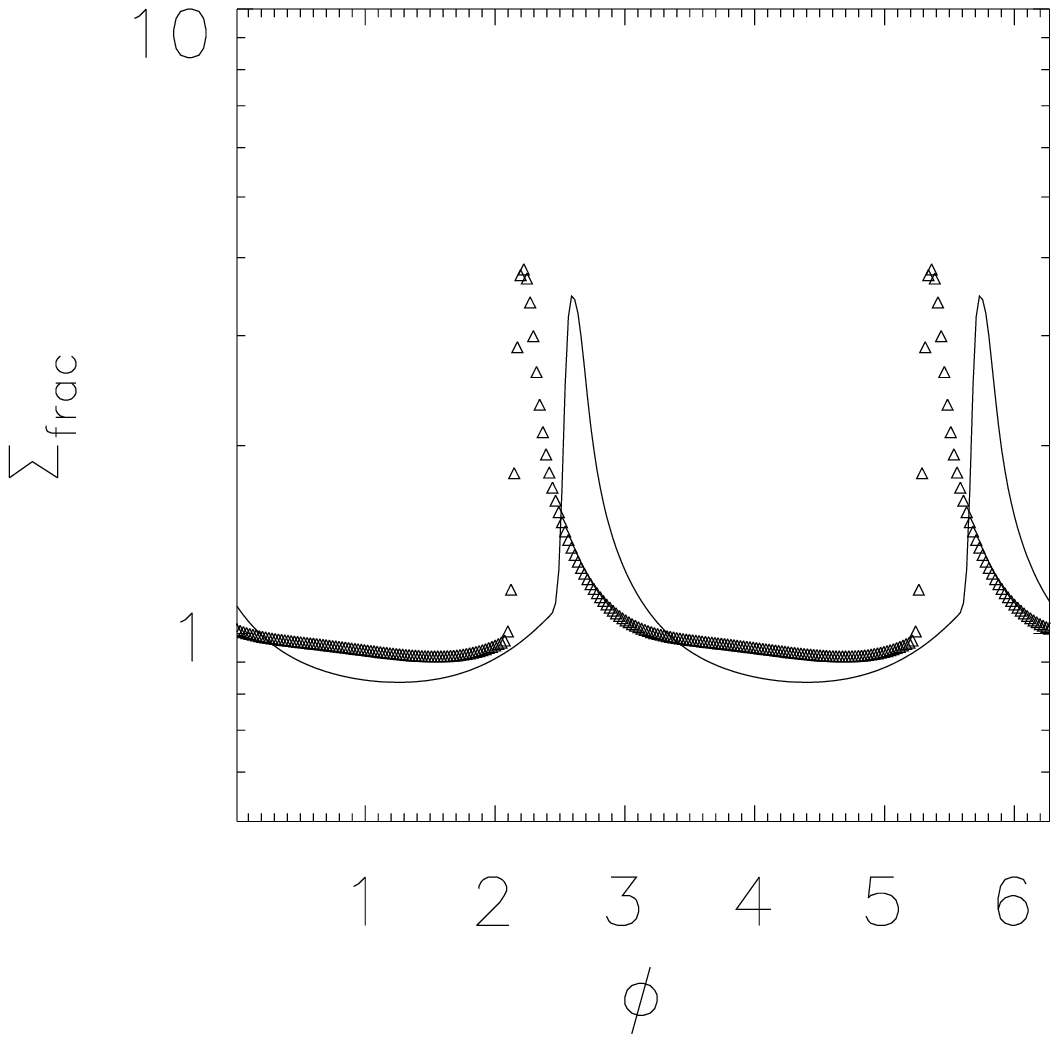,height=2.5in,width=2.5in}}
\end{center}
\caption{(a)-(c) Snapshot of model H6 at 95 Myr, 318 Myr, and 477 Myr,
(d)-(f) Azimuthal cuts at $\xi$=1.3 and $\xi$=1.17 at 95 Myr, 318 Myr, and 477 Myr}
\end{figure}

\begin{figure}[h] \begin{center}
%\centerline{\psfig{file=f11a.eps,height=2.5in,width=2.5in}
%\psfig{file=f11b.eps,height=2.5in,width=2.5in}
%\psfig{file=f11c.eps,height=2.5in,width=2.5in}}

\centerline{\psfig{file=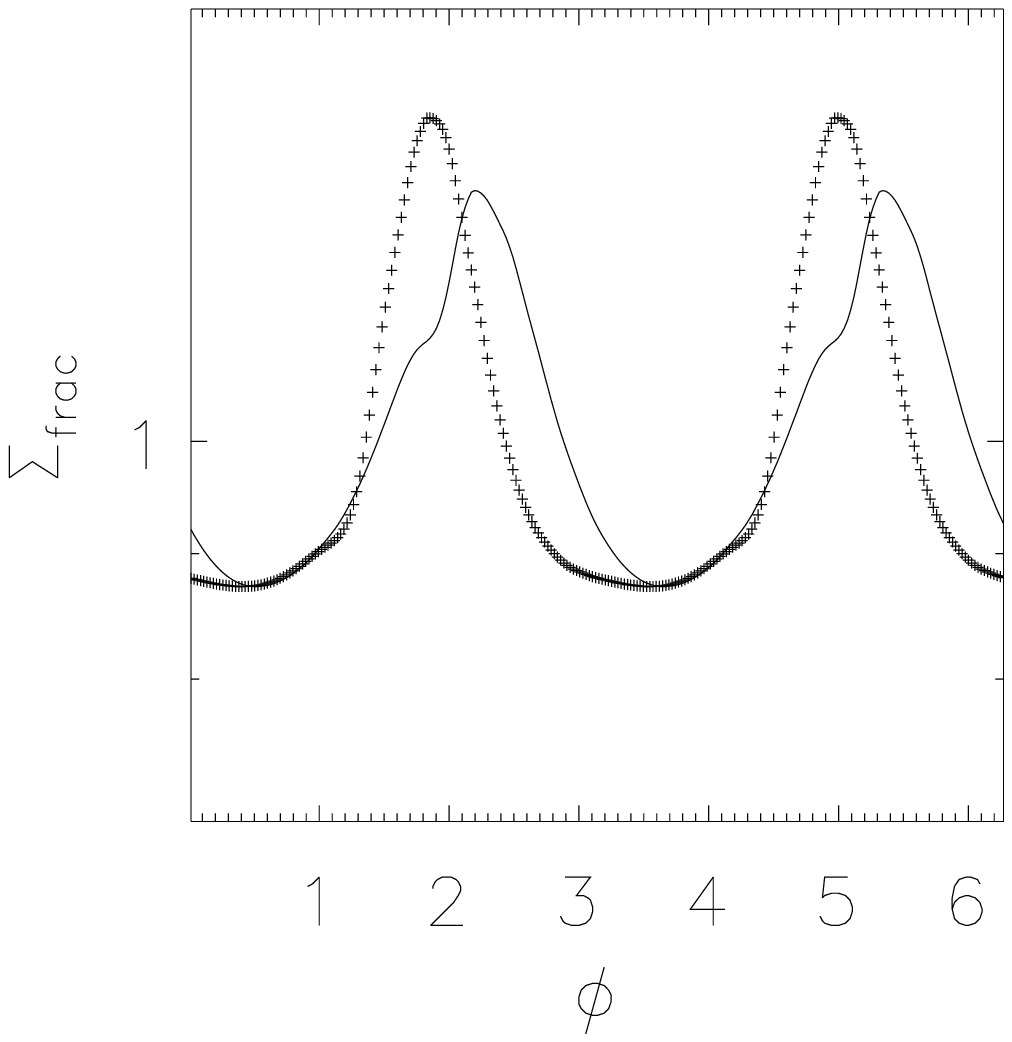,height=2.5in,width=2.5in}
\psfig{file=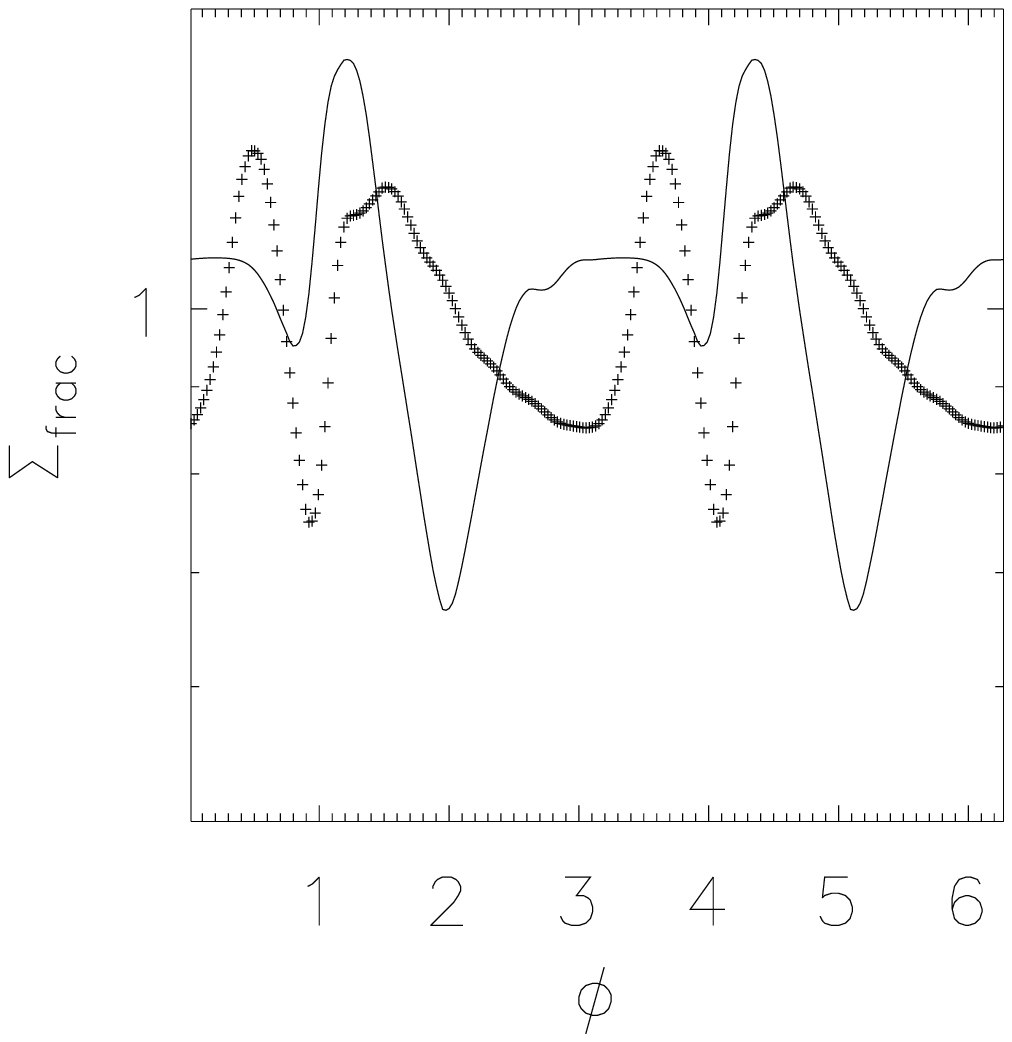,height=2.5in,width=2.5in}
\psfig{file=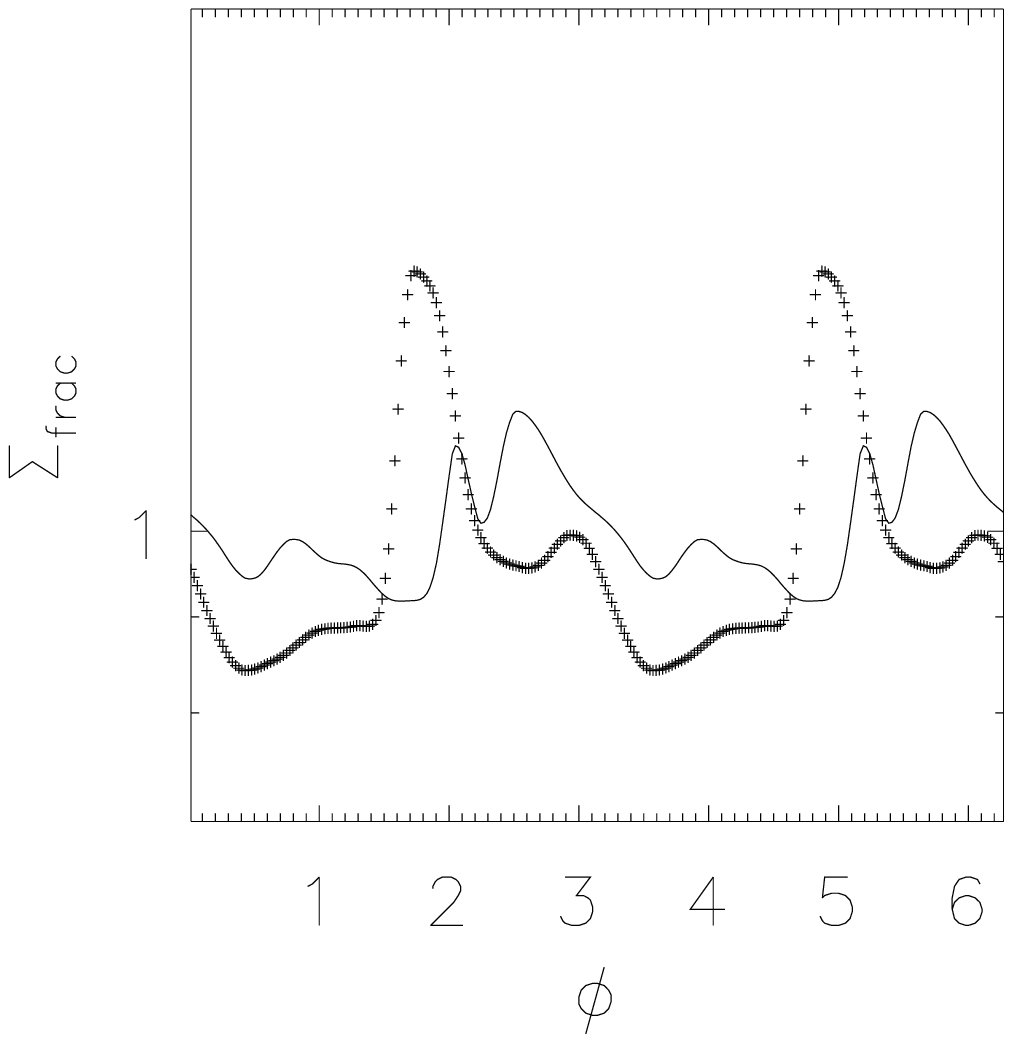,height=2.5in,width=2.5in}}
\end{center}
\caption{(a)-(c) Snapshot of model L1 at 573 Myr, 1.91 Gyr, and 2.87 Gyr,
(d)-(f) Azimuthal cuts at $\xi$=1.7 and $\xi$=1.56 at 573 Myr, 1.91 Gyr, and 2.87 Gyr}
\end{figure}

\begin{figure}[h] \begin{center}
%\centerline{\psfig{file=f12a.eps,height=2.5in,width=2.5in}
%\psfig{file=f12b.eps,height=2.5in,width=2.5in}
%\psfig{file=f12c.eps,height=2.5in,width=2.5in}}

\centerline{\psfig{file=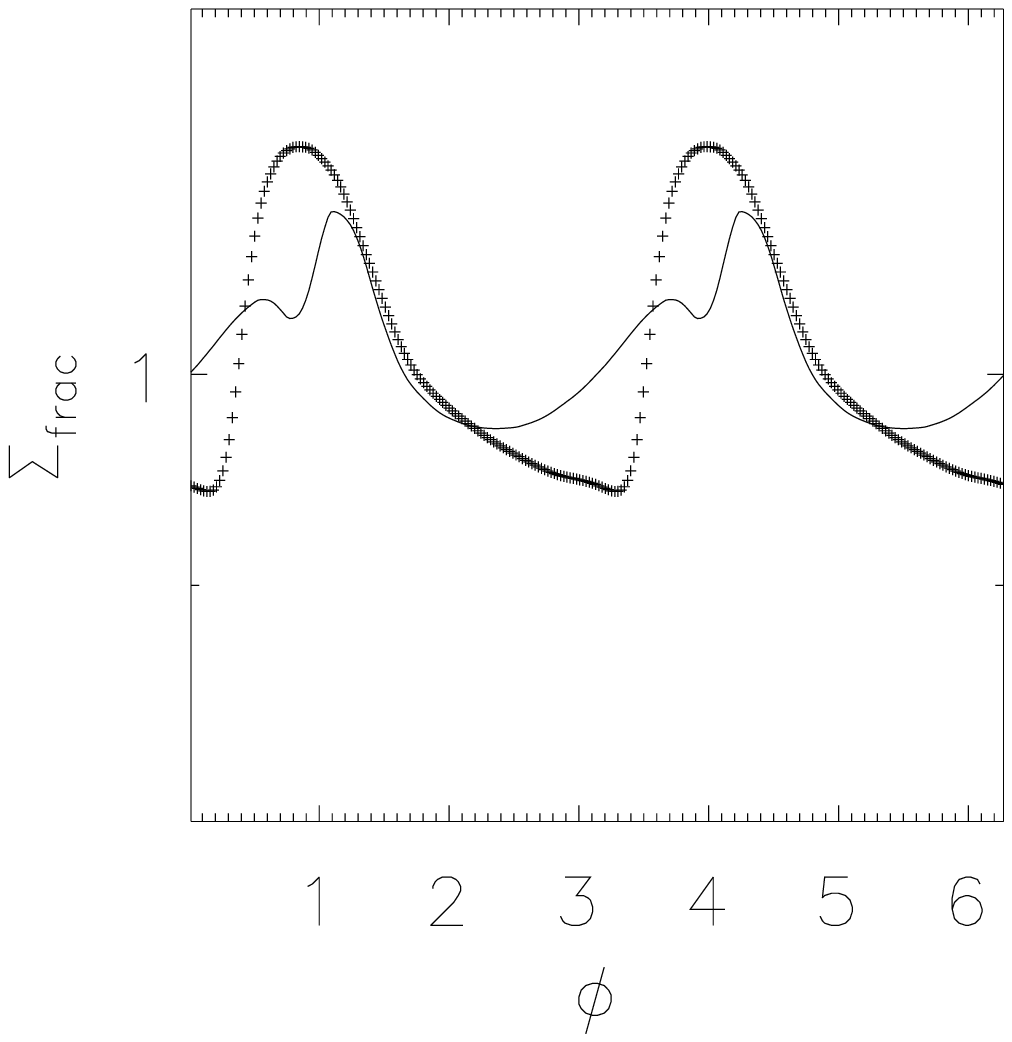,height=2.5in,width=2.5in}
\psfig{file=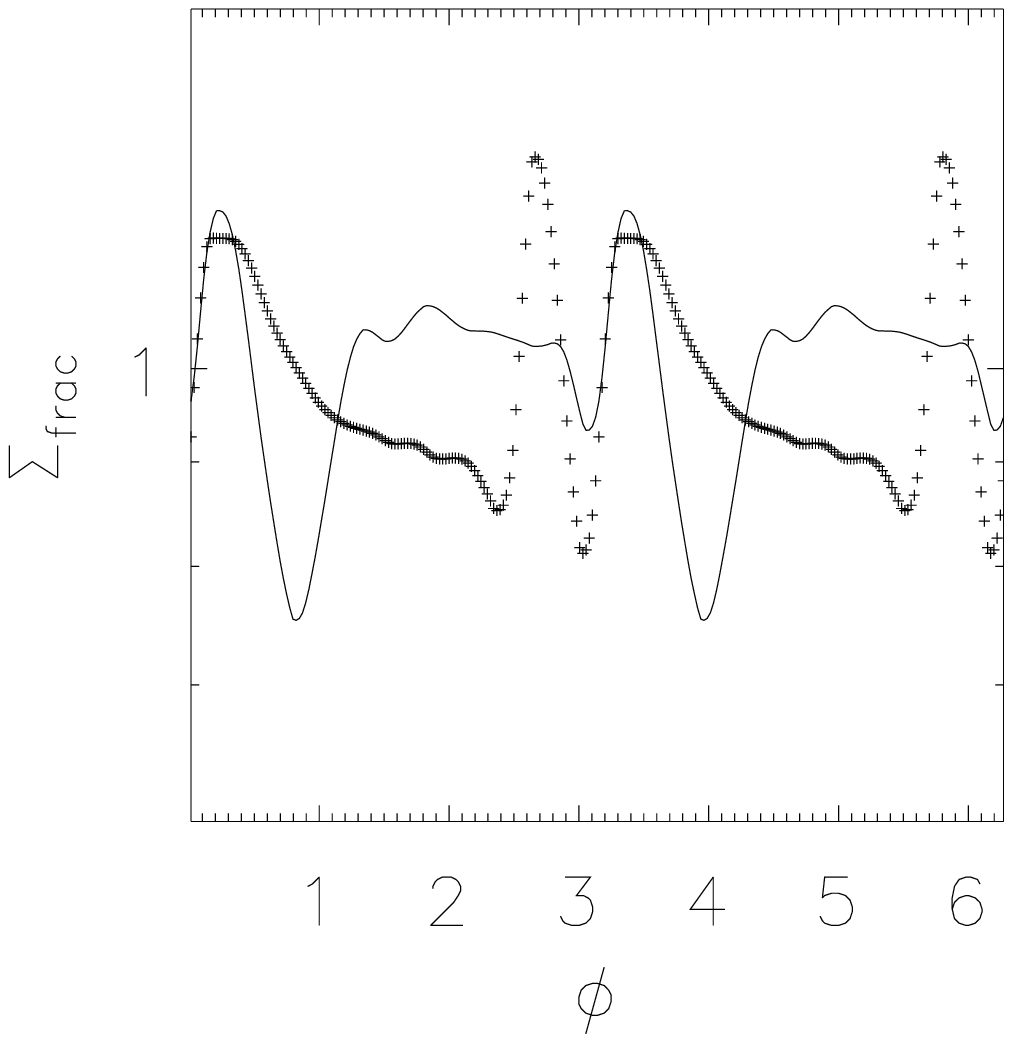,height=2.5in,width=2.5in}
\psfig{file=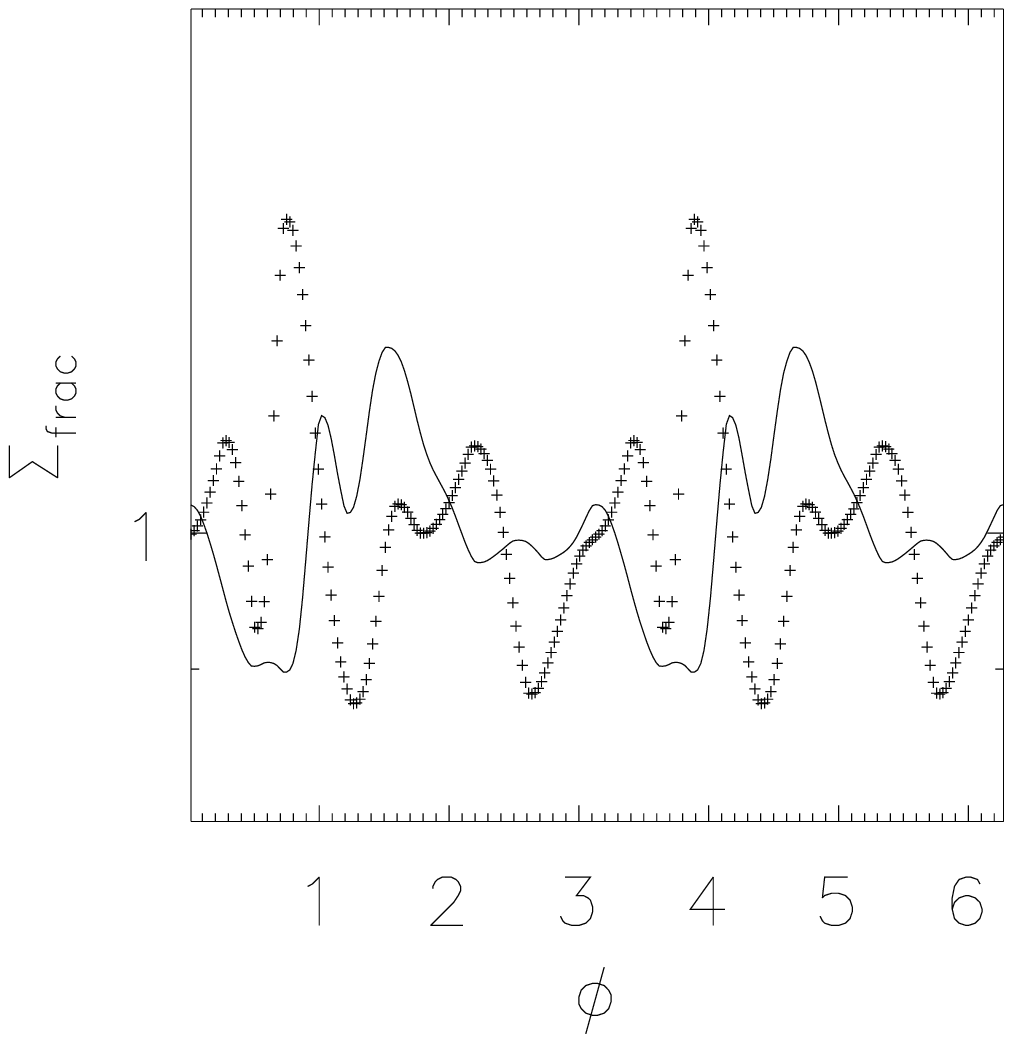,height=2.5in,width=2.5in}}
\end{center}
\caption{(a)-(c) Snapshot of model LSYL1 at 573 Myr, 1.91 Gyr, and 2.87 Gyr,
(d)-(f) Azimuthal cuts at $\xi$=1.7 and $\xi$=1.56 at 573 Myr, 1.91 Gyr, and 2.87 Gyr}
\end{figure}

\begin{figure}[h] \begin{center}
%\centerline{\psfig{file=f13a.eps,height=2.5in,width=2.5in}
%\psfig{file=f13b.eps,height=2.5in,width=2.5in}
%\psfig{file=f13c.eps,height=2.5in,width=2.5in}}

\centerline{\psfig{file=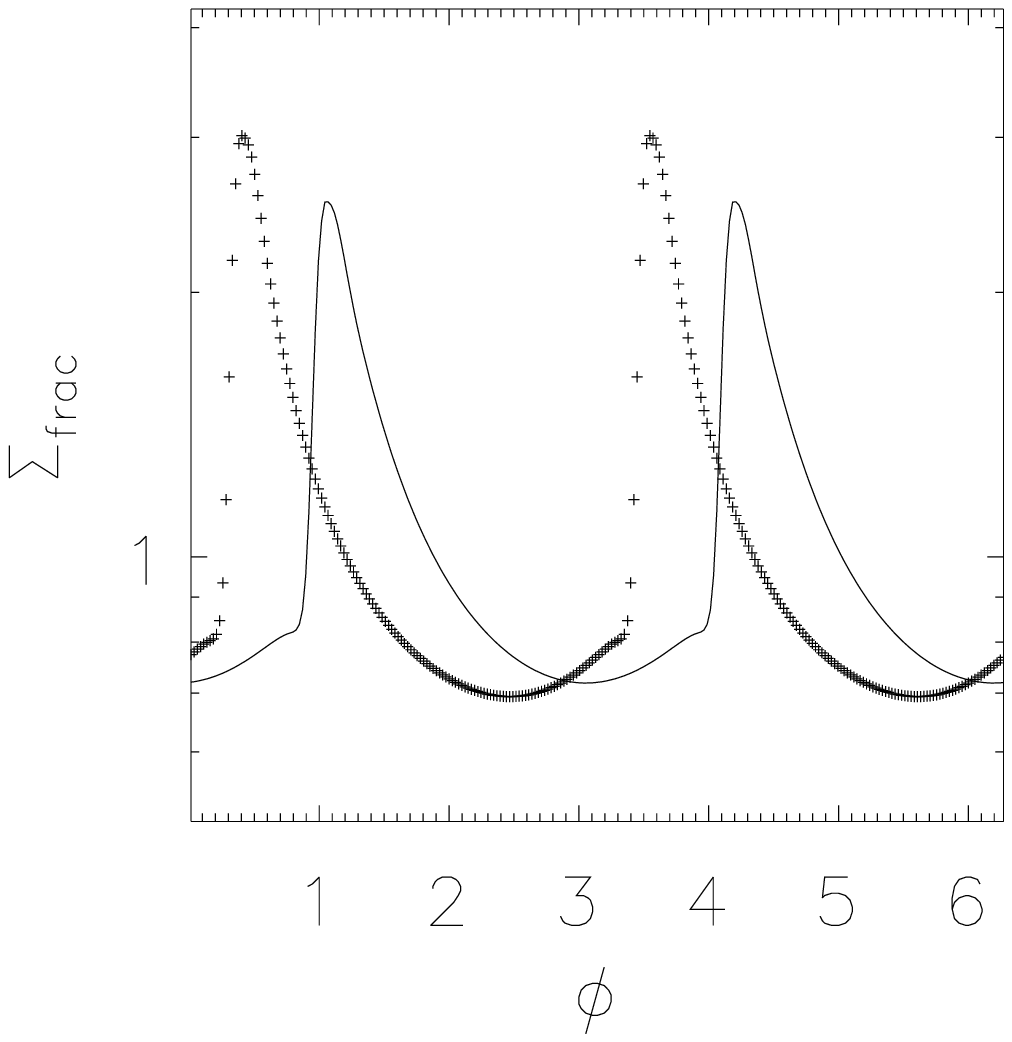,height=2.5in,width=2.5in}
\psfig{file=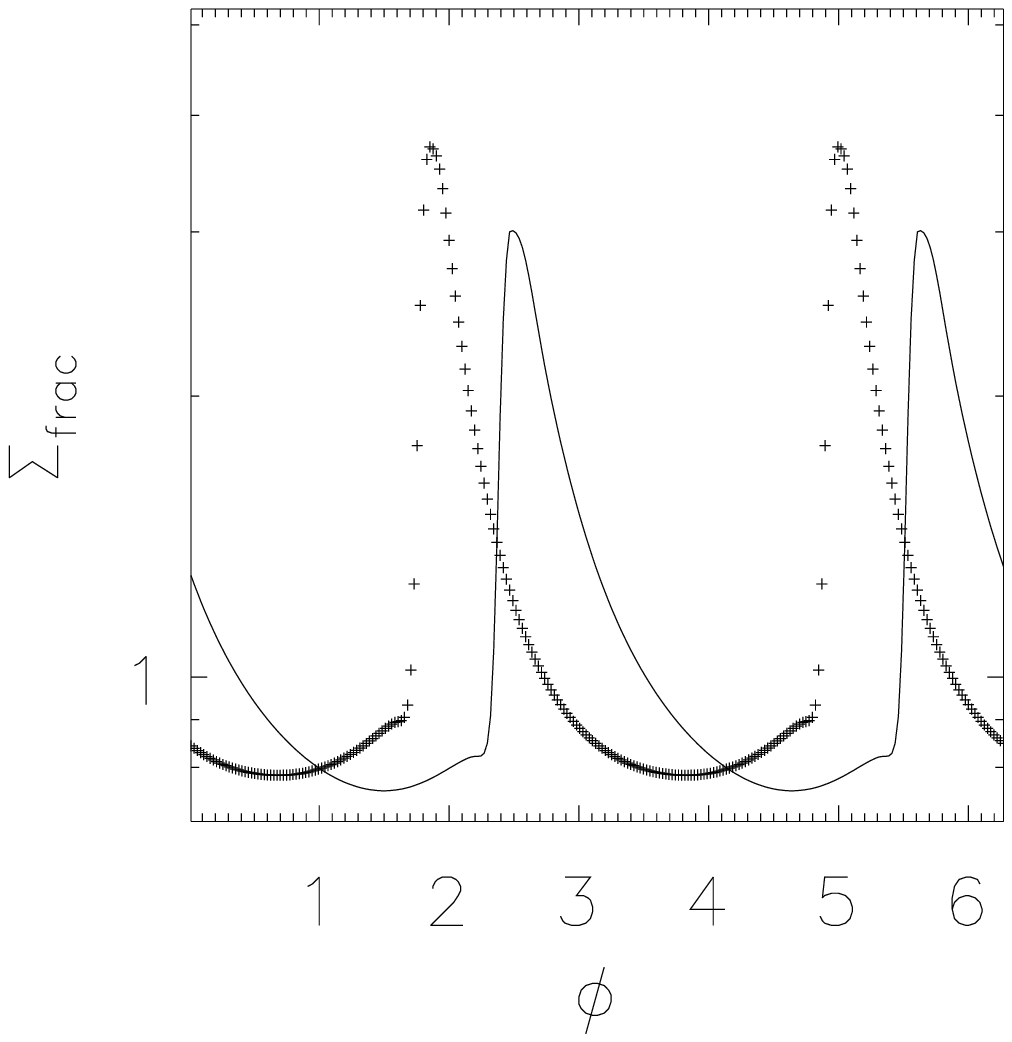,height=2.5in,width=2.5in}
\psfig{file=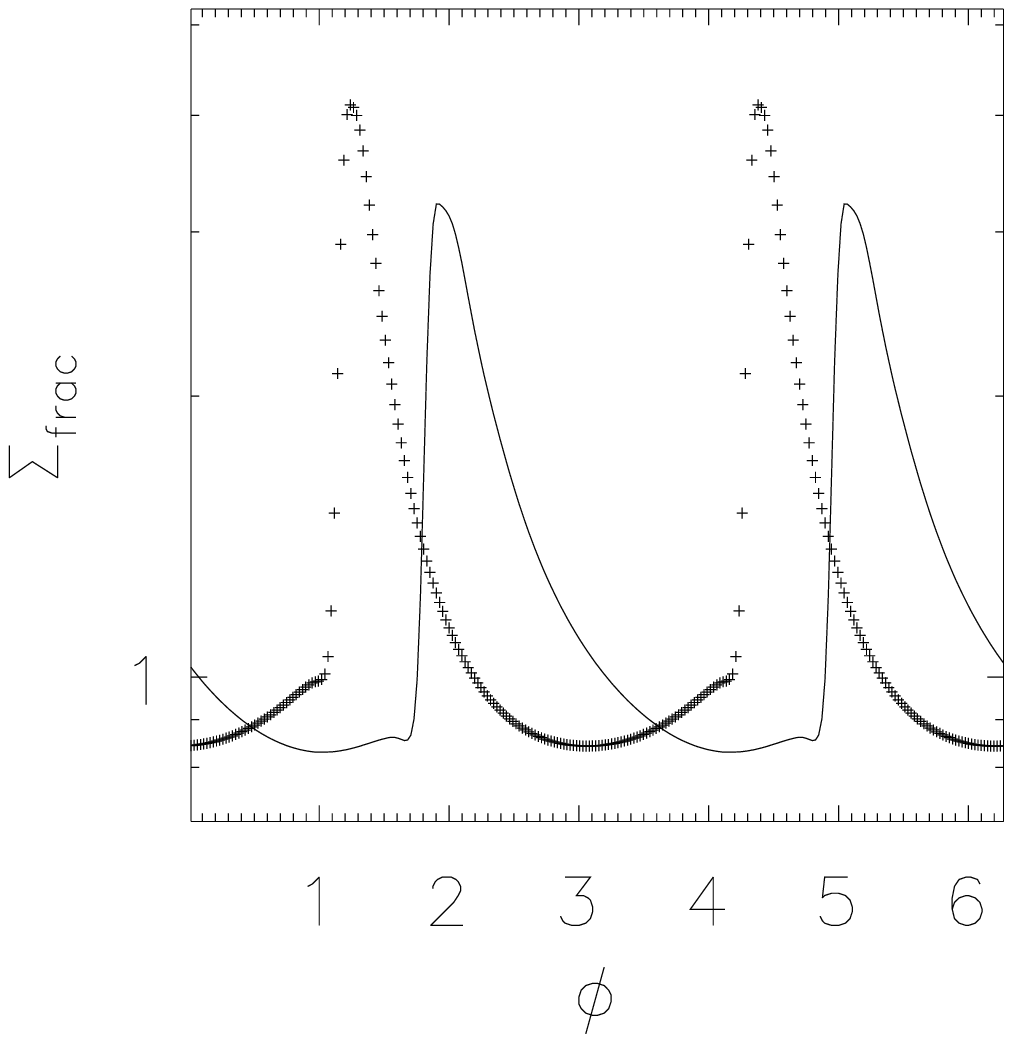,height=2.5in,width=2.5in}}
\end{center}
\caption{(a)-(c) Snapshot of model L2 at 573 Myr, 1.91 Gyr, and 2.87 Gyr,
(d)-(f) Azimuthal cuts at $\xi$=1.3 and $\xi$=1.17 at 573 Myr, 1.91 Gyr, and 2.87 Gyr}
\end{figure}

\begin{figure}[h] \begin{center}
%\centerline{\psfig{file=f14a.eps,height=2.5in,width=2.5in}
%\psfig{file=f14b.eps,height=2.5in,width=2.5in}
%\psfig{file=f14c.eps,height=2.5in,width=2.5in}}

\centerline{\psfig{file=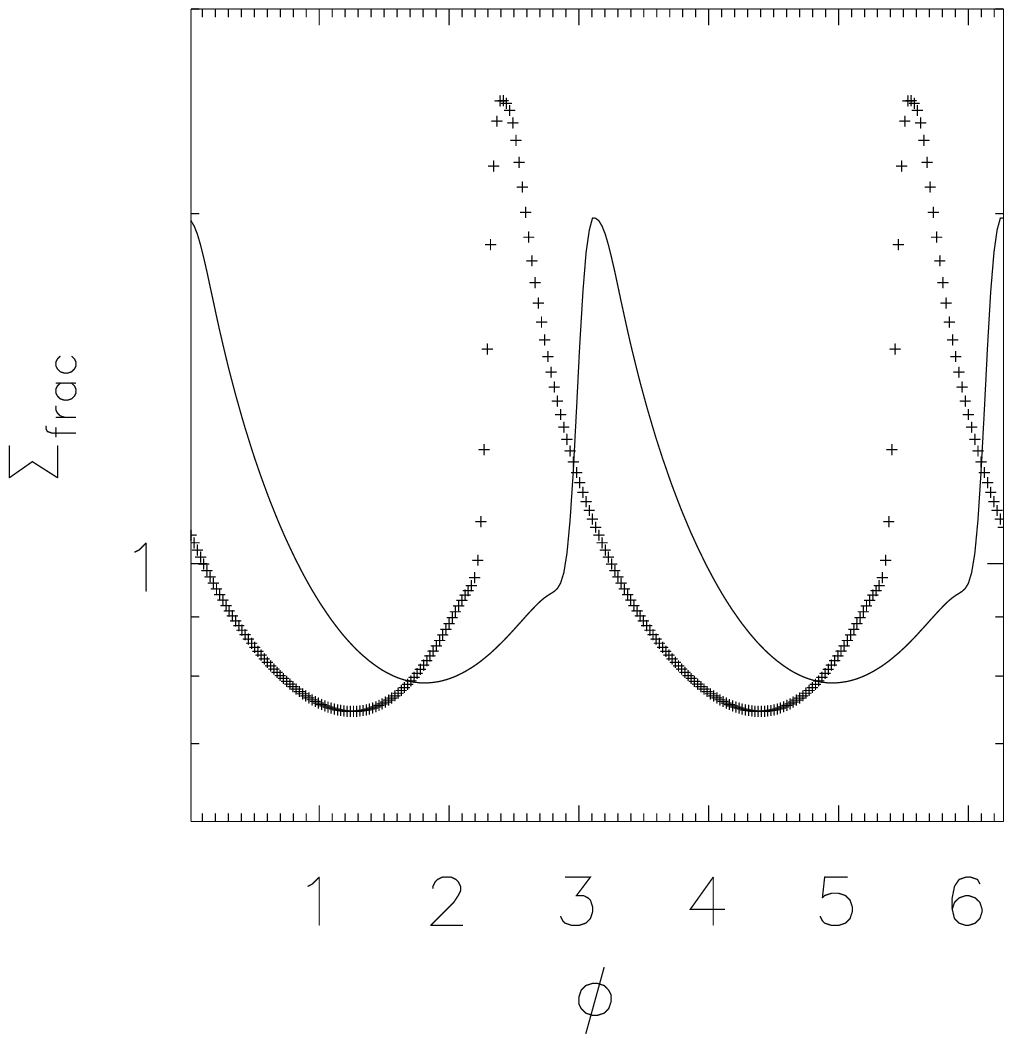,height=2.5in,width=2.5in}
\psfig{file=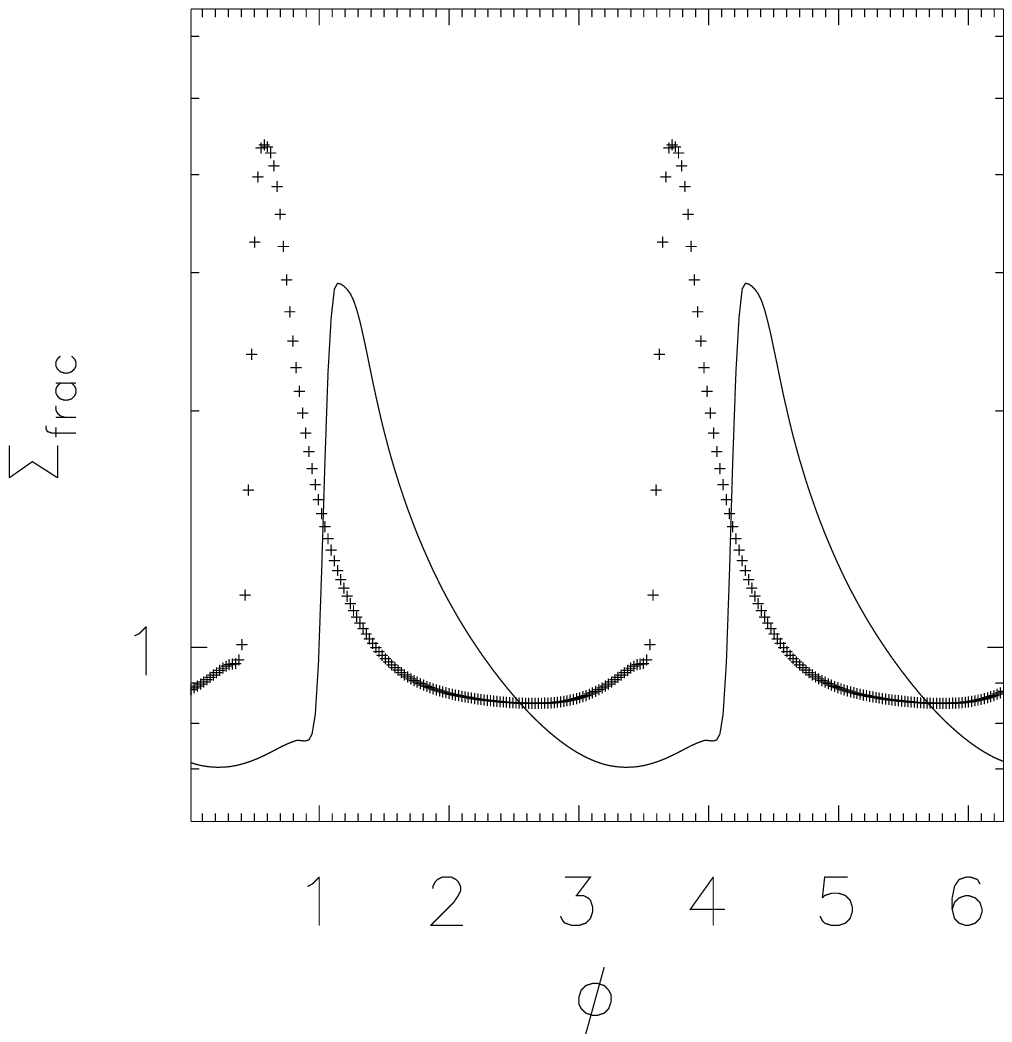,height=2.5in,width=2.5in}
\psfig{file=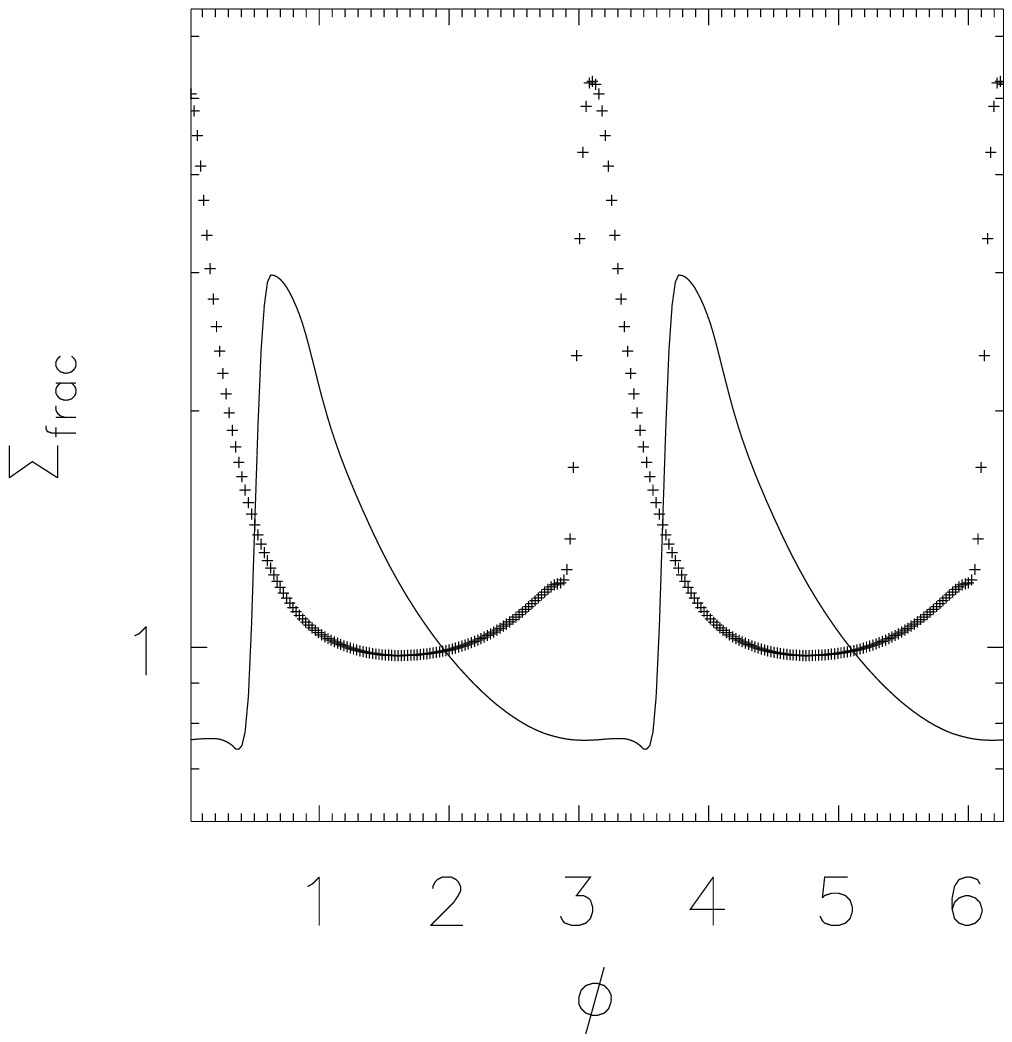,height=2.5in,width=2.5in}}
\end{center}
\caption{(a)-(c) Snapshot of model LSYL2 at 573 Myr, 1.91 Gyr, and 2.87 Gyr,
(d)-(f) Azimuthal cuts at $\xi$=1.3 and $\xi$=1.17 at 573 Myr, 1.91 Gyr, and 2.87 Gyr}
\end{figure}

\begin{figure}[h] \begin{center}
%\centerline{\psfig{file=f15a.eps,height=2.5in,width=2.5in}
%\psfig{file=f15b.eps,height=2.5in,width=2.5in}
%\psfig{file=f15c.eps,height=2.5in,width=2.5in}}

\centerline{\psfig{file=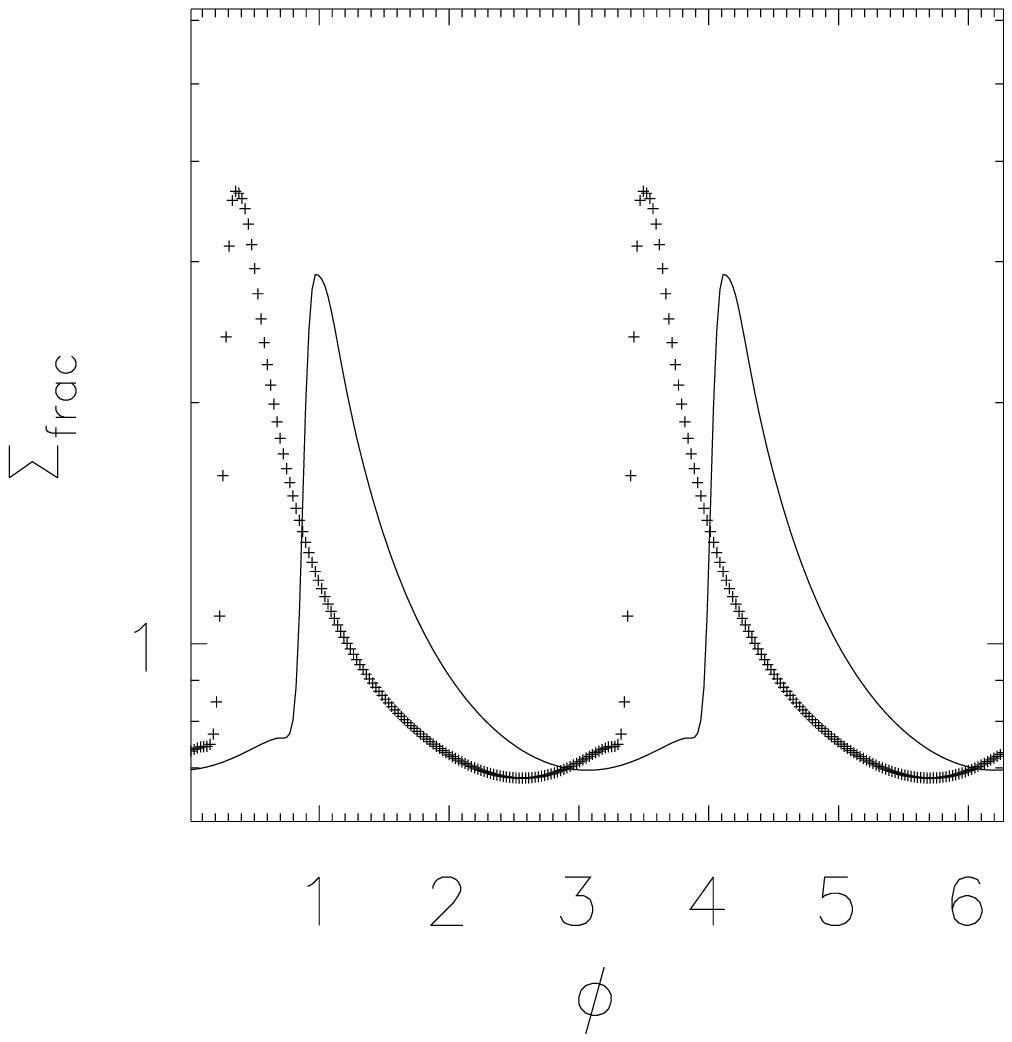,height=2.5in,width=2.5in}
\psfig{file=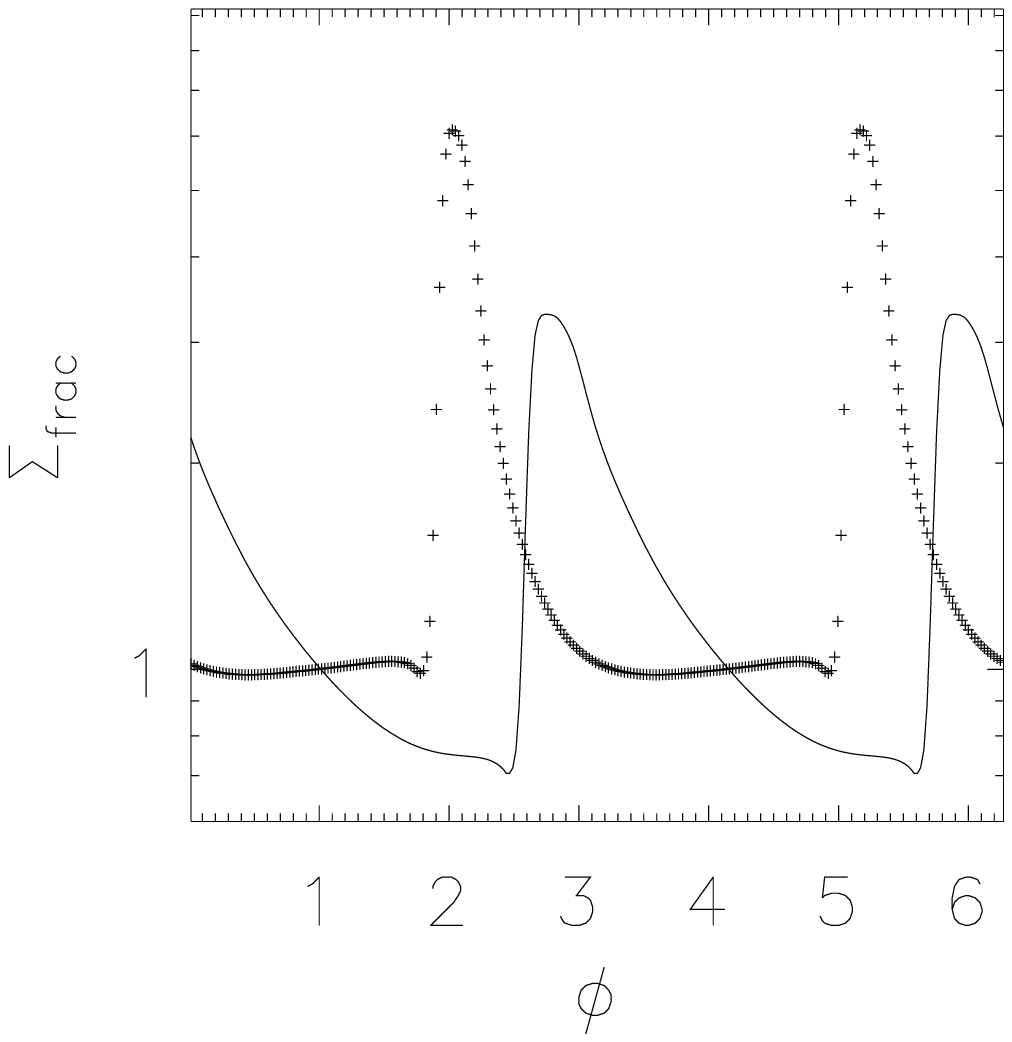,height=2.5in,width=2.5in}
\psfig{file=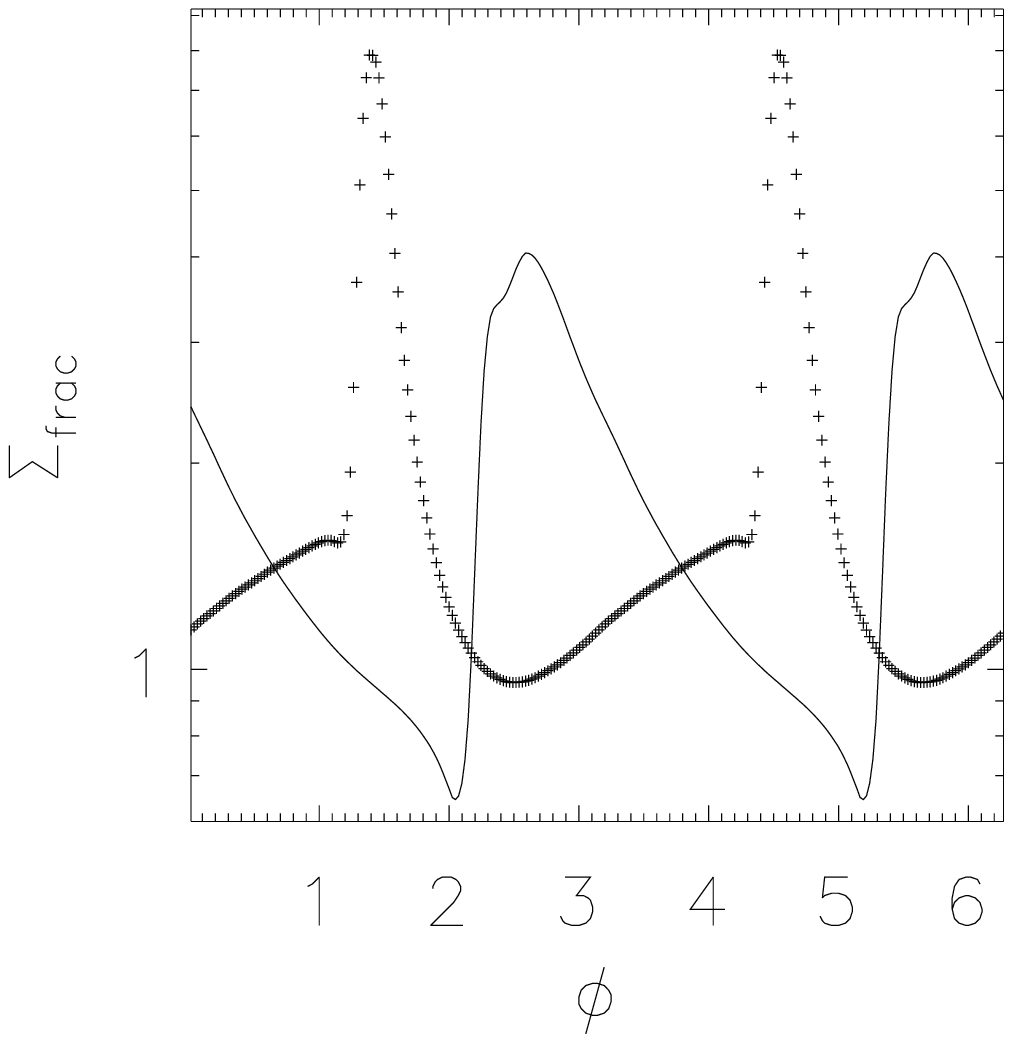,height=2.5in,width=2.5in}}
\end{center}
\caption{(a)-(c) Snapshot of model L3 at 573 Myr, 1.91 Gyr, and 2.87 Gyr,
(d)-(f) Azimuthal cuts at $\xi$=1.3 and $\xi$=1.17 at 573 Myr, 1.91 Gyr, and 2.87 Gyr}
\end{figure}

\begin{figure}[h] \begin{center}
%\centerline{\psfig{file=f16a.eps,height=2.5in,width=2.5in}
%\psfig{file=f16b.eps,height=2.5in,width=2.5in}
%\psfig{file=f16c.eps,height=2.5in,width=2.5in}}

\centerline{\psfig{file=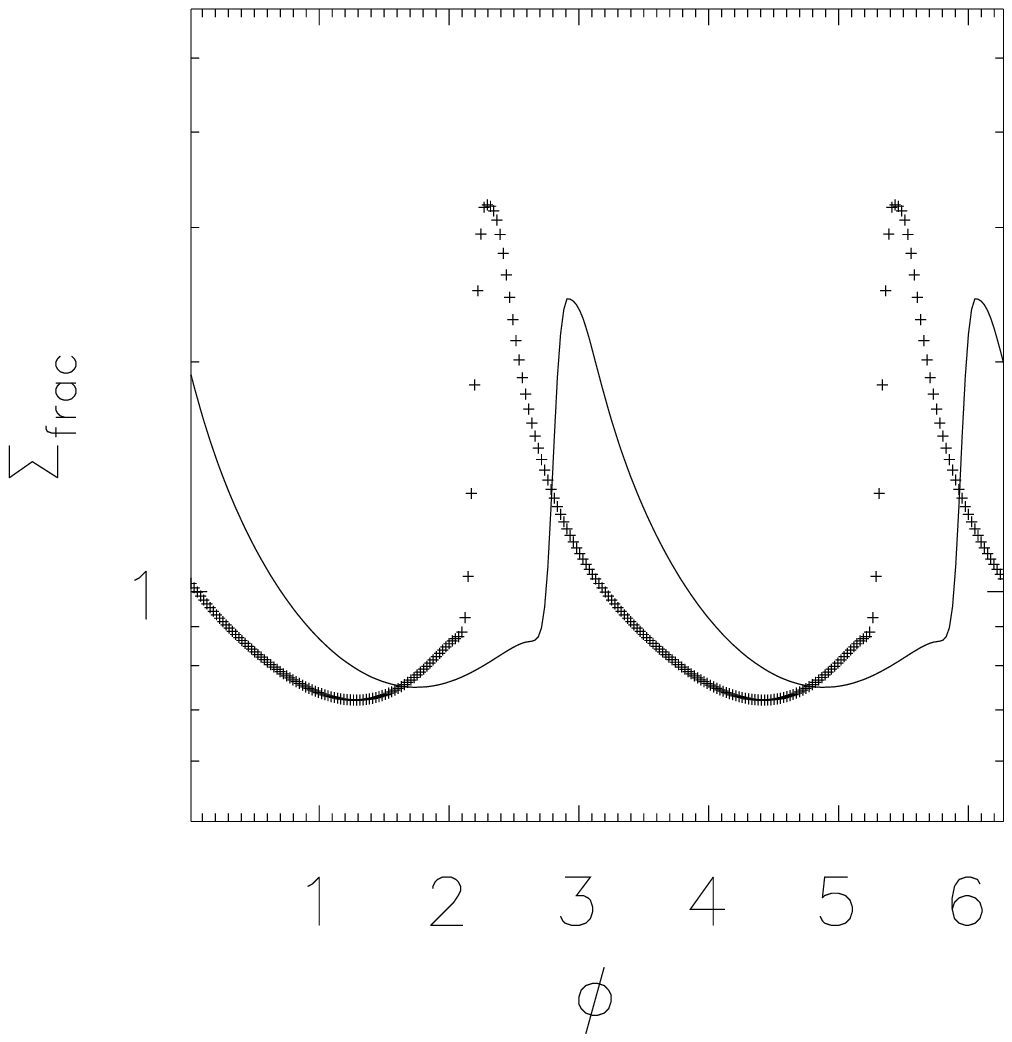,height=2.5in,width=2.5in}
\psfig{file=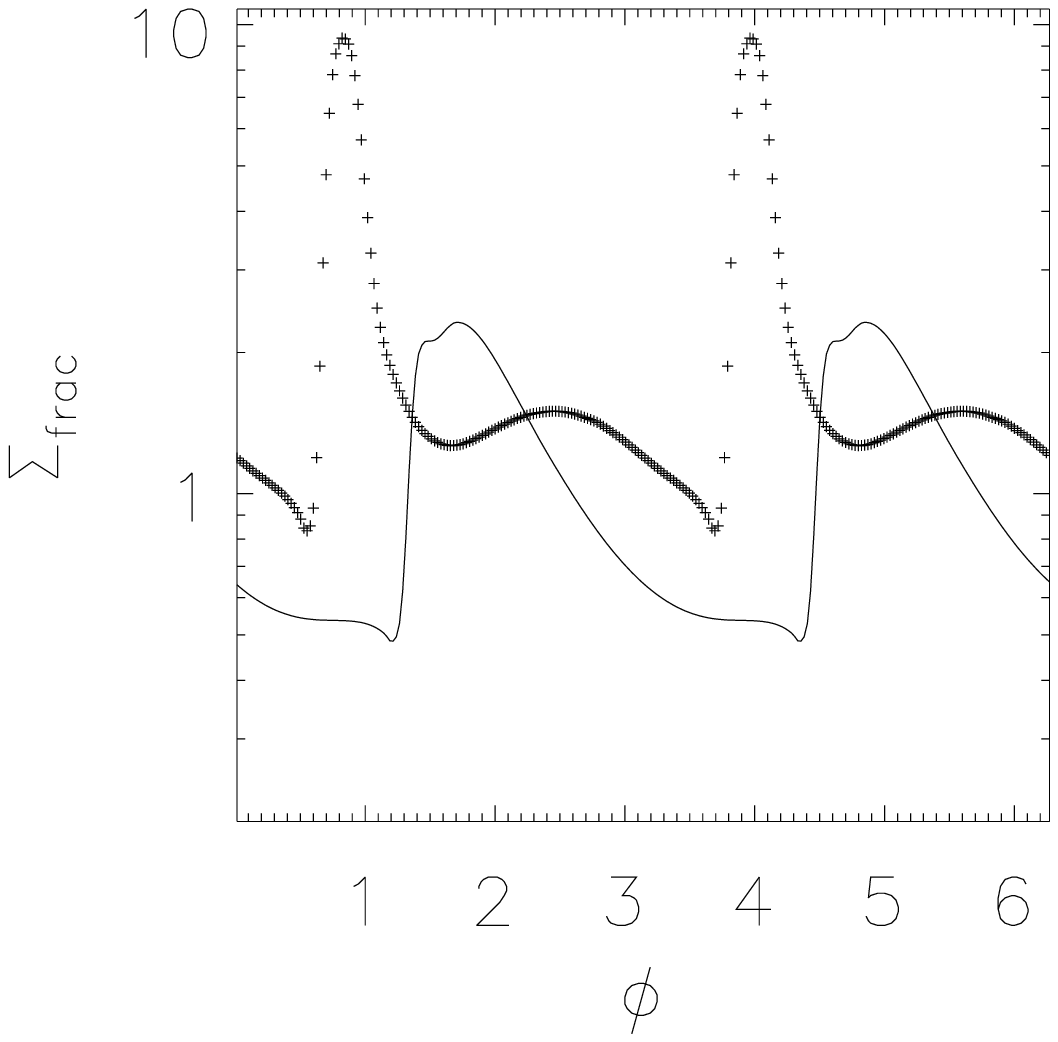,height=2.5in,width=2.5in}
\psfig{file=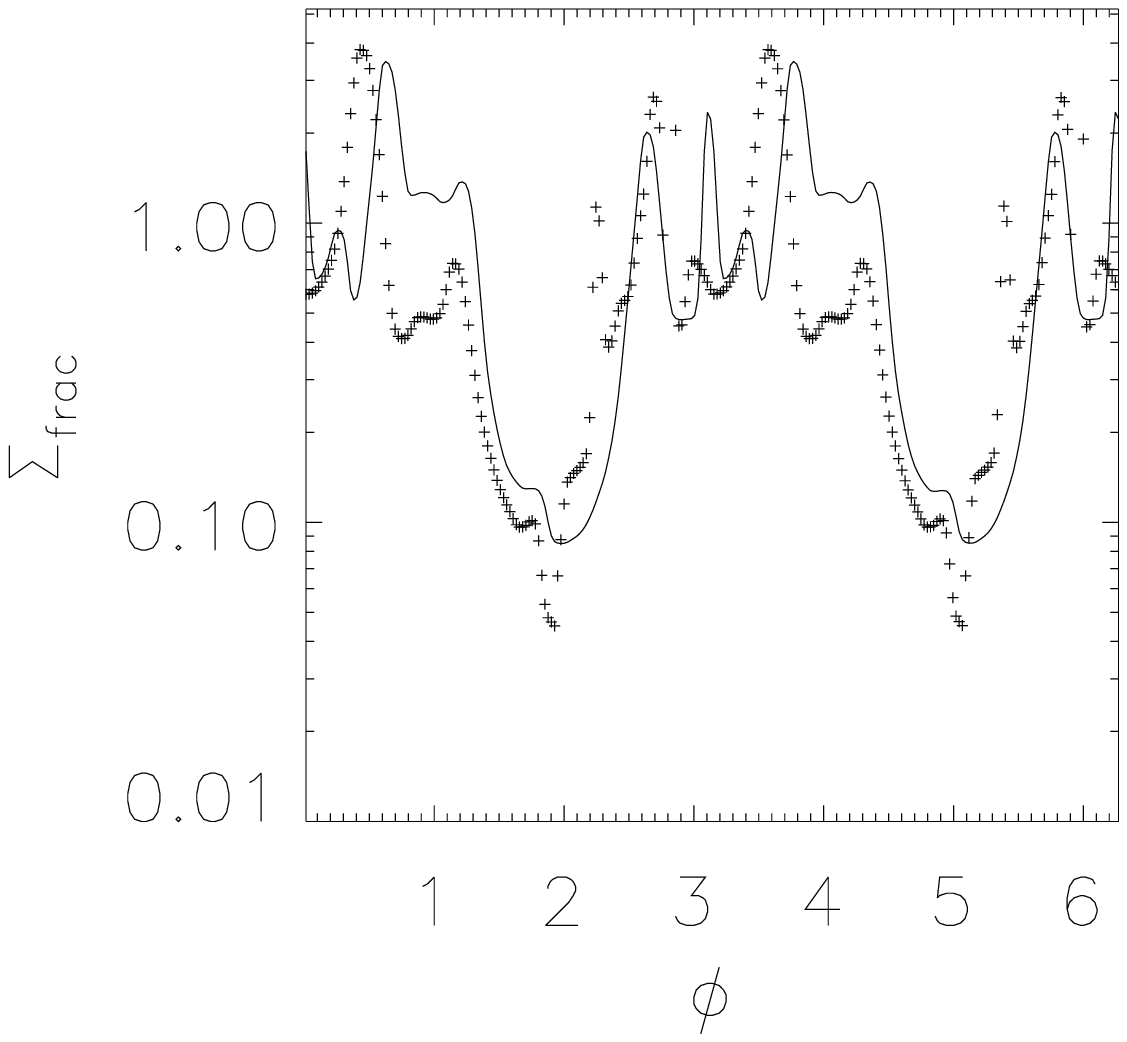,height=2.5in,width=2.5in}}
\end{center}
\caption{(a)-(c) Snapshot of model LSYL3 at 573 Myr, 1.91 Gyr, and 2.87 Gyr,
(d)-(f) Azimuthal cuts at $\xi$=1.3 and $\xi$=1.17 at 573 Myr, 1.91 Gyr, and 2.87 Gyr}
\end{figure}

\begin{deluxetable}{lcclc}
\tablewidth{0pt}
\tablecolumns{5}
\tablecaption{Parameters of Short-Term Simulations}
\tablehead{\colhead{Model} &
\colhead{$Q_{o}$} & \colhead{F} & \colhead{$\mathcal{F}$} & \colhead{$c_{g}/v_{rot}$}}
\startdata
H1   & 1.3        & 0.1 & 2.5\% & 4.6\% \\
H2   & 1.3        & 0.1 & 5\%   & 4.6\% \\
H3   & $\infty$   & 0.1 & 5\%   & 4.6\% \\
HSYL & 2.48       & 0.1 & 7\%   & 8.8\% \\
H5   & 2.48       & 0.1 & 15\%  & 8.8\% \\
H6   & $\infty$   & 0.1 & 15\%  & 8.8\% \\
\enddata
\end{deluxetable}

\begin{deluxetable}{lcclc} 
\tablewidth{0pt}
\tablecolumns{5}
\tablecaption{Parameters of Long-Term Simulations}
\tablehead{\colhead{Model} & \colhead{$Q_{o}$} & \colhead{F} & \colhead{$\mathcal{F}$} & \colhead{$c_{g}/v_{rot}$}}
\startdata
L1 & 1.3 & 0.1 & 1.3\% & 4.6\%\\
LSYL1 & 1.3 & 0.1 & 1.5\%  & 4.6\%\\  
L2 & 2.48 & 0.1 & 3.5\%  & 8.8\% \\ 
LSYL2 & 2.48 & 0.1 & 3.5\%  & 8.8\% \\   
L3 & 2.48 & 0.1 & 5\% & 8.8\%\\
LSYL3 & 2.48 & 0.1 & 5\% & 8.8\% 
\enddata
\end{deluxetable}

\begin{deluxetable}{lc}
\tablewidth{0pt}
\tablecolumns{2}
\tablecaption{Resonance Parameters for low-Q models}
\tablehead{\colhead{$(\frac{m(\Omega_{p}-\Omega)}{\kappa})^{2}$} & \colhead{r}} 
\startdata
0.16 & 1.5  \\
0.18 & 1.52 \\
0.21 & 1.54 \\
0.24 & 1.56 \\
0.26 & 1.58 \\
0.29 & 1.6  \\
0.31 & 1.62 \\
0.35 & 1.64 \\
0.39 & 1.66 \\
0.42 & 1.68 \\
0.45 & 1.7  
\enddata
\end{deluxetable}

\begin{deluxetable}{lc}
\tablewidth{0pt}
\tablecolumns{2}
\tablecaption{Resonance Parameters for high-Q models}
\tablehead{\colhead{$(\frac{m(\Omega_{p}-\Omega)}{\kappa})^{2}$} & \colhead{r}}  
\startdata
0.25 & 1.17  \\
0.24 & 1.18 \\
0.23 & 1.2 \\
0.22 & 1.21 \\
0.20 & 1.23 \\
0.19 & 1.24  \\
0.17 & 1.26 \\
0.16 & 1.28 \\
0.15 & 1.29 \\
0.14 & 1.3 \\
0.13 & 1.32  
\enddata
\end{deluxetable}

\end{document}